\documentclass[submission,copyright,creativecommons]{eptcs}

\providecommand{\event}{MSFP 2020}

\usepackage[style=numeric,natbib=true]{biblatex}
\makeatletter
\@ifundefined{lhs2tex.lhs2tex.sty.read}%
  {\@namedef{lhs2tex.lhs2tex.sty.read}{}%
   \newcommand\SkipToFmtEnd{}%
   \newcommand\EndFmtInput{}%
   \long\def\SkipToFmtEnd#1\EndFmtInput{}%
  }\SkipToFmtEnd

\newcommand\ReadOnlyOnce[1]{\@ifundefined{#1}{\@namedef{#1}{}}\SkipToFmtEnd}
\usepackage{amstext}
\usepackage{amssymb}
\usepackage{stmaryrd}
\DeclareFontFamily{OT1}{cmtex}{}
\DeclareFontShape{OT1}{cmtex}{m}{n}
  {<5><6><7><8>cmtex8
   <9>cmtex9
   <10><10.95><12><14.4><17.28><20.74><24.88>cmtex10}{}
\DeclareFontShape{OT1}{cmtex}{m}{it}
  {<-> ssub * cmtt/m/it}{}

\DeclareFontShape{OT1}{cmtt}{bx}{n}
  {<5><6><7><8>cmtt8
   <9>cmbtt9
   <10><10.95><12><14.4><17.28><20.74><24.88>cmbtt10}{}
\DeclareFontShape{OT1}{cmtex}{bx}{n}
  {<-> ssub * cmtt/bx/n}{}

\newcommand{\Conid}[1]{\mathit{#1}}
\newcommand{\Varid}[1]{\mathit{#1}}
\newcommand{\anonymous}{\kern0.06em \vbox{\hrule\@width.5em}}
\newcommand{\plus}{\mathbin{+\!\!\!+}}
\newcommand{\bind}{\mathbin{>\!\!\!>\mkern-6.7mu=}}

\usepackage{polytable}

\@ifundefined{mathindent}%
  {\newdimen\mathindent\mathindent\leftmargini}%
  {}%

\def\resethooks{%
  \global\let\SaveRestoreHook\empty
  \global\let\ColumnHook\empty}
\newcommand*{\savecolumns}[1][default]%
  {\g@addto@macro\SaveRestoreHook{\savecolumns[#1]}}
\newcommand*{\restorecolumns}[1][default]%
  {\g@addto@macro\SaveRestoreHook{\restorecolumns[#1]}}
\newcommand*{\aligncolumn}[2]%
  {\g@addto@macro\ColumnHook{\column{#1}{#2}}}

\resethooks

\newcommand{\onelinecommentchars}{\quad-{}- }
\newcommand{\commentbeginchars}{\enskip\{-}
\newcommand{\commentendchars}{-\}\enskip}

\newcommand{\visiblecomments}{%
  \let\onelinecomment=\onelinecommentchars
  \let\commentbegin=\commentbeginchars
  \let\commentend=\commentendchars}

\newcommand{\invisiblecomments}{%
  \let\onelinecomment=\empty
  \let\commentbegin=\empty
  \let\commentend=\empty}

\visiblecomments

\newlength{\blanklineskip}
\newcommand{\hsindent}[1]{\quad}%
\let\hspre\empty
\let\hspost\empty

\EndFmtInput
\makeatother
\ReadOnlyOnce{polycode.fmt}%
\makeatletter

\newcommand{\hsnewpar}[1]%
  {{\parskip=0pt\parindent=0pt\par\vskip #1\noindent}}

\newcommand{\hscodestyle}{}

\newcommand{\sethscode}[1]%
  {\expandafter\let\expandafter\hscode\csname #1\endcsname
   \expandafter\let\expandafter\endhscode\csname end#1\endcsname}

  {\par\noindent
   \advance\leftskip\mathindent
   \hscodestyle
   \let\\=\@normalcr
   \let\hspre\(\let\hspost\)%
   \pboxed}%
  {\endpboxed\)%
   \par\noindent
   \ignorespacesafterend}

  {\hsnewpar\abovedisplayskip
   \advance\leftskip\mathindent
   \hscodestyle
   \let\hspre\(\let\hspost\)%
   \pboxed}%
  {\endpboxed%
   \hsnewpar\belowdisplayskip
   \ignorespacesafterend}

  {\hsnewpar\abovedisplayskip
   \advance\leftskip\mathindent
   \hscodestyle
   \let\\=\@normalcr
   \(\pboxed}%
  {\endpboxed\)%
   \hsnewpar\belowdisplayskip
   \ignorespacesafterend}

\newcommand{\plainhs}{\sethscode{plainhscode}}

\plainhs

  {\hsnewpar\abovedisplayskip
   \advance\leftskip\mathindent
   \hscodestyle
   \let\\=\@normalcr
   \(\parray}%
  {\endparray\)%
   \hsnewpar\belowdisplayskip
   \ignorespacesafterend}

  {\parray}{\endparray}

  {\(\parray}{\endparray\)}

\def\codeframewidth{\arrayrulewidth}
\RequirePackage{calc}

  {\parskip=\abovedisplayskip\par\noindent
   \hscodestyle
   \arrayrulewidth=\codeframewidth
   \tabular{@{}|p{\linewidth-2\arraycolsep-2\arrayrulewidth-2pt}|@{}}%
   \hline\framedhslinecorrect\\{-1.5ex}%
   \let\endoflinesave=\\
   \let\\=\@normalcr
   \(\pboxed}%
  {\endpboxed\)%
   \framedhslinecorrect\endoflinesave{.5ex}\hline
   \endtabular
   \parskip=\belowdisplayskip\par\noindent
   \ignorespacesafterend}

\newcommand{\framedhslinecorrect}[2]%
  {#1[#2]}

  {\(\def\column##1##2{}%
   \let\>\undefined\let\<\undefined\let\\\undefined
   \newcommand\>[1][]{}\newcommand\<[1][]{}\newcommand\\[1][]{}%
   \def\fromto##1##2##3{##3}%
   }{\) }%

  {\let\orighscode=\hscode
   \let\origendhscode=\endhscode
   \def\endhscode{\def\hscode{\endgroup\def\@currenvir{hscode}\\}\begingroup}
   \orighscode\def\hscode{\endgroup\def\@currenvir{hscode}}}%
  {\origendhscode
   \global\let\hscode=\orighscode
   \global\let\endhscode=\origendhscode}%

\makeatother
\EndFmtInput
\ReadOnlyOnce{agda.fmt}%

\newcommand\Keyword[1]{\mathsf{\mathbf{#1}}}
\EndFmtInput

\usepackage{amsmath}
\usepackage{xcolor}

\usepackage{newunicodechar}
\newunicodechar{⊥}{\ensuremath{\bot}}
\newunicodechar{·}{\ensuremath{\cdot}}
\newunicodechar{∀}{\ensuremath{\forall}}
\newunicodechar{∈}{\ensuremath{\in}}
\newunicodechar{λ}{\ensuremath{\lambda}}
\newunicodechar{∣}{\ensuremath{\mid}}
\newunicodechar{¬}{\ensuremath{\neg}}
\newunicodechar{⊑}{\ensuremath{\sqsubseteq}}
\newunicodechar{≟}{\ensuremath{\stackrel{?}{=}}}
\newunicodechar{⊤}{\ensuremath{\top}}
\newunicodechar{∧}{\ensuremath{\wedge}}
\newunicodechar{₁}{\ensuremath{_1}}
\newunicodechar{₂}{\ensuremath{_2}}

\newcommand{\N}{\mathbb{N}}

\newtheorem{Def}{Definition}

\begin{document}

\title{Combining predicate transformer semantics for effects:\\a case study in parsing regular languages}
\def\titlerunning{Combining predicate transformer semantics for effects}
\def\authorrunning{Anne Baanen and Wouter Swierstra}
\author{Anne Baanen
  \institute{Vrije Universiteit Amsterdam}
    \and
    Wouter Swierstra
  \institute{Utrecht University}
  }
\maketitle              %

\begin{abstract}
This paper describes how to verify a parser for regular expressions in a functional programming language using predicate transformer semantics for a variety of effects.
Where our previous work in this area focused on the semantics for a
single effect, parsing requires a combination of effects:
non-determinism, general recursion and mutable state.  Reasoning about
such combinations of effects is notoriously difficult, yet our
approach using predicate transformers enables the careful separation
of program syntax, correctness proofs and termination proofs.
\end{abstract}

\section{Introduction}
\label{sec:intro}

There is a significant body of work on parsing using combinators
in functional programming langua\-ges~\cite{list-of-successes, hutton, functional-parsers, swierstra-duponcheel, leijen2001parsec, efficient-combinator-parsers, parser-combinators-for-left-recursive, parsing-with-derivatives}, among many others.
Yet how can we ensure that these parsers are correct? There is notably
less work that attempts to  answer this
question~\cite{total-parser-combinators, firsov-certification-context-free-grammars}.

Reasoning about such parser combinators is not at all trivial. They
use a variety of effects: state to store the string being parsed;
non-determinism to handle backtracking; and general recursion to deal
with recursive grammars. Proof techniques, such as equational
reasoning, that are commonly used when reasoning about pure functional programs, are less
suitable when verifying effectful programs~\cite{just-do-it,hutton2008reasoning}.

In this paper, we explore a novel approach, drawing inspiration from recent
work on algebraic effects~\cite{eff, effect-handlers-in-scope,
McBride-totally-free}.  We demonstrate how to reason about all parsers
uniformly using \emph{predicate transformers}~\cite{pt-semantics-for-effects}.
We extend our previous work that uses predicate transformer semantics to reason
about a single effect, to handle the combinations of effects used by parsers.
Our semantics is modular, meaning we can introduce new effects (\ensuremath{\Conid{Rec}} in
Section \ref{sec:combinations}), semantics (\ensuremath{\Varid{hParser}} in Section \ref{sec:dmatch}) and specifications (\ensuremath{\mathit{terminates\text{-}in}} in Section \ref{sec:dmatch-correct}) when they are
needed, without having to
rework the previous definitions.  In particular, our careful treatment of
general recursion lets us separate partial correctness
from the proof of termination cleanly. Most existing proofs require combinators to
guarantee that the string being parsed decreases, conflating these two issues.

In particular, the sections of this paper make the following contributions:
\begin{itemize}
\item After quickly revisiting our previous work on predicate
  transformer semantics for effects (Section~\ref{sec:recap}), we show how the
  non-recursive fragment of regular expressions can be correctly
  parsed using non-determinism (Section \ref{sec:regex-nondet}).
\item By combining non-determinism with general recursion (Section \ref{sec:combinations}),
  support for the Kleene star can be added without compromising our previous definitions.
\item Although the resulting parser is not guaranteed to terminate,
  we can define another implementation using Brzozowski derivatives (Section \ref{sec:dmatch}),
  introducing an additional effect and its semantics in the process.
\item Finally, we show that the derivative-based implementation terminates
  and \emph{refines} the original parser (Section \ref{sec:dmatch-correct}).

\ifdefined\includeCFGs
\item Next, we show how this approach may be extended to handle
  context-free languages. To do so, we show how to write parsers using
  algebraic effects (Section \ref{sec:parser}), and map grammars to parsers (Section
  \ref{sec:fromProductions}). Once again, we can cleanly separate the proofs of partial
  correctness (Section \ref{sec:partialCorrectness}) and termination (Section \ref{sec:fromProds-terminates}).
\fi
\end{itemize}

The goal of our work is not so much the verification of a parser for regular languages,
which has been done before~\cite{harper-regex, intrinsic-verification-regex}.
Instead, we aim to illustrate the steps of incrementally developing and verifying a parser
using predicate transformers and algebraic effects.
This work is in the spirit of a Theoretical Pearl~\cite{harper-regex}:
we begin by defining a \ensuremath{\Varid{match}} function that does not terminate. The remainder of the paper
then shows how to fix this function, without having to redo the complete proof of correctness.

All the programs and proofs in this paper are written in the dependently typed language Agda~\cite{agda-thesis}.
The full source code, including lemmas we have chosen to omit for sake of readability,
is available online.\footnote{\url{https://github.com/Vierkantor/refinement-parsers}}
Apart from postulating function extensionality,
we remain entirely within Agda's default theory.

\section{Recap: algebraic effects and predicate transformers}
\label{sec:recap}
Algebraic effects separate the \emph{syntax} and \emph{semantics} of
effectful operations. In this paper, we will model them by taking the
free monad over a given signature~\cite{extensible-effects}, describing certain
operations. Signatures are represented by the type \ensuremath{\Conid{Sig}}, as follows:
\begin{hscode}\SaveRestoreHook
\column{B}{@{}>{\hspre}l<{\hspost}@{}}%
\column{3}{@{}>{\hspre}l<{\hspost}@{}}%
\column{5}{@{}>{\hspre}l<{\hspost}@{}}%
\column{E}{@{}>{\hspre}l<{\hspost}@{}}%
\>[B]{}\Keyword{record}\;\Conid{Sig}\;\mathbin{:}\;\Conid{Set}\;\Keyword{where}{}\<[E]%
\\
\>[B]{}\hsindent{3}{}\<[3]%
\>[3]{}\Keyword{constructor}\;\Varid{mkSig}{}\<[E]%
\\
\>[B]{}\hsindent{3}{}\<[3]%
\>[3]{}\Keyword{field}{}\<[E]%
\\
\>[3]{}\hsindent{2}{}\<[5]%
\>[5]{}\Conid{C}\;\mathbin{:}\;\Conid{Set}{}\<[E]%
\\
\>[3]{}\hsindent{2}{}\<[5]%
\>[5]{}\Conid{R}\;\mathbin{:}\;\Conid{C}\;\to\;\Conid{Set}{}\<[E]%
\ColumnHook
\end{hscode}\resethooks
This is Agda syntax for defining a type \ensuremath{\Conid{Sig}} with constructor \ensuremath{\Varid{mkSig}\;\mathbin{:}\;(\Conid{C}\;\mathbin{:}\;\Conid{Set})\;\to\;(\Conid{C}\;\to\;\Conid{Set})\;\to\;\Conid{Sig}} and two fields, \ensuremath{\Conid{C}\;\mathbin{:}\;\Conid{Sig}\;\to\;\Conid{Set}} and \ensuremath{\Conid{R}\;\mathbin{:}\;(\Varid{e}\;\mathbin{:}\;\Conid{Sig})\;\to\;\Conid{C}\;\Varid{e}\;\to\;\Conid{Set}}.
Here the type \ensuremath{\Conid{C}} contains the `commands', or effectful operations
that a given effect supports. For each command \ensuremath{\Varid{c}\;\mathbin{:}\;\Conid{C}}, the type \ensuremath{\Conid{R}\;\Varid{c}}
describes the possible responses.
The structure on a signature is that of a \emph{container}~\cite{categories-of-containers}.
The following signature describes two commands: the
non-deterministic choice between two values, \ensuremath{\Conid{Choice}}; and a failure
operator, \ensuremath{\Conid{Fail}}. The response type \ensuremath{\Conid{RNondet}} is defined by pattern matching on the command.
If the command is \ensuremath{\Conid{Choice}}, the response is a \ensuremath{\Conid{Bool}}; the \ensuremath{\Conid{Fail}} command gives no response, indicated by the empty type \ensuremath{\bot}.
\begin{hscode}\SaveRestoreHook
\column{B}{@{}>{\hspre}l<{\hspost}@{}}%
\column{3}{@{}>{\hspre}l<{\hspost}@{}}%
\column{11}{@{}>{\hspre}l<{\hspost}@{}}%
\column{17}{@{}>{\hspre}l<{\hspost}@{}}%
\column{E}{@{}>{\hspre}l<{\hspost}@{}}%
\>[B]{}\Keyword{data}\;\Conid{CNondet}\;\mathbin{:}\;\Conid{Set}\;\Keyword{where}{}\<[E]%
\\
\>[B]{}\hsindent{3}{}\<[3]%
\>[3]{}\Conid{Choice}\;{}\<[11]%
\>[11]{}\mathbin{:}\;\Conid{CNondet}{}\<[E]%
\\
\>[B]{}\hsindent{3}{}\<[3]%
\>[3]{}\Conid{Fail}\;{}\<[11]%
\>[11]{}\mathbin{:}\;\Conid{CNondet}{}\<[E]%
\\[\blanklineskip]%
\>[B]{}\Conid{RNondet}\;\mathbin{:}\;\Conid{CNondet}\;\to\;\Conid{Set}{}\<[E]%
\\
\>[B]{}\Conid{RNondet}\;\Conid{Choice}\;{}\<[17]%
\>[17]{}\mathrel{=}\;\Conid{Bool}{}\<[E]%
\\
\>[B]{}\Conid{RNondet}\;\Conid{Fail}\;{}\<[17]%
\>[17]{}\mathrel{=}\;\bot{}\<[E]%
\\[\blanklineskip]%
\>[B]{}\Conid{Nondet}\;\mathrel{=}\;\Varid{mkSig}\;\Conid{CNondet}\;\Conid{RNondet}{}\<[E]%
\ColumnHook
\end{hscode}\resethooks
We represent effectful programs that use a particular effect using the
corresponding \emph{free monad}:
\begin{hscode}\SaveRestoreHook
\column{B}{@{}>{\hspre}l<{\hspost}@{}}%
\column{3}{@{}>{\hspre}l<{\hspost}@{}}%
\column{5}{@{}>{\hspre}l<{\hspost}@{}}%
\column{10}{@{}>{\hspre}l<{\hspost}@{}}%
\column{E}{@{}>{\hspre}l<{\hspost}@{}}%
\>[3]{}\Keyword{data}\;\Conid{Free}\;(\Varid{e}\;\mathbin{:}\;\Conid{Sig})\;(\Varid{a}\;\mathbin{:}\;\Conid{Set})\;\mathbin{:}\;\Conid{Set}\;\Keyword{where}{}\<[E]%
\\
\>[3]{}\hsindent{2}{}\<[5]%
\>[5]{}\Conid{Pure}\;\mathbin{:}\;\Varid{a}\;\to\;\Conid{Free}\;\Varid{e}\;\Varid{a}{}\<[E]%
\\
\>[3]{}\hsindent{2}{}\<[5]%
\>[5]{}\Conid{Op}\;{}\<[10]%
\>[10]{}\mathbin{:}\;(\Varid{c}\;\mathbin{:}\;\Conid{C}\;\Varid{e})\;\to\;(\Conid{R}\;\Varid{e}\;\Varid{c}\;\to\;\Conid{Free}\;\Varid{e}\;\Varid{a})\;\to\;\Conid{Free}\;\Varid{e}\;\Varid{a}{}\<[E]%
\ColumnHook
\end{hscode}\resethooks
This gives a monad, with the bind operator defined as follows.
\begin{hscode}\SaveRestoreHook
\column{B}{@{}>{\hspre}l<{\hspost}@{}}%
\column{3}{@{}>{\hspre}l<{\hspost}@{}}%
\column{13}{@{}>{\hspre}l<{\hspost}@{}}%
\column{E}{@{}>{\hspre}l<{\hspost}@{}}%
\>[3]{}\_\!\bind\!\_\;\mathbin{:}\;\!\!\;\Conid{Free}\;\Varid{e}\;\Varid{a}\;\to\;(\Varid{a}\;\to\;\Conid{Free}\;\Varid{e}\;\Varid{b})\;\to\;\Conid{Free}\;\Varid{e}\;\Varid{b}{}\<[E]%
\\
\>[3]{}\Conid{Pure}\;\Varid{x}\;{}\<[13]%
\>[13]{}\bind \;\Varid{f}\;\mathrel{=}\;\Varid{f}\;\Varid{x}{}\<[E]%
\\
\>[3]{}\Conid{Op}\;\Varid{c}\;\Varid{k}\;{}\<[13]%
\>[13]{}\bind \;\Varid{f}\;\mathrel{=}\;\Conid{Op}\;\Varid{c}\;(\lambda\;\Varid{x}\;\to\;\Varid{k}\;\Varid{x}\;\bind \;\Varid{f}){}\<[E]%
\ColumnHook
\end{hscode}\resethooks
To facilitate programming with effects, we define the following smart
constructors, sometimes referred to as \emph{generic effects} in the
literature~\cite{algebraic-operations-and-generic-effects}:
\begin{hscode}\SaveRestoreHook
\column{B}{@{}>{\hspre}l<{\hspost}@{}}%
\column{3}{@{}>{\hspre}l<{\hspost}@{}}%
\column{E}{@{}>{\hspre}l<{\hspost}@{}}%
\>[3]{}\Varid{fail}\;\mathbin{:}\;\!\!\;\Conid{Free}\;\Conid{Nondet}\;\Varid{a}{}\<[E]%
\\
\>[3]{}\Varid{fail}\;\mathrel{=}\;\Conid{Op}\;\Conid{Fail}\;(\lambda\;()){}\<[E]%
\\
\>[3]{}\Varid{choice}\;\mathbin{:}\;\!\!\;\Conid{Free}\;\Conid{Nondet}\;\Varid{a}\;\to\;\Conid{Free}\;\Conid{Nondet}\;\Varid{a}\;\to\;\Conid{Free}\;\Conid{Nondet}\;\Varid{a}{}\<[E]%
\\
\>[3]{}\Varid{choice}\;\Conid{S₁}\;\Conid{S₂}\;\mathrel{=}\;\Conid{Op}\;\Conid{Choice}\;(\lambda\;\Varid{b}\;\to\;\textrm{\bfseries if}\;\Varid{b}\;\textrm{\bfseries then}\;\Conid{S₁}\;\textrm{\bfseries else}\;\Conid{S₂}){}\<[E]%
\ColumnHook
\end{hscode}\resethooks
The empty parentheses \ensuremath{()} in the definition of \ensuremath{\Varid{fail}} are Agda syntax for an argument
in an uninhabited type, hence no body for the lambda is provided.

In this paper, we will assign \emph{semantics} to effectful programs
by mapping them to \emph{predicate transformers}.
Each semantics will be computed by a fold over the free monad, mapping
some predicate \ensuremath{\Conid{P}\;\mathbin{:}\;\Varid{a}\;\to\;\Conid{Set}} to a predicate of the entire computation of type \ensuremath{\Conid{Free}\;(\Varid{mkSig}\;\Conid{C}\;\Conid{R})\;\Varid{a}\;\to\;\Conid{Set}}.
\begin{hscode}\SaveRestoreHook
\column{B}{@{}>{\hspre}l<{\hspost}@{}}%
\column{3}{@{}>{\hspre}l<{\hspost}@{}}%
\column{24}{@{}>{\hspre}l<{\hspost}@{}}%
\column{E}{@{}>{\hspre}l<{\hspost}@{}}%
\>[3]{}\llbracket \_ \rrbracket\;\mathbin{:}\;\!\!\;\!\!\;\!\!\;(\Varid{alg}\;\mathbin{:}\;(\Varid{c}\;\mathbin{:}\;\Conid{C})\;\to\;(\Conid{R}\;\Varid{c}\;\to\;\Conid{Set})\;\to\;\Conid{Set})\;\to\;\Conid{Free}\;(\Varid{mkSig}\;\Conid{C}\;\Conid{R})\;\Varid{a}\;\to\;(\Varid{a}\;\to\;\Conid{Set})\;\to\;\Conid{Set}{}\<[E]%
\\
\>[3]{}\llbracket\Conid{Pure}\;\Varid{x}\rrbracket_{\Varid{alg}}\;\Conid{P}\;{}\<[24]%
\>[24]{}\mathrel{=}\;\Conid{P}\;\Varid{x}{}\<[E]%
\\
\>[3]{}\llbracket\Conid{Op}\;\Varid{c}\;\Varid{k}\rrbracket_{\Varid{alg}}\;\Conid{P}\;{}\<[24]%
\>[24]{}\mathrel{=}\;\Varid{alg}\;\Varid{c}\;(\lambda\;\Varid{x}\;\to\;\llbracket\Varid{k}\;\Varid{x}\rrbracket_{\Varid{alg}}\;\Conid{P}){}\<[E]%
\ColumnHook
\end{hscode}\resethooks
The \emph{predicate transformer} nature of these semantics
becomes evident when we assume the type of responses \ensuremath{\Conid{R}} does not depend on the command \ensuremath{\Varid{c}\;\mathbin{:}\;\Conid{C}}.
The type of \ensuremath{\Varid{alg}\;\mathbin{:}\;(\Varid{c}\;\mathbin{:}\;\Conid{C})\;\to\;(\Conid{R}\;\Varid{c}\;\to\;\Conid{Set})\;\to\;\Conid{Set}} then becomes \ensuremath{\Conid{C}\;\to\;(\Conid{R}\;\to\;\Conid{Set})\;\to\;\Conid{Set}},
which is isomorphic to \ensuremath{(\Conid{R}\;\to\;\Conid{Set})\;\to\;(\Conid{C}\;\to\;\Conid{Set})}.
Thus, \ensuremath{\Varid{alg}} has the form of a predicate transformer from postconditions of type
\ensuremath{\Conid{R}\;\to\;\Conid{Set}} into preconditions of type \ensuremath{\Conid{C}\;\to\;\Conid{Set}}.

Two considerations lead us to define the types as \ensuremath{\Varid{alg}\;\mathbin{:}\;(\Varid{c}\;\mathbin{:}\;\Conid{C})\;\to\;(\Conid{R}\;\Varid{c}\;\to\;\Conid{Set})\;\to\;\Conid{Set}}
and \ensuremath{\llbracket \_ \rrbracket_{\Varid{alg}}\;\mathbin{:}\;\Conid{Free}\;(\Varid{mkSig}\;\Conid{C}\;\Conid{R})\;\Varid{a}\;\to\;(\Varid{a}\;\to\;\Conid{Set})\;\to\;\Conid{Set}}.
By passing the command \ensuremath{\Varid{c}\;\mathbin{:}\;\Conid{C}} as first argument to \ensuremath{\Varid{alg}}, we allow \ensuremath{\Conid{R}} to depend on \ensuremath{\Varid{c}}.
Moreover, \ensuremath{\llbracket \_ \rrbracket_{\Varid{alg}}} computes semantics,
so it should take a program \ensuremath{\Conid{S}\;\mathbin{:}\;\Conid{Free}\;(\Varid{mkSig}\;\Conid{C}\;\Conid{R})\;\Varid{a}} as its argument
and return the semantics of \ensuremath{\Conid{S}}, which is then of type \ensuremath{(\Varid{a}\;\to\;\Conid{Set})\;\to\;\Conid{Set}}.

In the case of non-determinism, for example, we may want to require that a given
predicate \ensuremath{\Conid{P}} holds for all possible results that may be returned:
\begin{hscode}\SaveRestoreHook
\column{B}{@{}>{\hspre}l<{\hspost}@{}}%
\column{3}{@{}>{\hspre}l<{\hspost}@{}}%
\column{16}{@{}>{\hspre}l<{\hspost}@{}}%
\column{19}{@{}>{\hspre}l<{\hspost}@{}}%
\column{E}{@{}>{\hspre}l<{\hspost}@{}}%
\>[3]{}\Varid{ptAll}\;\mathbin{:}\;(\Varid{c}\;\mathbin{:}\;\Conid{CNondet})\;\to\;(\Conid{RNondet}\;\Varid{c}\;\to\;\Conid{Set})\;\to\;\Conid{Set}{}\<[E]%
\\
\>[3]{}\Varid{ptAll}\;\Conid{Fail}\;{}\<[16]%
\>[16]{}\Conid{P}\;{}\<[19]%
\>[19]{}\mathrel{=}\;\top{}\<[E]%
\\
\>[3]{}\Varid{ptAll}\;\Conid{Choice}\;\Conid{P}\;{}\<[19]%
\>[19]{}\mathrel{=}\;\Conid{P}\;\Conid{True}\;\mathbin{\wedge}\;\Conid{P}\;\Conid{False}{}\<[E]%
\ColumnHook
\end{hscode}\resethooks
A different semantics may instead require that \ensuremath{\Conid{P}} holds on any of the return values:
\begin{hscode}\SaveRestoreHook
\column{B}{@{}>{\hspre}l<{\hspost}@{}}%
\column{3}{@{}>{\hspre}l<{\hspost}@{}}%
\column{16}{@{}>{\hspre}l<{\hspost}@{}}%
\column{19}{@{}>{\hspre}l<{\hspost}@{}}%
\column{E}{@{}>{\hspre}l<{\hspost}@{}}%
\>[3]{}\Varid{ptAny}\;\mathbin{:}\;(\Varid{c}\;\mathbin{:}\;\Conid{CNondet})\;\to\;(\Conid{RNondet}\;\Varid{c}\;\to\;\Conid{Set})\;\to\;\Conid{Set}{}\<[E]%
\\
\>[3]{}\Varid{ptAny}\;\Conid{Fail}\;{}\<[16]%
\>[16]{}\Conid{P}\;{}\<[19]%
\>[19]{}\mathrel{=}\;\bot{}\<[E]%
\\
\>[3]{}\Varid{ptAny}\;\Conid{Choice}\;\Conid{P}\;{}\<[19]%
\>[19]{}\mathrel{=}\;\Conid{P}\;\Conid{True}\;\mathbin{\vee}\;\Conid{P}\;\Conid{False}{}\<[E]%
\ColumnHook
\end{hscode}\resethooks

Predicate transformers provide a single semantic domain to relate
programs and specifications~\cite{prog-from-spec}.
Throughout this paper, we will consider specifications consisting of a
pre- and postcondition:
\begin{samepage}
\begin{hscode}\SaveRestoreHook
\column{B}{@{}>{\hspre}l<{\hspost}@{}}%
\column{3}{@{}>{\hspre}l<{\hspost}@{}}%
\column{5}{@{}>{\hspre}l<{\hspost}@{}}%
\column{7}{@{}>{\hspre}l<{\hspost}@{}}%
\column{E}{@{}>{\hspre}l<{\hspost}@{}}%
\>[3]{}\Keyword{record}\;\Conid{Spec}\;(\Varid{a}\;\mathbin{:}\;\Conid{Set})\;\mathbin{:}\;\Conid{Set}\;\Keyword{where}{}\<[E]%
\\
\>[3]{}\hsindent{2}{}\<[5]%
\>[5]{}\Keyword{constructor}\;[\_,\_]{}\<[E]%
\\
\>[3]{}\hsindent{2}{}\<[5]%
\>[5]{}\Keyword{field}{}\<[E]%
\\
\>[5]{}\hsindent{2}{}\<[7]%
\>[7]{}\Varid{pre}\;\mathbin{:}\;\Conid{Set}{}\<[E]%
\\
\>[5]{}\hsindent{2}{}\<[7]%
\>[7]{}\Varid{post}\;\mathbin{:}\;\Varid{a}\;\to\;\Conid{Set}{}\<[E]%
\ColumnHook
\end{hscode}\resethooks
\end{samepage}
Inspired by work on the refinement calculus, we can assign a predicate
transformer semantics to specifications as follows:
\begin{hscode}\SaveRestoreHook
\column{B}{@{}>{\hspre}l<{\hspost}@{}}%
\column{3}{@{}>{\hspre}l<{\hspost}@{}}%
\column{E}{@{}>{\hspre}l<{\hspost}@{}}%
\>[3]{}\llbracket \_ , \_ \rrbracket_{\text{spec}}\;\mathbin{:}\;\!\!\;\Conid{Spec}\;\Varid{a}\;\to\;(\Varid{a}\;\to\;\Conid{Set})\;\to\;\Conid{Set}{}\<[E]%
\\
\>[3]{}\llbracket\Varid{pre},\Varid{post}\rrbracket_{\text{spec}}\;\Conid{P}\;\mathrel{=}\;\Varid{pre}\;\mathbin{\wedge}\;(\forall\;\Varid{o}\;\to\;\Varid{post}\;\Varid{o}\;\to\;\Conid{P}\;\Varid{o}){}\<[E]%
\ColumnHook
\end{hscode}\resethooks
This computes the `weakest precondition' necessary for a specification
to imply that the desired postcondition \ensuremath{\Conid{P}} holds. In particular, the
precondition \ensuremath{\Varid{pre}} should hold and any possible result satisfying the
postcondition \ensuremath{\Varid{post}} should imply the postcondition \ensuremath{\Conid{P}}.

Finally, we use the \emph{refinement relation} to compare programs and specifications:
\begin{hscode}\SaveRestoreHook
\column{B}{@{}>{\hspre}l<{\hspost}@{}}%
\column{3}{@{}>{\hspre}l<{\hspost}@{}}%
\column{E}{@{}>{\hspre}l<{\hspost}@{}}%
\>[3]{}\_\!\sqsubseteq\!\_\;\mathbin{:}\;\!\!\;(\Varid{pt}_1\;\Varid{pt}_2\;\mathbin{:}\;(\Varid{a}\;\to\;\Conid{Set})\;\to\;\Conid{Set})\;\to\;\Conid{Set}{}\<[E]%
\\
\>[3]{}\Varid{pt}_1\;\sqsubseteq\;\Varid{pt}_2\;\mathrel{=}\;\forall\;\Conid{P}\;\to\;\Varid{pt}_1\;\Conid{P}\;\to\;\Varid{pt}_2\;\Conid{P}{}\<[E]%
\ColumnHook
\end{hscode}\resethooks
Together with the predicate transformer semantics we have defined
above, this refinement relation can be used to relate programs to
their specifications. The refinement relation is both transitive and
reflexive.

\section{Regular languages without recursion} \label{sec:regex-nondet}
To illustrate how to reason about non-deterministic code, we will
define and verify a regular expression matcher. Initially, we will restrict
ourselves to non-recursive regular expressions; we will add recursion
in the next section.

We begin by defining the \ensuremath{\Conid{Regex}} datatype for regular expressions.
An element of this type represents the syntax of a regular expression.
\begin{hscode}\SaveRestoreHook
\column{B}{@{}>{\hspre}l<{\hspost}@{}}%
\column{3}{@{}>{\hspre}l<{\hspost}@{}}%
\column{14}{@{}>{\hspre}l<{\hspost}@{}}%
\column{E}{@{}>{\hspre}l<{\hspost}@{}}%
\>[B]{}\Keyword{data}\;\Conid{Regex}\;\mathbin{:}\;\Conid{Set}\;\Keyword{where}{}\<[E]%
\\
\>[B]{}\hsindent{3}{}\<[3]%
\>[3]{}\Conid{Empty}\;{}\<[14]%
\>[14]{}\mathbin{:}\;\Conid{Regex}{}\<[E]%
\\
\>[B]{}\hsindent{3}{}\<[3]%
\>[3]{}\Conid{Epsilon}\;{}\<[14]%
\>[14]{}\mathbin{:}\;\Conid{Regex}{}\<[E]%
\\
\>[B]{}\hsindent{3}{}\<[3]%
\>[3]{}\Conid{Singleton}\;{}\<[14]%
\>[14]{}\mathbin{:}\;\Conid{Char}\;\to\;\Conid{Regex}{}\<[E]%
\\
\>[B]{}\hsindent{3}{}\<[3]%
\>[3]{}\_\!\mathbin{\mid}\!\_\;{}\<[14]%
\>[14]{}\mathbin{:}\;\Conid{Regex}\;\to\;\Conid{Regex}\;\to\;\Conid{Regex}{}\<[E]%
\\
\>[B]{}\hsindent{3}{}\<[3]%
\>[3]{}\_\!\cdot\!\_\;{}\<[14]%
\>[14]{}\mathbin{:}\;\Conid{Regex}\;\to\;\Conid{Regex}\;\to\;\Conid{Regex}{}\<[E]%
\\
\>[B]{}\hsindent{3}{}\<[3]%
\>[3]{}\_\!\mathbin{\star}\;{}\<[14]%
\>[14]{}\mathbin{:}\;\Conid{Regex}\;\to\;\Conid{Regex}{}\<[E]%
\ColumnHook
\end{hscode}\resethooks
The \ensuremath{\Conid{Empty}} regular expression corresponds to the empty
language, while the \ensuremath{\Conid{Epsilon}} expression only matches the empty
string. Furthermore, our language for regular expressions is closed
under choice (\ensuremath{\_\!\mathbin{\mid}\!\_}), concatenation (\ensuremath{\_\!\cdot\!\_}) and linear repetition
denoted by the Kleene star (\ensuremath{\_\!\mathbin{\star}}).

The input to the regular expression matcher will be a \ensuremath{\Conid{String}} together with a \ensuremath{\Conid{Regex}} denoting the language to match the string against.
What should our matcher return?  A Boolean value is
not particularly informative; yet we also choose not to provide an
intrinsically correct definition, instead performing extrinsic
verification using our predicate transformer semantics. The \ensuremath{\mathit{Tree}}
data type below captures a potential parse tree associated with a
given regular expression:
\begin{hscode}\SaveRestoreHook
\column{B}{@{}>{\hspre}l<{\hspost}@{}}%
\column{21}{@{}>{\hspre}l<{\hspost}@{}}%
\column{E}{@{}>{\hspre}l<{\hspost}@{}}%
\>[B]{}\mathit{Tree}\;\mathbin{:}\;\Conid{Regex}\;\to\;\Conid{Set}{}\<[E]%
\\
\>[B]{}\mathit{Tree}\;\Conid{Empty}\;{}\<[21]%
\>[21]{}\mathrel{=}\;\bot{}\<[E]%
\\
\>[B]{}\mathit{Tree}\;\Conid{Epsilon}\;{}\<[21]%
\>[21]{}\mathrel{=}\;\top{}\<[E]%
\\
\>[B]{}\mathit{Tree}\;(\Conid{Singleton}\;\anonymous )\;{}\<[21]%
\>[21]{}\mathrel{=}\;\Conid{Char}{}\<[E]%
\\
\>[B]{}\mathit{Tree}\;(\Varid{l}\;\mathbin{\mid}\;\Varid{r})\;{}\<[21]%
\>[21]{}\mathrel{=}\;\Conid{Either}\;(\mathit{Tree}\;\Varid{l})\;(\mathit{Tree}\;\Varid{r}){}\<[E]%
\\
\>[B]{}\mathit{Tree}\;(\Varid{l}\;\cdot\;\Varid{r})\;{}\<[21]%
\>[21]{}\mathrel{=}\;\Conid{Pair}\;(\mathit{Tree}\;\Varid{l})\;(\mathit{Tree}\;\Varid{r}){}\<[E]%
\\
\>[B]{}\mathit{Tree}\;(\Varid{r}\;\mathbin{\star})\;{}\<[21]%
\>[21]{}\mathrel{=}\;\Conid{List}\;(\mathit{Tree}\;\Varid{r}){}\<[E]%
\ColumnHook
\end{hscode}\resethooks
In the remainder of this section, we will develop a regular expression
matcher with the following type:
\begin{hscode}\SaveRestoreHook
\column{B}{@{}>{\hspre}l<{\hspost}@{}}%
\column{3}{@{}>{\hspre}l<{\hspost}@{}}%
\column{E}{@{}>{\hspre}l<{\hspost}@{}}%
\>[3]{}\Varid{match}\;\mathbin{:}\;(\Varid{r}\;\mathbin{:}\;\Conid{Regex})\;(\Varid{xs}\;\mathbin{:}\;\Conid{String})\;\to\;\Conid{Free}\;\Conid{Nondet}\;(\mathit{Tree}\;\Varid{r}){}\<[E]%
\ColumnHook
\end{hscode}\resethooks
Before we do so, however, we will complete our specification. Although
the type above guarantees that we return a parse tree matching the
regular expression \ensuremath{\Varid{r}}, there is no relation between the tree and the
input string. To capture this relation, we define the following
\ensuremath{\Conid{Match}} data type. A value of type \ensuremath{\Conid{Match}\;\Varid{r}\;\Varid{xs}\;\Varid{t}} states that the
string \ensuremath{\Varid{xs}} is in the language given by the regular expression \ensuremath{\Varid{r}} as
witnessed by the parse tree \ensuremath{\Varid{t}}:

\begin{hscode}\SaveRestoreHook
\column{B}{@{}>{\hspre}l<{\hspost}@{}}%
\column{3}{@{}>{\hspre}l<{\hspost}@{}}%
\column{15}{@{}>{\hspre}l<{\hspost}@{}}%
\column{118}{@{}>{\hspre}l<{\hspost}@{}}%
\column{E}{@{}>{\hspre}l<{\hspost}@{}}%
\>[B]{}\Keyword{data}\;\Conid{Match}\;\mathbin{:}\;(\Varid{r}\;\mathbin{:}\;\Conid{Regex})\;\to\;\Conid{String}\;\to\;\mathit{Tree}\;\Varid{r}\;\to\;\Conid{Set}\;\Keyword{where}{}\<[E]%
\\
\>[B]{}\hsindent{3}{}\<[3]%
\>[3]{}\Conid{Epsilon}\;{}\<[15]%
\>[15]{}\mathbin{:}\;{}\<[118]%
\>[118]{}\Conid{Match}\;\Conid{Epsilon}\;\Conid{Nil}\;\Varid{tt}{}\<[E]%
\\
\>[B]{}\hsindent{3}{}\<[3]%
\>[3]{}\Conid{Singleton}\;{}\<[15]%
\>[15]{}\mathbin{:}\;\!\!\;{}\<[118]%
\>[118]{}\Conid{Match}\;(\Conid{Singleton}\;\Varid{x})\;(\Varid{x}\;::\;\Conid{Nil})\;\Varid{x}{}\<[E]%
\\
\>[B]{}\hsindent{3}{}\<[3]%
\>[3]{}\Conid{OrLeft}\;{}\<[15]%
\>[15]{}\mathbin{:}\;\!\!\;\!\!\;\!\!\;{}\<[118]%
\>[118]{}\Conid{Match}\;\Varid{l}\;\Varid{xs}\;\Varid{x}\;\to\;\Conid{Match}\;(\Varid{l}\;\mathbin{\mid}\;\Varid{r})\;\Varid{xs}\;(\Conid{Inl}\;\Varid{x}){}\<[E]%
\\
\>[B]{}\hsindent{3}{}\<[3]%
\>[3]{}\Conid{OrRight}\;{}\<[15]%
\>[15]{}\mathbin{:}\;\!\!\;\!\!\;\!\!\;{}\<[118]%
\>[118]{}\Conid{Match}\;\Varid{r}\;\Varid{xs}\;\Varid{x}\;\to\;\Conid{Match}\;(\Varid{l}\;\mathbin{\mid}\;\Varid{r})\;\Varid{xs}\;(\Conid{Inr}\;\Varid{x}){}\<[E]%
\\
\>[B]{}\hsindent{3}{}\<[3]%
\>[3]{}\Conid{Concat}\;{}\<[15]%
\>[15]{}\mathbin{:}\;\!\!\;\!\!\;\!\!\;\!\!\;{}\<[118]%
\>[118]{}\Conid{Match}\;\Varid{l}\;\Varid{ys}\;\Varid{y}\;\to\;\Conid{Match}\;\Varid{r}\;\Varid{zs}\;\Varid{z}\;\to\;\Conid{Match}\;(\Varid{l}\;\cdot\;\Varid{r})\;(\Varid{ys}\;\plus \;\Varid{zs})\;(\Varid{y}\;\Varid{,}\;\Varid{z}){}\<[E]%
\\
\>[B]{}\hsindent{3}{}\<[3]%
\>[3]{}\Conid{StarNil}\;{}\<[15]%
\>[15]{}\mathbin{:}\;\!\!\;{}\<[118]%
\>[118]{}\Conid{Match}\;(\Varid{r}\;\mathbin{\star})\;\Conid{Nil}\;\Conid{Nil}{}\<[E]%
\\
\>[B]{}\hsindent{3}{}\<[3]%
\>[3]{}\Conid{StarConcat}\;{}\<[15]%
\>[15]{}\mathbin{:}\;\!\!\;\!\!\;\!\!\;\!\!\;{}\<[118]%
\>[118]{}\Conid{Match}\;(\Varid{r}\;\cdot\;(\Varid{r}\;\mathbin{\star}))\;\Varid{xs}\;(\Varid{y}\;\Varid{,}\;\Varid{ys})\;\to\;\Conid{Match}\;(\Varid{r}\;\mathbin{\star})\;\Varid{xs}\;(\Varid{y}\;::\;\Varid{ys}){}\<[E]%
\ColumnHook
\end{hscode}\resethooks
Note that there is no constructor for \ensuremath{\Conid{Match}\;\Conid{Empty}\;\Varid{xs}\;\Varid{ms}} for any \ensuremath{\Varid{xs}}
or \ensuremath{\Varid{ms}}, as there is no way to match the \ensuremath{\Conid{Empty}} language with a
string \ensuremath{\Varid{xs}}.  Similarly, the only constructor for \ensuremath{\Conid{Match}\;\Conid{Epsilon}\;\Varid{xs}\;\Varid{ms}} is where \ensuremath{\Varid{xs}} is the empty string \ensuremath{\Conid{Nil}}. There are two
constructors that produce a \ensuremath{\Conid{Match}} for a regular expression of the
form \ensuremath{\Varid{l}\;\mathbin{\mid}\;\Varid{r}}, corresponding to the choice of matching either \ensuremath{\Varid{l}} or
\ensuremath{\Varid{r}}.

The cases for concatenation and iteration are more
interesting. Crucially the \ensuremath{\Conid{Concat}} constructor constructs a match on
the concatenation of the strings \ensuremath{\Varid{ys}} and \ensuremath{\Varid{zs}} -- although there may
be many possible ways to decompose a string into two
substrings. Finally, the two constructors for the Kleene star, \ensuremath{\Varid{r}\;\mathbin{\star}},
match zero (\ensuremath{\Conid{StarNil}}) or many (\ensuremath{\Conid{StarConcat}}) repetitions of \ensuremath{\Varid{r}}.

We will now turn our attention to the \ensuremath{\Varid{match}} function. The complete
definition, by induction on the argument regular expression, can be
found in Figure~\ref{fig:match}. Most of the cases are
straightforward---the most difficult case is that for concatenation,
where we non-deterministically consider all possible splittings of the
input string \ensuremath{\Varid{xs}} into a pair of strings \ensuremath{\Varid{ys}} and \ensuremath{\Varid{zs}}. The
\ensuremath{\Varid{allSplits}} function, defined below, computes all possible splittings:

\begin{hscode}\SaveRestoreHook
\column{B}{@{}>{\hspre}l<{\hspost}@{}}%
\column{3}{@{}>{\hspre}l<{\hspost}@{}}%
\column{5}{@{}>{\hspre}l<{\hspost}@{}}%
\column{E}{@{}>{\hspre}l<{\hspost}@{}}%
\>[3]{}\Varid{allSplits}\;\mathbin{:}\;\!\!\;(\Varid{xs}\;\mathbin{:}\;\Conid{List}\;\Varid{a})\;\to\;\Conid{Free}\;\Conid{Nondet}\;(\Conid{List}\;\Varid{a}\;\times\;\Conid{List}\;\Varid{a}){}\<[E]%
\\
\>[3]{}\Varid{allSplits}\;\Conid{Nil}\;\mathrel{=}\;\Conid{Pure}\;(\Conid{Nil}\;\Varid{,}\;\Conid{Nil}){}\<[E]%
\\
\>[3]{}\Varid{allSplits}\;(\Varid{x}\;::\;\Varid{xs})\;\mathrel{=}\;\Varid{choice}\;{}\<[E]%
\\
\>[3]{}\hsindent{2}{}\<[5]%
\>[5]{}(\Conid{Pure}\;(\Conid{Nil}\;\Varid{,}\;(\Varid{x}\;::\;\Varid{xs})))\;{}\<[E]%
\\
\>[3]{}\hsindent{2}{}\<[5]%
\>[5]{}(\Varid{allSplits}\;\Varid{xs}\;\bind \;\lambda\;\{\mskip1.5mu (\Varid{ys}\;\Varid{,}\;\Varid{zs})\;\to\;\Conid{Pure}\;((\Varid{x}\;::\;\Varid{ys})\;\Varid{,}\;\Varid{zs})\mskip1.5mu\}){}\<[E]%
\ColumnHook
\end{hscode}\resethooks
\begin{figure}
\begin{hscode}\SaveRestoreHook
\column{B}{@{}>{\hspre}l<{\hspost}@{}}%
\column{3}{@{}>{\hspre}l<{\hspost}@{}}%
\column{24}{@{}>{\hspre}l<{\hspost}@{}}%
\column{39}{@{}>{\hspre}l<{\hspost}@{}}%
\column{45}{@{}>{\hspre}l<{\hspost}@{}}%
\column{51}{@{}>{\hspre}l<{\hspost}@{}}%
\column{E}{@{}>{\hspre}l<{\hspost}@{}}%
\>[3]{}\Varid{match}\;\mathbin{:}\;(\Varid{r}\;\mathbin{:}\;\Conid{Regex})\;(\Varid{xs}\;\mathbin{:}\;\Conid{String})\;\to\;\Conid{Free}\;\Conid{Nondet}\;(\mathit{Tree}\;\Varid{r}){}\<[E]%
\\
\>[3]{}\Varid{match}\;\Conid{Empty}\;{}\<[24]%
\>[24]{}\Varid{xs}\;{}\<[39]%
\>[39]{}\mathrel{=}\;\Varid{fail}{}\<[E]%
\\
\>[3]{}\Varid{match}\;\Conid{Epsilon}\;{}\<[24]%
\>[24]{}\Conid{Nil}\;{}\<[39]%
\>[39]{}\mathrel{=}\;\Conid{Pure}\;\Varid{tt}{}\<[E]%
\\
\>[3]{}\Varid{match}\;\Conid{Epsilon}\;{}\<[24]%
\>[24]{}(\anonymous \;::\;\anonymous )\;{}\<[39]%
\>[39]{}\mathrel{=}\;\Varid{fail}{}\<[E]%
\\
\>[3]{}\Varid{match}\;(\Conid{Singleton}\;\Varid{c})\;{}\<[24]%
\>[24]{}\Conid{Nil}\;{}\<[39]%
\>[39]{}\mathrel{=}\;\Varid{fail}{}\<[E]%
\\
\>[3]{}\Varid{match}\;(\Conid{Singleton}\;\Varid{c})\;{}\<[24]%
\>[24]{}(\Varid{x}\;::\;\Conid{Nil})\;{}\<[39]%
\>[39]{}\Keyword{with}\;\Varid{c}\;\mathbin{\smash{\overset{\raisebox{-0.2em}{\tiny ?}}{=}}}\;\Varid{x}{}\<[E]%
\\
\>[3]{}\Varid{match}\;(\Conid{Singleton}\;\Varid{c})\;{}\<[24]%
\>[24]{}(\Varid{c}\;::\;\Conid{Nil})\;{}\<[39]%
\>[39]{}\mid \;\Varid{yes}\;\Varid{refl}\;{}\<[51]%
\>[51]{}\mathrel{=}\;\Conid{Pure}\;\Varid{c}{}\<[E]%
\\
\>[3]{}\Varid{match}\;(\Conid{Singleton}\;\Varid{c})\;{}\<[24]%
\>[24]{}(\Varid{x}\;::\;\Conid{Nil})\;{}\<[39]%
\>[39]{}\mid \;\Varid{no}\;\Varid{¬p}\;{}\<[51]%
\>[51]{}\mathrel{=}\;\Varid{fail}{}\<[E]%
\\
\>[3]{}\Varid{match}\;(\Conid{Singleton}\;\Varid{c})\;{}\<[24]%
\>[24]{}(\anonymous \;::\;\anonymous \;::\;\anonymous )\;{}\<[39]%
\>[39]{}\mathrel{=}\;\Varid{fail}{}\<[E]%
\\
\>[3]{}\Varid{match}\;(\Varid{l}\;\mathbin{\mid}\;\Varid{r})\;{}\<[24]%
\>[24]{}\Varid{xs}\;{}\<[39]%
\>[39]{}\mathrel{=}\;\Varid{choice}\;(\Conid{Inl}\;\langle\$\rangle\;\Varid{match}\;\Varid{l}\;\Varid{xs})\;(\Conid{Inr}\;\langle\$\rangle\;\Varid{match}\;\Varid{r}\;\Varid{xs}){}\<[E]%
\\
\>[3]{}\Varid{match}\;(\Varid{l}\;\cdot\;\Varid{r})\;{}\<[24]%
\>[24]{}\Varid{xs}\;{}\<[39]%
\>[39]{}\mathrel{=}\;\textrm{\bfseries do}\;{}\<[45]%
\>[45]{}(\Varid{ys}\;\Varid{,}\;\Varid{zs})\;\leftarrow \;\Varid{allSplits}\;\Varid{xs}\;{}\<[E]%
\\
\>[45]{}\Varid{y}\;\leftarrow \;\Varid{match}\;\Varid{l}\;\Varid{ys}\;{}\<[E]%
\\
\>[45]{}\Varid{z}\;\leftarrow \;\Varid{match}\;\Varid{r}\;\Varid{zs}\;{}\<[E]%
\\
\>[45]{}\Conid{Pure}\;(\Varid{y}\;\Varid{,}\;\Varid{z}){}\<[E]%
\\
\>[3]{}\Varid{match}\;(\Varid{r}\;\mathbin{\star})\;\Varid{xs}\;{}\<[39]%
\>[39]{}\mathrel{=}\;\Varid{fail}{}\<[E]%
\ColumnHook
\end{hscode}\resethooks
  \caption{The definition of the \ensuremath{\Varid{match}} function}
  \label{fig:match}
\end{figure}

Finally, we cannot yet implement the case for the Kleene star.  We could
attempt to mimic the case for concatenation, attempting to match \ensuremath{\Varid{r}\;\cdot\;(\Varid{r}\;\mathbin{\star})}.
This definition, however, is rejected by Agda as it is not structurally
recursive. For now we choose to simply fail on all such regular expressions.
In Section \ref{sec:combinations} we will fix this issue, after introducing
the auxiliary definitions.

Still, we can prove that the \ensuremath{\Varid{match}} function behaves correctly on all
regular expressions that do not contain iteration. We introduce a \ensuremath{\mathit{hasNo\star}}
predicate, which holds of all such iteration-free regular expressions:
\begin{hscode}\SaveRestoreHook
\column{B}{@{}>{\hspre}l<{\hspost}@{}}%
\column{3}{@{}>{\hspre}l<{\hspost}@{}}%
\column{E}{@{}>{\hspre}l<{\hspost}@{}}%
\>[3]{}\mathit{hasNo\star}\;\mathbin{:}\;\Conid{Regex}\;\to\;\Conid{Set}{}\<[E]%
\ColumnHook
\end{hscode}\resethooks
To verify our matcher is correct, we need to prove that it satisfies
the specification consisting of the following pre- and postcondition:
\begin{hscode}\SaveRestoreHook
\column{B}{@{}>{\hspre}l<{\hspost}@{}}%
\column{3}{@{}>{\hspre}l<{\hspost}@{}}%
\column{E}{@{}>{\hspre}l<{\hspost}@{}}%
\>[3]{}\Varid{pre}\;\mathbin{:}\;(\Varid{r}\;\mathbin{:}\;\Conid{Regex})\;(\Varid{xs}\;\mathbin{:}\;\Conid{String})\;\to\;\Conid{Set}{}\<[E]%
\\
\>[3]{}\Varid{pre}\;\Varid{r}\;\Varid{xs}\;\mathrel{=}\;\mathit{hasNo\star}\;\Varid{r}{}\<[E]%
\\
\>[3]{}\Varid{post}\;\mathbin{:}\;(\Varid{r}\;\mathbin{:}\;\Conid{Regex})\;(\Varid{xs}\;\mathbin{:}\;\Conid{String})\;\to\;\mathit{Tree}\;\Varid{r}\;\to\;\Conid{Set}{}\<[E]%
\\
\>[3]{}\Varid{post}\;\mathrel{=}\;\Conid{Match}{}\<[E]%
\ColumnHook
\end{hscode}\resethooks
The main correctness result can now be formulated as follows:
\begin{hscode}\SaveRestoreHook
\column{B}{@{}>{\hspre}l<{\hspost}@{}}%
\column{3}{@{}>{\hspre}l<{\hspost}@{}}%
\column{E}{@{}>{\hspre}l<{\hspost}@{}}%
\>[3]{}\Varid{matchSound}\;\mathbin{:}\;\forall\;\Varid{r}\;\Varid{xs}\;\to\;\llbracket(\Varid{pre}\;\Varid{r}\;\Varid{xs}),(\Varid{post}\;\Varid{r}\;\Varid{xs})\rrbracket_{\text{spec}}\;\sqsubseteq\;\llbracket\Varid{match}\;\Varid{r}\;\Varid{xs}\rrbracket_{\Varid{ptAll}}{}\<[E]%
\ColumnHook
\end{hscode}\resethooks
This lemma guarantees that all the parse trees computed by the \ensuremath{\Varid{match}}
function satisfy the \ensuremath{\Conid{Match}} relation, provided the input regular
expression does not contain iteration. The proof goes by induction on the
regular expression \ensuremath{\Varid{r}}. Although we have omitted the proof, we will sketch the
key lemmas and definitions that are necessary to complete it.

In most of the cases for \ensuremath{\Varid{r}}, the definition of \ensuremath{\Varid{match}\;\Varid{r}} is uncomplicated and
the proof is similarly simple. As soon as we need to reason about programs
composed using the monadic bind operator, we quickly run into issues. In
particular, when verifying the case for \ensuremath{\Varid{l}\;\cdot\;\Varid{r}}, we would like to use
our induction hypotheses on two recursive calls. To do so, we prove the
following lemma that allows us to replace the semantics of a composite
program built using the monadic bind operation with the composition of
the underlying predicate transformers:
\begin{hscode}\SaveRestoreHook
\column{B}{@{}>{\hspre}l<{\hspost}@{}}%
\column{3}{@{}>{\hspre}l<{\hspost}@{}}%
\column{5}{@{}>{\hspre}l<{\hspost}@{}}%
\column{E}{@{}>{\hspre}l<{\hspost}@{}}%
\>[3]{}\Varid{consequence}\;\mathbin{:}\;\!\!\;\forall\;\Varid{pt}\;(\Varid{mx}\;\mathbin{:}\;\Conid{Free}\;\Varid{es}\;\Varid{a})\;(\Varid{f}\;\mathbin{:}\;\Varid{a}\;\to\;\Conid{Free}\;\Varid{es}\;\Varid{b})\;\to\;{}\<[E]%
\\
\>[3]{}\hsindent{2}{}\<[5]%
\>[5]{}\llbracket\Varid{mx}\rrbracket_{\Varid{pt}}\;(\lambda\;\Varid{x}\;\to\;\llbracket\Varid{f}\;\Varid{x}\rrbracket_{\Varid{pt}}\;\Conid{P})\;\equiv\;\llbracket\Varid{mx}\;\bind \;\Varid{f}\rrbracket_{\Varid{pt}}\;\Conid{P}{}\<[E]%
\ColumnHook
\end{hscode}\resethooks
Substituting along this equality gives us the lemmas we need to deal with the \ensuremath{\_\!\bind\!\_} operator:
\begin{hscode}\SaveRestoreHook
\column{B}{@{}>{\hspre}l<{\hspost}@{}}%
\column{3}{@{}>{\hspre}l<{\hspost}@{}}%
\column{5}{@{}>{\hspre}l<{\hspost}@{}}%
\column{E}{@{}>{\hspre}l<{\hspost}@{}}%
\>[3]{}\Varid{wpToBind}\;\mathbin{:}\;\!\!\;(\Varid{mx}\;\mathbin{:}\;\Conid{Free}\;\Varid{es}\;\Varid{a})\;(\Varid{f}\;\mathbin{:}\;\Varid{a}\;\to\;\Conid{Free}\;\Varid{es}\;\Varid{b})\;\to\;{}\<[E]%
\\
\>[3]{}\hsindent{2}{}\<[5]%
\>[5]{}\llbracket\Varid{mx}\rrbracket_{\Varid{pt}}\;(\lambda\;\Varid{x}\;\to\;\llbracket\Varid{f}\;\Varid{x}\rrbracket_{\Varid{pt}}\;\Conid{P})\;\to\;\llbracket\Varid{mx}\;\bind \;\Varid{f}\rrbracket_{\Varid{pt}}\;\Conid{P}{}\<[E]%
\\
\>[3]{}\Varid{wpFromBind}\;\mathbin{:}\;\!\!\;(\Varid{mx}\;\mathbin{:}\;\Conid{Free}\;\Varid{es}\;\Varid{a})\;(\Varid{f}\;\mathbin{:}\;\Varid{a}\;\to\;\Conid{Free}\;\Varid{es}\;\Varid{b})\;\to\;{}\<[E]%
\\
\>[3]{}\hsindent{2}{}\<[5]%
\>[5]{}\llbracket\Varid{mx}\;\bind \;\Varid{f}\rrbracket_{\Varid{pt}}\;\Conid{P}\;\to\;\llbracket\Varid{mx}\rrbracket_{\Varid{pt}}\;(\lambda\;\Varid{x}\;\to\;\llbracket\Varid{f}\;\Varid{x}\rrbracket_{\Varid{pt}}\;\Conid{P}){}\<[E]%
\ColumnHook
\end{hscode}\resethooks

Not only does \ensuremath{\Varid{match}\;(\Varid{l}\;\cdot\;\Varid{r})} result in two recursive calls,
it also makes a call to a helper function \ensuremath{\Varid{allSplits}}.
Thus, we also need to formulate and prove the correctness of that function, as follows:
\begin{hscode}\SaveRestoreHook
\column{B}{@{}>{\hspre}l<{\hspost}@{}}%
\column{3}{@{}>{\hspre}l<{\hspost}@{}}%
\column{E}{@{}>{\hspre}l<{\hspost}@{}}%
\>[3]{}\Varid{allSplitsPost}\;\mathbin{:}\;\Conid{String}\;\to\;\Conid{String}\;\times\;\Conid{String}\;\to\;\Conid{Set}{}\<[E]%
\\
\>[3]{}\Varid{allSplitsPost}\;\Varid{xs}\;(\Varid{ys}\;\Varid{,}\;\Varid{zs})\;\mathrel{=}\;\Varid{xs}\;\equiv\;\Varid{ys}\;\plus \;\Varid{zs}{}\<[E]%
\\
\>[3]{}\Varid{allSplitsSound}\;\mathbin{:}\;\forall\;\Varid{xs}\;\to\;\llbracket\top,(\Varid{allSplitsPost}\;\Varid{xs})\rrbracket_{\text{spec}}\;\sqsubseteq\;\llbracket\Varid{allSplits}\;\Varid{xs}\rrbracket_{\Varid{ptAll}}{}\<[E]%
\ColumnHook
\end{hscode}\resethooks
Using \ensuremath{\Varid{wpToBind}}, we can incorporate the correctness proof of \ensuremath{\Varid{allSplits}}
in the correctness proof of \ensuremath{\Varid{match}}.
We refer to the accompanying code for the complete details of these
proofs.

\section{General recursion and non-determinism} \label{sec:combinations}
The matcher we have defined in the previous section is incomplete,
since it fails to handle regular expressions that use the Kleene star.
The fundamental issue is that the Kleene star allows for arbitrarily many matches in certain cases, that
in turn, leads to problems with Agda's termination checker.
For example, matching \ensuremath{\Conid{Epsilon}\;\mathbin{\star}} with the empty string \ensuremath{\text{\ttfamily \char34 \char34}} may unfold the Kleene star infinitely often
without ever terminating.
As a result, we cannot implement \ensuremath{\Varid{match}} for the Kleene star using recursion directly.

Instead, we will deal with this (possibly unbounded) recursion by introducing a new \emph{effect}.
We will represent a recursively defined dependent function of type \ensuremath{(\Varid{i}\;\mathbin{:}\;\Conid{I})\;\to\;\Conid{O}\;\Varid{i}}
as an element of the type \ensuremath{(\Varid{i}\;\mathbin{:}\;\Conid{I})\;\to\;\Conid{Free}\;(\Conid{Rec}\;\Conid{I}\;\Conid{O})\;(\Conid{O}\;\Varid{i})}.
Here \ensuremath{\Conid{Rec}\;\Conid{I}\;\Conid{O}} is a synonym of the the signature type we saw previously~\cite{McBride-totally-free}:
\begin{hscode}\SaveRestoreHook
\column{B}{@{}>{\hspre}l<{\hspost}@{}}%
\column{E}{@{}>{\hspre}l<{\hspost}@{}}%
\>[B]{}\Conid{Rec}\;\mathbin{:}\;(\Conid{I}\;\mathbin{:}\;\Conid{Set})\;(\Conid{O}\;\mathbin{:}\;\Conid{I}\;\to\;\Conid{Set})\;\to\;\Conid{Sig}{}\<[E]%
\\
\>[B]{}\Conid{Rec}\;\Conid{I}\;\Conid{O}\;\mathrel{=}\;\Varid{mkSig}\;\Conid{I}\;\Conid{O}{}\<[E]%
\ColumnHook
\end{hscode}\resethooks
Intuitively, you may want to think of values of type \ensuremath{(\Varid{i}\;\mathbin{:}\;\Conid{I})\;\to\;\Conid{Free}\;(\Conid{Rec}\;\Conid{I}\;\Conid{O})\;(\Conid{O}\;\Varid{i})} as computing a (finite) call graph for every input \ensuremath{\Varid{i}\;\mathbin{:}\;\Conid{I}}. Instead of
recurring directly, the `effects' that this signature supports require an input
\ensuremath{\Varid{i}\;\mathbin{:}\;\Conid{I}} corresponding to the argument of the recursive call; the continuation
abstracts over a value of type \ensuremath{\Conid{O}\;\Varid{i}}, corresponding to the result of a
recursive call. Note that the functions defined in this style are \emph{not}
recursive; instead we will need to write handlers to unfold the function
definition or prove termination separately. A handler for the \ensuremath{\Conid{Rec}} effect,
under the intended semantics, thus behaves like a fixed-point combinator,
introducing recursion to an otherwise recursion-free language by
substituting the function body in each recursive call.

We cannot, however, define a \ensuremath{\Varid{match}} function of the form \ensuremath{\Conid{Free}\;(\Conid{Rec}\;\anonymous \;\anonymous )} directly, as our previous definition also used non-determinism. To
account for both non-determinism and unbounded recursion, we need a
way to combine effects. Fortunately, free monads are known to be
closed under coproducts; there is a substantial body of work that
exploits this to (syntactically) compose separate
effects~\cite{effect-handlers-in-scope, la-carte}.

Rather than restrict ourselves to the binary composition using
coproducts, we modify the \ensuremath{\Conid{Free}} monad to take a \emph{list} of
signatures as its argument, taking the coproduct of the elements of
the list as its signature functor.  The \ensuremath{\Conid{Pure}} constructor remains
unchanged, while the \ensuremath{\Conid{Op}} constructor additionally takes an index into the
list to specify the effect that is invoked.
\begin{hscode}\SaveRestoreHook
\column{B}{@{}>{\hspre}l<{\hspost}@{}}%
\column{3}{@{}>{\hspre}l<{\hspost}@{}}%
\column{5}{@{}>{\hspre}l<{\hspost}@{}}%
\column{E}{@{}>{\hspre}l<{\hspost}@{}}%
\>[3]{}\Keyword{data}\;\Conid{Free}\;(\Varid{es}\;\mathbin{:}\;\Conid{List}\;\Conid{Sig})\;(\Varid{a}\;\mathbin{:}\;\Conid{Set})\;\mathbin{:}\;\Conid{Set}\;\Keyword{where}{}\<[E]%
\\
\>[3]{}\hsindent{2}{}\<[5]%
\>[5]{}\Conid{Pure}\;\mathbin{:}\;\Varid{a}\;\to\;\Conid{Free}\;\Varid{es}\;\Varid{a}{}\<[E]%
\\
\>[3]{}\hsindent{2}{}\<[5]%
\>[5]{}\Conid{Op}\;\mathbin{:}\;\!\!\;(\Varid{i}\;\mathbin{:}\;\Varid{e}\;\in\;\Varid{es})\;(\Varid{c}\;\mathbin{:}\;\Conid{C}\;\Varid{e})\;(\Varid{k}\;\mathbin{:}\;\Conid{R}\;\Varid{e}\;\Varid{c}\;\to\;\Conid{Free}\;\Varid{es}\;\Varid{a})\;\to\;\Conid{Free}\;\Varid{es}\;\Varid{a}{}\<[E]%
\ColumnHook
\end{hscode}\resethooks
By using a list of effects instead of allowing arbitrary disjoint unions,
we have effectively chosen that the disjoint unions canonically associate to the right.
We choose to use the same names and (almost) the same syntax for this new definition of \ensuremath{\Conid{Free}},
since all the definitions that we have seen previously can be readily adapted to work with this data type instead.

Most of this bookkeeping involved with different effects can be inferred using Agda's \emph{instance arguments}~\cite{instance-arguments-agda}.
Instance arguments, marked using the double curly braces \ensuremath{\{\!\!\{\!\;\;\;\!\}\!\!\}}, are
automatically filled in by Agda, provided a unique value of the
required type can be found. For example, we can define the generic
effects that we saw previously as follows:
\begin{hscode}\SaveRestoreHook
\column{B}{@{}>{\hspre}l<{\hspost}@{}}%
\column{3}{@{}>{\hspre}l<{\hspost}@{}}%
\column{E}{@{}>{\hspre}l<{\hspost}@{}}%
\>[3]{}\Varid{fail}\;\mathbin{:}\;\!\!\;\{\!\!\{\!\;\Varid{iND}\;\mathbin{:}\;\Conid{Nondet}\;\in\;\Varid{es}\;\!\}\!\!\}\;\to\;\Conid{Free}\;\Varid{es}\;\Varid{a}{}\<[E]%
\\
\>[3]{}\Varid{fail}\;\{\!\!\{\!\;\Varid{iND}\;\!\}\!\!\}\;\mathrel{=}\;\Conid{Op}\;\Varid{iND}\;\Conid{Fail}\;(\lambda\;()){}\<[E]%
\\
\>[3]{}\Varid{choice}\;\mathbin{:}\;\!\!\;\{\!\!\{\!\;\Varid{iND}\;\mathbin{:}\;\Conid{Nondet}\;\in\;\Varid{es}\;\!\}\!\!\}\;\to\;\Conid{Free}\;\Varid{es}\;\Varid{a}\;\to\;\Conid{Free}\;\Varid{es}\;\Varid{a}\;\to\;\Conid{Free}\;\Varid{es}\;\Varid{a}{}\<[E]%
\\
\>[3]{}\Varid{choice}\;\{\!\!\{\!\;\Varid{iND}\;\!\}\!\!\}\;\Conid{S₁}\;\Conid{S₂}\;\mathrel{=}\;\Conid{Op}\;\Varid{iND}\;\Conid{Choice}\;(\lambda\;\Varid{b}\;\to\;\textrm{\bfseries if}\;\Varid{b}\;\textrm{\bfseries then}\;\Conid{S₁}\;\textrm{\bfseries else}\;\Conid{S₂}){}\<[E]%
\\[\blanklineskip]%
\>[3]{}\Varid{call}\;\mathbin{:}\;\!\!\;\{\!\!\{\!\;\Varid{iRec}\;\mathbin{:}\;\Conid{Rec}\;\Conid{I}\;\Conid{O}\;\in\;\Varid{es}\;\!\}\!\!\}\;\to\;(\Varid{i}\;\mathbin{:}\;\Conid{I})\;\to\;\Conid{Free}\;\Varid{es}\;(\Conid{O}\;\Varid{i}){}\<[E]%
\\
\>[3]{}\Varid{call}\;\{\!\!\{\!\;\Varid{iRec}\;\!\}\!\!\}\;\Varid{i}\;\mathrel{=}\;\Conid{Op}\;\Varid{iRec}\;\Varid{i}\;\Conid{Pure}{}\<[E]%
\ColumnHook
\end{hscode}\resethooks
These now operate over any free monad with effects given by \ensuremath{\Varid{es}},
provided we can show that the list \ensuremath{\Varid{es}} contains the \ensuremath{\Conid{Nondet}} and
\ensuremath{\Conid{Rec}} effects respectively.
For convenience of notation, we introduce the \ensuremath{\anonymous \overset{\Varid{es}}{\looparrowright}\anonymous } notation for the type of generally recursive functions with effects in \ensuremath{\Varid{es}},
i.e. Kleisli arrows into \ensuremath{\Conid{Free}\;(\Conid{Rec}\;\anonymous \;\anonymous \;::\;\Varid{es})}.
\begin{hscode}\SaveRestoreHook
\column{B}{@{}>{\hspre}l<{\hspost}@{}}%
\column{3}{@{}>{\hspre}l<{\hspost}@{}}%
\column{E}{@{}>{\hspre}l<{\hspost}@{}}%
\>[3]{}\_\overset{\_}{\looparrowright}\_\;\mathbin{:}\;(\Conid{I}\;\mathbin{:}\;\Conid{Set})\;(\Varid{es}\;\mathbin{:}\;\Conid{List}\;\Conid{Sig})\;(\Conid{O}\;\mathbin{:}\;\Conid{I}\;\to\;\Conid{Set})\;\to\;\Conid{Set}{}\<[E]%
\\
\>[3]{}\Conid{I}\overset{\Varid{es}}{\looparrowright}\Conid{O}\;\mathrel{=}\;(\Varid{i}\;\mathbin{:}\;\Conid{I})\;\to\;\Conid{Free}\;(\Conid{Rec}\;\Conid{I}\;\Conid{O}\;::\;\Varid{es})\;(\Conid{O}\;\Varid{i}){}\<[E]%
\ColumnHook
\end{hscode}\resethooks

With the syntax for combinations of effects defined, let us turn to semantics.
Since the weakest precondition predicate transformer for a single effect is given as a fold
over the effect's signature,
the semantics for a combination of effects can be given by a list of such semantics.
\begin{hscode}\SaveRestoreHook
\column{B}{@{}>{\hspre}l<{\hspost}@{}}%
\column{3}{@{}>{\hspre}l<{\hspost}@{}}%
\column{5}{@{}>{\hspre}l<{\hspost}@{}}%
\column{7}{@{}>{\hspre}l<{\hspost}@{}}%
\column{13}{@{}>{\hspre}l<{\hspost}@{}}%
\column{E}{@{}>{\hspre}l<{\hspost}@{}}%
\>[3]{}\Keyword{record}\;\Conid{PT}\;(\Varid{e}\;\mathbin{:}\;\Conid{Sig})\;\mathbin{:}\;\Conid{Set}\;\Keyword{where}{}\<[E]%
\\
\>[3]{}\hsindent{2}{}\<[5]%
\>[5]{}\Keyword{constructor}\;\Varid{mkPT}{}\<[E]%
\\
\>[3]{}\hsindent{2}{}\<[5]%
\>[5]{}\Keyword{field}{}\<[E]%
\\
\>[5]{}\hsindent{2}{}\<[7]%
\>[7]{}\Varid{pt}\;{}\<[13]%
\>[13]{}\mathbin{:}\;(\Varid{c}\;\mathbin{:}\;\Conid{C}\;\Varid{e})\;\to\;(\Conid{R}\;\Varid{e}\;\Varid{c}\;\to\;\Conid{Set})\;\to\;\Conid{Set}{}\<[E]%
\\
\>[5]{}\hsindent{2}{}\<[7]%
\>[7]{}\Varid{mono}\;{}\<[13]%
\>[13]{}\mathbin{:}\;\forall\;\Varid{c}\;\Conid{P}\;\Conid{P'}\;\to\;(\forall\;\Varid{x}\;\to\;\Conid{P}\;\Varid{x}\;\to\;\Conid{P'}\;\Varid{x})\;\to\;\Varid{pt}\;\Varid{c}\;\Conid{P}\;\to\;\Varid{pt}\;\Varid{c}\;\Conid{P'}{}\<[E]%
\ColumnHook
\end{hscode}\resethooks
\begin{samepage}
\begin{hscode}\SaveRestoreHook
\column{B}{@{}>{\hspre}l<{\hspost}@{}}%
\column{3}{@{}>{\hspre}l<{\hspost}@{}}%
\column{5}{@{}>{\hspre}l<{\hspost}@{}}%
\column{11}{@{}>{\hspre}l<{\hspost}@{}}%
\column{E}{@{}>{\hspre}l<{\hspost}@{}}%
\>[3]{}\Keyword{data}\;\Conid{PTs}\;\mathbin{:}\;\Conid{List}\;\Conid{Sig}\;\to\;\Conid{Set}\;\Keyword{where}{}\<[E]%
\\
\>[3]{}\hsindent{2}{}\<[5]%
\>[5]{}\Conid{Nil}\;{}\<[11]%
\>[11]{}\mathbin{:}\;\Conid{PTs}\;\Conid{Nil}{}\<[E]%
\\
\>[3]{}\hsindent{2}{}\<[5]%
\>[5]{}\_\!::\!\_\;{}\<[11]%
\>[11]{}\mathbin{:}\;\!\!\;\Conid{PT}\;\Varid{e}\;\to\;\Conid{PTs}\;\Varid{es}\;\to\;\Conid{PTs}\;(\Varid{e}\;::\;\Varid{es}){}\<[E]%
\ColumnHook
\end{hscode}\resethooks
\end{samepage}
The record type \ensuremath{\Conid{PT}} not only contains a predicate transformer \ensuremath{\Varid{pt}},
but also a proof that this predicate transformer is
\emph{monotone}. Several lemmas throughout this paper, such as the
\ensuremath{\mathit{terminates\text{-}fmap}} lemma of Section~\ref{sec:dmatch-correct}, rely on the monotonicity of the
underlying predicate transformers;
for each semantics we present, the proof of monotonicity is immediate.

Given such a list of predicate transformers,
defining the semantics of an effectful program is a straightforward generalization of the previously defined semantics.
The \ensuremath{\Conid{Pure}} case is identical, and in the \ensuremath{\Conid{Op}} case we can apply the predicate transformer returned by the \ensuremath{\Varid{lookupPT}} helper function.
\begin{hscode}\SaveRestoreHook
\column{B}{@{}>{\hspre}l<{\hspost}@{}}%
\column{3}{@{}>{\hspre}l<{\hspost}@{}}%
\column{25}{@{}>{\hspre}l<{\hspost}@{}}%
\column{36}{@{}>{\hspre}l<{\hspost}@{}}%
\column{E}{@{}>{\hspre}l<{\hspost}@{}}%
\>[3]{}\Varid{lookupPT}\;\mathbin{:}\;\!\!\;(\Varid{pts}\;\mathbin{:}\;\Conid{PTs}\;\Varid{es})\;(\Varid{i}\;\mathbin{:}\;\Varid{mkSig}\;\Conid{C}\;\Conid{R}\;\in\;\Varid{es})\;\to\;(\Varid{c}\;\mathbin{:}\;\Conid{C})\;\to\;(\Conid{R}\;\Varid{c}\;\to\;\Conid{Set})\;\to\;\Conid{Set}{}\<[E]%
\\
\>[3]{}\Varid{lookupPT}\;(\Varid{pt}\;::\;\Varid{pts})\;{}\<[25]%
\>[25]{}\in\!\mathit{Head}\;{}\<[36]%
\>[36]{}\mathrel{=}\;\Conid{PT.pt}\;\Varid{pt}{}\<[E]%
\\
\>[3]{}\Varid{lookupPT}\;(\Varid{pt}\;::\;\Varid{pts})\;{}\<[25]%
\>[25]{}(\in\!\mathit{Tail}\;\Varid{i})\;{}\<[36]%
\>[36]{}\mathrel{=}\;\Varid{lookupPT}\;\Varid{pts}\;\Varid{i}{}\<[E]%
\ColumnHook
\end{hscode}\resethooks
This results in the following definition of the semantics for combinations of effects.
\begin{hscode}\SaveRestoreHook
\column{B}{@{}>{\hspre}l<{\hspost}@{}}%
\column{3}{@{}>{\hspre}l<{\hspost}@{}}%
\column{24}{@{}>{\hspre}l<{\hspost}@{}}%
\column{27}{@{}>{\hspre}l<{\hspost}@{}}%
\column{E}{@{}>{\hspre}l<{\hspost}@{}}%
\>[3]{}\llbracket \_ \rrbracket\;\mathbin{:}\;\!\!\;(\Varid{pts}\;\mathbin{:}\;\Conid{PTs}\;\Varid{es})\;\to\;\Conid{Free}\;\Varid{es}\;\Varid{a}\;\to\;(\Varid{a}\;\to\;\Conid{Set})\;\to\;\Conid{Set}{}\<[E]%
\\
\>[3]{}\llbracket\Conid{Pure}\;\Varid{x}\rrbracket_{\Varid{pts}}\;{}\<[24]%
\>[24]{}\Conid{P}\;{}\<[27]%
\>[27]{}\mathrel{=}\;\Conid{P}\;\Varid{x}{}\<[E]%
\\
\>[3]{}\llbracket\Conid{Op}\;\Varid{i}\;\Varid{c}\;\Varid{k}\rrbracket_{\Varid{pts}}\;{}\<[24]%
\>[24]{}\Conid{P}\;{}\<[27]%
\>[27]{}\mathrel{=}\;\Varid{lookupPT}\;\Varid{pts}\;\Varid{i}\;\Varid{c}\;(\lambda\;\Varid{x}\;\to\;\llbracket\Varid{k}\;\Varid{x}\rrbracket_{\Varid{pts}}\;\Conid{P}){}\<[E]%
\ColumnHook
\end{hscode}\resethooks

The effects that we will use for our \ensuremath{\Varid{match}} function consist of a
combination of non-determinism and general recursion.  Although we can
reuse the \ensuremath{\Varid{ptAll}} semantics of non-determinism, we have not yet given
the semantics for recursion.  However, it is not as easy to give a
predicate transformer semantics for recursion in general, since the
intended semantics of a recursive call depend on the function that is
being defined. Instead, to give semantics to a recursive function, we
assume that we have been provided with a relation of the type \ensuremath{(\Varid{i}\;\mathbin{:}\;\Conid{I})\;\to\;\Conid{O}\;\Varid{i}\;\to\;\Conid{Set}}, reminiscent of a loop invariant in an imperative
program. The semantics then establishes whether or not the recursive
function adheres to this invariant or not:
\begin{hscode}\SaveRestoreHook
\column{B}{@{}>{\hspre}l<{\hspost}@{}}%
\column{3}{@{}>{\hspre}l<{\hspost}@{}}%
\column{12}{@{}>{\hspre}l<{\hspost}@{}}%
\column{42}{@{}>{\hspre}l<{\hspost}@{}}%
\column{E}{@{}>{\hspre}l<{\hspost}@{}}%
\>[3]{}\Varid{ptRec}\;\mathbin{:}\;\!\!\;((\Varid{i}\;\mathbin{:}\;\Conid{I})\;\to\;\Conid{O}\;\Varid{i}\;\to\;\Conid{Set})\;\to\;\Conid{PT}\;(\Conid{Rec}\;\Conid{I}\;\Conid{O}){}\<[E]%
\\
\>[3]{}\Conid{PT.pt}\;{}\<[12]%
\>[12]{}(\Varid{ptRec}\;\Conid{R})\;\Varid{i}\;\Conid{P}\;{}\<[42]%
\>[42]{}\mathrel{=}\;\forall\;\Varid{o}\;\to\;\Conid{R}\;\Varid{i}\;\Varid{o}\;\to\;\Conid{P}\;\Varid{o}{}\<[E]%
\ColumnHook
\end{hscode}\resethooks
As we shall see shortly, when revisiting the \ensuremath{\Varid{match}} function, the
\ensuremath{\Conid{Match}} relation defined previously will fulfill the role of this
`invariant.'

To deal with the Kleene star, we rewrite \ensuremath{\Varid{match}} as a generally recursive function using a combination of effects.
Since \ensuremath{\Varid{match}} makes use of \ensuremath{\Varid{allSplits}}, we also rewrite that function to use a combination of effects.
The types become:
\begin{hscode}\SaveRestoreHook
\column{B}{@{}>{\hspre}l<{\hspost}@{}}%
\column{3}{@{}>{\hspre}l<{\hspost}@{}}%
\column{39}{@{}>{\hspre}l<{\hspost}@{}}%
\column{E}{@{}>{\hspre}l<{\hspost}@{}}%
\>[3]{}\Varid{allSplits}\;\mathbin{:}\;\!\!\;\{\!\!\{\!\;\Varid{iND}\;\mathbin{:}\;{}\<[39]%
\>[39]{}\Conid{Nondet}\;\in\;\Varid{es}\;\!\}\!\!\}\;\to\;\Conid{List}\;\Varid{a}\;\to\;\Conid{Free}\;\Varid{es}\;(\Conid{List}\;\Varid{a}\;\times\;\Conid{List}\;\Varid{a}){}\<[E]%
\\
\>[3]{}\Varid{match}\;\mathbin{:}\;\!\!\;\{\!\!\{\!\;\Varid{iND}\;\mathbin{:}\;\Conid{Nondet}\;\in\;\Varid{es}\;\!\}\!\!\}\;\to\;(\Varid{x}:\Conid{Regex}\;\times\;\Conid{String})\overset{\Varid{es}}{\looparrowright}\mathit{Tree}\;(\Conid{Pair.fst}\;\Varid{x}){}\<[E]%
\ColumnHook
\end{hscode}\resethooks

Since the index argument to the smart constructor is inferred by Agda,
the only change in the definition of \ensuremath{\Varid{match}} and \ensuremath{\Varid{allSplits}} will be
that \ensuremath{\Varid{match}} now does have a meaningful branch for the Kleene star case:
\begin{samepage}
\begin{hscode}\SaveRestoreHook
\column{B}{@{}>{\hspre}l<{\hspost}@{}}%
\column{3}{@{}>{\hspre}l<{\hspost}@{}}%
\column{5}{@{}>{\hspre}l<{\hspost}@{}}%
\column{E}{@{}>{\hspre}l<{\hspost}@{}}%
\>[3]{}\Varid{match}\;((\Varid{r}\;\mathbin{\star})\;\Varid{,}\;\Conid{Nil})\;\mathrel{=}\;\Conid{Pure}\;\Conid{Nil}{}\<[E]%
\\
\>[3]{}\Varid{match}\;((\Varid{r}\;\mathbin{\star})\;\Varid{,}\;\Varid{xs@}\;(\anonymous \;::\;\anonymous ))\;\mathrel{=}\;\textrm{\bfseries do}\;{}\<[E]%
\\
\>[3]{}\hsindent{2}{}\<[5]%
\>[5]{}(\Varid{y}\;\Varid{,}\;\Varid{ys})\;\leftarrow \;\Varid{call}\;\!\!\;((\Varid{r}\;\cdot\;(\Varid{r}\;\mathbin{\star}))\;\Varid{,}\;\Varid{xs})\;{}\<[E]%
\\
\>[3]{}\hsindent{2}{}\<[5]%
\>[5]{}\Conid{Pure}\;(\Varid{y}\;::\;\Varid{ys}){}\<[E]%
\ColumnHook
\end{hscode}\resethooks
\end{samepage}

The effects we need to use for running \ensuremath{\Varid{match}} are a combination of non-determinism and general recursion.
As discussed, we first need to give the specification for \ensuremath{\Varid{match}} before we can verify a program that performs a recursive \ensuremath{\Varid{call}} to \ensuremath{\Varid{match}}.
\begin{hscode}\SaveRestoreHook
\column{B}{@{}>{\hspre}l<{\hspost}@{}}%
\column{3}{@{}>{\hspre}l<{\hspost}@{}}%
\column{5}{@{}>{\hspre}l<{\hspost}@{}}%
\column{E}{@{}>{\hspre}l<{\hspost}@{}}%
\>[3]{}\Varid{matchSpec}\;\mathbin{:}\;(\Varid{r,xs}\;\mathbin{:}\;\Conid{Pair}\;\Conid{Regex}\;\Conid{String})\;\to\;\mathit{Tree}\;(\Conid{Pair.fst}\;\Varid{r,xs})\;\to\;\Conid{Set}{}\<[E]%
\\
\>[3]{}\Varid{matchSpec}\;(\Varid{r}\;\Varid{,}\;\Varid{xs})\;\Varid{ms}\;\mathrel{=}\;\Conid{Match}\;\Varid{r}\;\Varid{xs}\;\Varid{ms}{}\<[E]%
\\[\blanklineskip]%
\>[3]{}\llbracket \_ \rrbracket_{\text{match}}\;\mathbin{:}\;\!\!\;\Conid{Free}\;(\Conid{Rec}\;(\Conid{Pair}\;\Conid{Regex}\;\Conid{String})\;(\mathit{Tree}\;\circ\;\Conid{Pair.fst})\;::\;\Conid{Nondet}\;::\;\Conid{Nil})\;\Varid{a}\;\to\;{}\<[E]%
\\
\>[3]{}\hsindent{2}{}\<[5]%
\>[5]{}(\Varid{a}\;\to\;\Conid{Set})\;\to\;\Conid{Set}{}\<[E]%
\\
\>[3]{}\llbracket\Conid{S}\rrbracket_{\text{match}}\;\mathrel{=}\;\llbracket\Conid{S}\rrbracket_{\Varid{ptRec}\;\Varid{matchSpec}\;::\;\Varid{ptAll}\;::\;\Conid{Nil}}{}\<[E]%
\ColumnHook
\end{hscode}\resethooks

We can reuse exactly our proof that \ensuremath{\Varid{allSplits}} is correct,
since we use the same semantics for the non-determinism used in the definition of \ensuremath{\Varid{allSplits}}.
Similarly, the partial correctness proof of \ensuremath{\Varid{match}} will be the same on all cases except the Kleene star.
Now we are able to prove correctness of \ensuremath{\Varid{match}} on a Kleene star.
\begin{hscode}\SaveRestoreHook
\column{B}{@{}>{\hspre}l<{\hspost}@{}}%
\column{3}{@{}>{\hspre}l<{\hspost}@{}}%
\column{35}{@{}>{\hspre}l<{\hspost}@{}}%
\column{57}{@{}>{\hspre}l<{\hspost}@{}}%
\column{E}{@{}>{\hspre}l<{\hspost}@{}}%
\>[3]{}\Varid{matchSound}\;((\Varid{r}\;\mathbin{\star})\;\Varid{,}\;\Conid{Nil})\;{}\<[35]%
\>[35]{}\Conid{P}\;(\Varid{preH}\;\Varid{,}\;\Varid{postH})\;{}\<[57]%
\>[57]{}\mathrel{=}\;\Varid{postH}\;\anonymous \;\Conid{StarNil}{}\<[E]%
\\
\>[3]{}\Varid{matchSound}\;((\Varid{r}\;\mathbin{\star})\;\Varid{,}\;(\Varid{x}\;::\;\Varid{xs}))\;{}\<[35]%
\>[35]{}\Conid{P}\;(\Varid{preH}\;\Varid{,}\;\Varid{postH})\;\Varid{o}\;\Conid{H}\;{}\<[57]%
\>[57]{}\mathrel{=}\;\Varid{postH}\;\anonymous \;(\Conid{StarConcat}\;\Conid{H}){}\<[E]%
\ColumnHook
\end{hscode}\resethooks

At this point, we have defined a matcher for regular languages and
formally proven that when it succeeds in recognizing a given string,
this string is indeed in the language generated by the argument
regular expression.  However, the \ensuremath{\Varid{match}} function does not
necessarily terminate: if \ensuremath{\Varid{r}} is a regular expression that accepts the
empty string, then calling \ensuremath{\Varid{match}} on \ensuremath{\Varid{r}\;\mathbin{\star}} and a string \ensuremath{\Varid{xs}} will
diverge. In the next section, we will write a new parser that is guaranteed to
terminate and show that this parser refines the \ensuremath{\Varid{match}} function
defined above.

\section{Derivatives and handlers} \label{sec:dmatch}
Since recursion on the structure of a regular expression does not guarantee
termination of the parser, we can instead perform recursion on the string to be
parsed, changing the regular expression to be matched based on the characters
we have seen.

The \emph{Brzozowski derivative} of a formal language \ensuremath{\Conid{L}} with respect to a character \ensuremath{\Varid{x}} consists of all strings \ensuremath{\Varid{xs}} such that \ensuremath{\Varid{x}\;::\;\Varid{xs}\;\in\;\Conid{L}}~\cite{Brzozowski}.
Crucially, if \ensuremath{\Conid{L}} is regular, so are all its derivatives.
Thus, let \ensuremath{\Varid{r}} be a regular expression, and \ensuremath{\Varid{d}\;\Varid{r}\;\Varid{/d}\;\Varid{x}} an expression for the derivative with respect to \ensuremath{\Varid{x}},
then \ensuremath{\Varid{r}} matches a string \ensuremath{\Varid{x}\;::\;\Varid{xs}} if and only if \ensuremath{\Varid{d}\;\Varid{r}\;\Varid{/d}\;\Varid{x}} matches \ensuremath{\Varid{xs}}.
This suggests the following implementation of matching an expression \ensuremath{\Varid{r}} with a string \ensuremath{\Varid{xs}}:
if \ensuremath{\Varid{xs}} is empty, check whether \ensuremath{\Varid{r}} matches the empty string;
otherwise remove the head \ensuremath{\Varid{x}} of the string and try to match \ensuremath{\Varid{d}\;\Varid{r}\;\Varid{/d}\;\Varid{x}}.

The first step in implementing a parser using the Brzozowski derivative is to compute the derivative for a given regular expression.
Following \citet{Brzozowski}, we use a helper function \ensuremath{\varepsilon?} that decides whether an expression matches the empty string.
\begin{hscode}\SaveRestoreHook
\column{B}{@{}>{\hspre}l<{\hspost}@{}}%
\column{3}{@{}>{\hspre}l<{\hspost}@{}}%
\column{E}{@{}>{\hspre}l<{\hspost}@{}}%
\>[3]{}\varepsilon?\;\mathbin{:}\;(\Varid{r}\;\mathbin{:}\;\Conid{Regex})\;\to\;\Conid{Dec}\;(\sum\;(\mathit{Tree}\;\Varid{r})\;(\Conid{Match}\;\Varid{r}\;\Conid{Nil})){}\<[E]%
\ColumnHook
\end{hscode}\resethooks
The definition of \ensuremath{\varepsilon?} is given by structural recursion on the regular expression, just as the derivative operator is:
\begin{samepage}
\begin{hscode}\SaveRestoreHook
\column{B}{@{}>{\hspre}l<{\hspost}@{}}%
\column{3}{@{}>{\hspre}l<{\hspost}@{}}%
\column{18}{@{}>{\hspre}l<{\hspost}@{}}%
\column{26}{@{}>{\hspre}l<{\hspost}@{}}%
\column{35}{@{}>{\hspre}l<{\hspost}@{}}%
\column{E}{@{}>{\hspre}l<{\hspost}@{}}%
\>[3]{}\Varid{d\char95 /d\char95 }\;\mathbin{:}\;\Conid{Regex}\;\to\;\Conid{Char}\;\to\;\Conid{Regex}{}\<[E]%
\\
\>[3]{}\Varid{d}\;\Conid{Empty}\;{}\<[18]%
\>[18]{}\Varid{/d}\;\Varid{c}\;{}\<[26]%
\>[26]{}\mathrel{=}\;\Conid{Empty}{}\<[E]%
\\
\>[3]{}\Varid{d}\;\Conid{Epsilon}\;{}\<[18]%
\>[18]{}\Varid{/d}\;\Varid{c}\;{}\<[26]%
\>[26]{}\mathrel{=}\;\Conid{Empty}{}\<[E]%
\\
\>[3]{}\Varid{d}\;\Conid{Singleton}\;\Varid{x}\;{}\<[18]%
\>[18]{}\Varid{/d}\;\Varid{c}\;{}\<[26]%
\>[26]{}\Keyword{with}\;\Varid{c}\;\mathbin{\smash{\overset{\raisebox{-0.2em}{\tiny ?}}{=}}}\;\Varid{x}{}\<[E]%
\\
\>[3]{}\Varid{...}\;{}\<[26]%
\>[26]{}\mid \;\Varid{yes}\;\Varid{p}\;{}\<[35]%
\>[35]{}\mathrel{=}\;\Conid{Epsilon}{}\<[E]%
\\
\>[3]{}\Varid{...}\;{}\<[26]%
\>[26]{}\mid \;\Varid{no}\;\Varid{¬p}\;{}\<[35]%
\>[35]{}\mathrel{=}\;\Conid{Empty}{}\<[E]%
\\
\>[3]{}\Varid{d}\;\Varid{l}\;\cdot\;\Varid{r}\;{}\<[18]%
\>[18]{}\Varid{/d}\;\Varid{c}\;{}\<[26]%
\>[26]{}\Keyword{with}\;\varepsilon?\;\Varid{l}{}\<[E]%
\\
\>[3]{}\Varid{...}\;{}\<[26]%
\>[26]{}\mid \;\Varid{yes}\;\Varid{p}\;{}\<[35]%
\>[35]{}\mathrel{=}\;((\Varid{d}\;\Varid{l}\;\Varid{/d}\;\Varid{c})\;\cdot\;\Varid{r})\;\mathbin{\mid}\;(\Varid{d}\;\Varid{r}\;\Varid{/d}\;\Varid{c}){}\<[E]%
\\
\>[3]{}\Varid{...}\;{}\<[26]%
\>[26]{}\mid \;\Varid{no}\;\Varid{¬p}\;{}\<[35]%
\>[35]{}\mathrel{=}\;(\Varid{d}\;\Varid{l}\;\Varid{/d}\;\Varid{c})\;\cdot\;\Varid{r}{}\<[E]%
\\
\>[3]{}\Varid{d}\;\Varid{l}\;\mathbin{\mid}\;\Varid{r}\;{}\<[18]%
\>[18]{}\Varid{/d}\;\Varid{c}\;{}\<[26]%
\>[26]{}\mathrel{=}\;(\Varid{d}\;\Varid{l}\;\Varid{/d}\;\Varid{c})\;\mathbin{\mid}\;(\Varid{d}\;\Varid{r}\;\Varid{/d}\;\Varid{c}){}\<[E]%
\\
\>[3]{}\Varid{d}\;\Varid{r}\;\mathbin{\star}\;{}\<[18]%
\>[18]{}\Varid{/d}\;\Varid{c}\;{}\<[26]%
\>[26]{}\mathrel{=}\;(\Varid{d}\;\Varid{r}\;\Varid{/d}\;\Varid{c})\;\cdot\;(\Varid{r}\;\mathbin{\star}){}\<[E]%
\ColumnHook
\end{hscode}\resethooks
\end{samepage}

To use the derivative of \ensuremath{\Varid{r}} to compute a parse tree for \ensuremath{\Varid{r}},
we need to be able to convert a tree for \ensuremath{\Varid{d}\;\Varid{r}\;\Varid{/d}\;\Varid{x}} to a tree for \ensuremath{\Varid{r}}.
As this function `inverts' the result of differentiation, we name it \ensuremath{\Varid{integralTree}}:
\begin{hscode}\SaveRestoreHook
\column{B}{@{}>{\hspre}l<{\hspost}@{}}%
\column{3}{@{}>{\hspre}l<{\hspost}@{}}%
\column{E}{@{}>{\hspre}l<{\hspost}@{}}%
\>[3]{}\Varid{integralTree}\;\mathbin{:}\;\!\!\;(\Varid{r}\;\mathbin{:}\;\Conid{Regex})\;\to\;\mathit{Tree}\;(\Varid{d}\;\Varid{r}\;\Varid{/d}\;\Varid{x})\;\to\;\mathit{Tree}\;\Varid{r}{}\<[E]%
\ColumnHook
\end{hscode}\resethooks
Its definition closely follows the pattern matching performed in the definition of \ensuremath{\Varid{d\char95 /d\char95 }}.

The description of a derivative-based matcher is stateful:
we perform a step by \emph{removing} a character from the input string.
To match the description, we introduce new effect \ensuremath{\Conid{Parser}} which provides a parser-specific interface to this state.
The \ensuremath{\Conid{Parser}} effect has one command \ensuremath{\Conid{Symbol}} that returns a \ensuremath{\Conid{Maybe}\;\Conid{Char}}.
Calling \ensuremath{\Conid{Symbol}} will return \ensuremath{\Varid{just}} the head of the unparsed remainder (advancing the string by one character) or \ensuremath{\Varid{nothing}} if the string has been totally consumed.
\begin{hscode}\SaveRestoreHook
\column{B}{@{}>{\hspre}l<{\hspost}@{}}%
\column{3}{@{}>{\hspre}l<{\hspost}@{}}%
\column{5}{@{}>{\hspre}l<{\hspost}@{}}%
\column{E}{@{}>{\hspre}l<{\hspost}@{}}%
\>[3]{}\Keyword{data}\;\Conid{CParser}\;\mathbin{:}\;\Conid{Set}\;\Keyword{where}{}\<[E]%
\\
\>[3]{}\hsindent{2}{}\<[5]%
\>[5]{}\Conid{Symbol}\;\mathbin{:}\;\Conid{CParser}{}\<[E]%
\\
\>[3]{}\Conid{RParser}\;\mathbin{:}\;\Conid{CParser}\;\to\;\Conid{Set}{}\<[E]%
\\
\>[3]{}\Conid{RParser}\;\Conid{Symbol}\;\mathrel{=}\;\Conid{Maybe}\;\Conid{Char}{}\<[E]%
\\
\>[3]{}\Conid{Parser}\;\mathrel{=}\;\Varid{mkSig}\;\Conid{CParser}\;\Conid{RParser}{}\<[E]%
\\[\blanklineskip]%
\>[3]{}\Varid{symbol}\;\mathbin{:}\;\!\!\;\{\!\!\{\!\;\Varid{iP}\;\mathbin{:}\;\Conid{Parser}\;\in\;\Varid{es}\;\!\}\!\!\}\;\to\;\Conid{Free}\;\Varid{es}\;(\Conid{Maybe}\;\Conid{Char}){}\<[E]%
\\
\>[3]{}\Varid{symbol}\;\{\!\!\{\!\;\Varid{iP}\;\!\}\!\!\}\;\mathrel{=}\;\Conid{Op}\;\Varid{iP}\;\Conid{Symbol}\;\Conid{Pure}{}\<[E]%
\ColumnHook
\end{hscode}\resethooks

The code for the new parser, \ensuremath{\Varid{dmatch}}, is now only a few lines long.  When the
input contains at least one character, we use the derivative to match the first
character and recurse; when the input string is empty, we check that the
expression matches the empty string.
\begin{hscode}\SaveRestoreHook
\column{B}{@{}>{\hspre}l<{\hspost}@{}}%
\column{3}{@{}>{\hspre}l<{\hspost}@{}}%
\column{5}{@{}>{\hspre}l<{\hspost}@{}}%
\column{69}{@{}>{\hspre}l<{\hspost}@{}}%
\column{E}{@{}>{\hspre}l<{\hspost}@{}}%
\>[3]{}\Varid{dmatch}\;\mathbin{:}\;\!\!\;\{\!\!\{\!\;\Varid{iP}\;\mathbin{:}\;\Conid{Parser}\;\in\;\Varid{es}\;\!\}\!\!\}\;\{\!\!\{\!\;\Varid{iND}\;\mathbin{:}\;\Conid{Nondet}\;\in\;\Varid{es}\;\!\}\!\!\}\;{}\<[69]%
\>[69]{}\to\;\Conid{Regex}\overset{\Varid{es}}{\looparrowright}\mathit{Tree}{}\<[E]%
\\
\>[3]{}\Varid{dmatch}\;\Varid{r}\;\mathrel{=}\;\Varid{symbol}\;\bind \;\Varid{maybe}\;{}\<[E]%
\\
\>[3]{}\hsindent{2}{}\<[5]%
\>[5]{}(\lambda\;\Varid{x}\;\to\;\Varid{integralTree}\;\Varid{r}\;\langle\$\rangle\;\Varid{call}\;\!\!\;(\Varid{d}\;\Varid{r}\;\Varid{/d}\;\Varid{x}))\;{}\<[E]%
\\
\>[3]{}\hsindent{2}{}\<[5]%
\>[5]{}(\textrm{\bfseries if}\;\Varid{p}\;\leftarrow \;\varepsilon?\;\Varid{r}\;\textrm{\bfseries then}\;\Conid{Pure}\;(\Conid{Sigma.fst}\;\Varid{p})\;\textrm{\bfseries else}\;\Varid{fail}){}\<[E]%
\ColumnHook
\end{hscode}\resethooks
Here, \ensuremath{\Varid{maybe}\;\Varid{f}\;\Varid{y}} takes a \ensuremath{\Conid{Maybe}} value and applies \ensuremath{\Varid{f}} to the value in \ensuremath{\Varid{just}}, or returns \ensuremath{\Varid{y}} if it is \ensuremath{\Varid{nothing}}.
Although the parser is easily seen to terminate in the intended semantics
(since a character is removed from the input string between each recursive
call), a semantics where the call to \ensuremath{\Varid{symbol}} always returns \ensuremath{\Varid{just}} a character
causes \ensuremath{\Varid{dmatch}} to diverge. The termination of \ensuremath{\Varid{dmatch}} is not a syntactical
property, as reflected by the use of the recursive \ensuremath{\Varid{call}} in its definition,
and the custom arrow used in the type of functions defined using general recursion.

Adding the new effect \ensuremath{\Conid{Parser}} to our repertoire thus requires specifying its semantics.
We gave the effects \ensuremath{\Conid{Nondet}} and \ensuremath{\Conid{Rec}} predicate transformer semantics in the form of a \ensuremath{\Conid{PT}} record.
After introducing the \ensuremath{\Conid{Parser}} effect, the pre- and postcondition become more complicated:
not only do they reference the `pure' arguments and return values (here of type \ensuremath{\Varid{r}\;\mathbin{:}\;\Conid{Regex}} and \ensuremath{\Conid{Tree}\;\Varid{r}} respectively),
there is also the current state, containing a \ensuremath{\Conid{String}}, to keep track of.
With these augmented predicates, the predicate transformer semantics for the \ensuremath{\Conid{Parser}} effect can be given as:
\begin{hscode}\SaveRestoreHook
\column{B}{@{}>{\hspre}l<{\hspost}@{}}%
\column{3}{@{}>{\hspre}l<{\hspost}@{}}%
\column{32}{@{}>{\hspre}l<{\hspost}@{}}%
\column{46}{@{}>{\hspre}l<{\hspost}@{}}%
\column{E}{@{}>{\hspre}l<{\hspost}@{}}%
\>[3]{}\Varid{ptParser}\;\mathbin{:}\;(\Varid{c}\;\mathbin{:}\;\Conid{CParser})\;\to\;(\Conid{RParser}\;\Varid{c}\;\to\;\Conid{String}\;\to\;\Conid{Set})\;\to\;\Conid{String}\;\to\;\Conid{Set}{}\<[E]%
\\
\>[3]{}\Varid{ptParser}\;\Conid{Symbol}\;\Conid{P}\;\Conid{Nil}\;{}\<[32]%
\>[32]{}\mathrel{=}\;\Conid{P}\;\Varid{nothing}\;{}\<[46]%
\>[46]{}\Conid{Nil}{}\<[E]%
\\
\>[3]{}\Varid{ptParser}\;\Conid{Symbol}\;\Conid{P}\;(\Varid{x}\;::\;\Varid{xs})\;{}\<[32]%
\>[32]{}\mathrel{=}\;\Conid{P}\;(\Varid{just}\;\Varid{x})\;{}\<[46]%
\>[46]{}\Varid{xs}{}\<[E]%
\ColumnHook
\end{hscode}\resethooks

In this article, we want to demonstrate the modularity of predicate transformer semantics,
allowing us to introduce new notions without having to rework existing constructions.
To illustrate how the semantics mesh well with other forms of semantics,
we do \emph{not} use \ensuremath{\Varid{ptParser}} as semantics for \ensuremath{\Conid{Parser}} in the remainder.
We give denotational semantics, in the form of an \emph{effect handler} for \ensuremath{\Conid{Parser}}~\cite{algebraic-effect-handlers,effect-handlers-in-scope}:
\begin{hscode}\SaveRestoreHook
\column{B}{@{}>{\hspre}l<{\hspost}@{}}%
\column{3}{@{}>{\hspre}l<{\hspost}@{}}%
\column{29}{@{}>{\hspre}l<{\hspost}@{}}%
\column{46}{@{}>{\hspre}l<{\hspost}@{}}%
\column{E}{@{}>{\hspre}l<{\hspost}@{}}%
\>[3]{}\Varid{hParser}\;\mathbin{:}\;\!\!\;\{\!\!\{\!\;\Varid{iND}\;\mathbin{:}\;\Conid{Nondet}\;\in\;\Varid{es}\;\!\}\!\!\}\;(\Varid{c}\;\mathbin{:}\;\Conid{CParser})\;\to\;\Conid{String}\;\to\;\Conid{Free}\;\Varid{es}\;(\Conid{RParser}\;\Varid{c}\;\times\;\Conid{String}){}\<[E]%
\\
\>[3]{}\Varid{hParser}\;\Conid{Symbol}\;\Conid{Nil}\;{}\<[29]%
\>[29]{}\mathrel{=}\;\Conid{Pure}\;(\Varid{nothing}\;{}\<[46]%
\>[46]{}\Varid{,}\;\Conid{Nil}){}\<[E]%
\\
\>[3]{}\Varid{hParser}\;\Conid{Symbol}\;(\Varid{x}\;::\;\Varid{xs})\;{}\<[29]%
\>[29]{}\mathrel{=}\;\Conid{Pure}\;(\Varid{just}\;\Varid{x}\;{}\<[46]%
\>[46]{}\Varid{,}\;\Varid{xs}){}\<[E]%
\ColumnHook
\end{hscode}\resethooks
The function \ensuremath{\Varid{handleRec}} folds a given handler over a recursive definition,
allowing us to handle the \ensuremath{\Conid{Parser}} effect in \ensuremath{\Varid{dmatch}}.
\begin{hscode}\SaveRestoreHook
\column{B}{@{}>{\hspre}l<{\hspost}@{}}%
\column{3}{@{}>{\hspre}l<{\hspost}@{}}%
\column{5}{@{}>{\hspre}l<{\hspost}@{}}%
\column{E}{@{}>{\hspre}l<{\hspost}@{}}%
\>[3]{}\Varid{handleRec}\;\mathbin{:}\;\!\!\;((\Varid{c}\;\mathbin{:}\;\Conid{C})\;\to\;\Varid{s}\;\to\;\Conid{Free}\;\Varid{es}\;(\Conid{R}\;\Varid{c}\;\times\;\Varid{s}))\;\to\;{}\<[E]%
\\
\>[3]{}\hsindent{2}{}\<[5]%
\>[5]{}\!\!\;\Varid{a}\overset{\Varid{mkSig}\;\Conid{C}\;\Conid{R}\;::\;\Varid{es}}{\looparrowright}\Varid{b}\;\to\;(\Varid{x}:\Varid{a}\;\times\;\Varid{s})\overset{\Varid{es}}{\looparrowright}\Varid{b}\;(\Conid{Pair.fst}\;\Varid{x}){}\<[E]%
\\
\>[3]{}\Varid{dmatch'}\;\mathbin{:}\;\!\!\;\{\!\!\{\!\;\Varid{iND}\;\mathbin{:}\;\Conid{Nondet}\;\in\;\Varid{es}\;\!\}\!\!\}\;\to\;(\Varid{x}:\Conid{Regex}\;\times\;\Conid{String})\overset{\Varid{es}}{\looparrowright}\mathit{Tree}\;(\Conid{Pair.fst}\;\Varid{x}){}\<[E]%
\\
\>[3]{}\Varid{dmatch'}\;\mathrel{=}\;\Varid{handleRec}\;\Varid{hParser}\;(\Varid{dmatch}\;\!\!){}\<[E]%
\ColumnHook
\end{hscode}\resethooks
Note that \ensuremath{\Varid{dmatch'}} has exactly the type of the previously defined \ensuremath{\Varid{match}},
conveniently allowing us to re-use the \ensuremath{\llbracket \_ \rrbracket_{\text{match}}} semantics.
In this way, the handler \ensuremath{\Varid{hParser}} ``hides'' the implementation detail that the
\ensuremath{\Conid{Parser}} effect was used.

\section{Proving total correctness} \label{sec:dmatch-correct}
We finish the development process by proving that \ensuremath{\Varid{dmatch}} is correct.
The first step in this proof is that \ensuremath{\Varid{dmatch}} always terminates.
To express the termination of a recursive computation, we define the following
predicate, \ensuremath{\mathit{terminates\text{-}in}}: 
\begin{hscode}\SaveRestoreHook
\column{B}{@{}>{\hspre}l<{\hspost}@{}}%
\column{3}{@{}>{\hspre}l<{\hspost}@{}}%
\column{5}{@{}>{\hspre}l<{\hspost}@{}}%
\column{43}{@{}>{\hspre}l<{\hspost}@{}}%
\column{44}{@{}>{\hspre}l<{\hspost}@{}}%
\column{53}{@{}>{\hspre}l<{\hspost}@{}}%
\column{E}{@{}>{\hspre}l<{\hspost}@{}}%
\>[3]{}\mathit{terminates\text{-}in}\;\mathbin{:}\;\!\!\;(\Varid{pts}\;\mathbin{:}\;\Conid{PTs}\;\Varid{es})\;(\Varid{f}\;\mathbin{:}\;\Conid{I}\overset{\Varid{es}}{\looparrowright}\Conid{O})\;(\Conid{S}\;\mathbin{:}\;\Conid{Free}\;(\Conid{Rec}\;\Conid{I}\;\Conid{O}\;::\;\Varid{es})\;\Varid{a})\;\to\;\N\;\to\;\Conid{Set}{}\<[E]%
\\
\>[3]{}\mathit{terminates\text{-}in}\;\Varid{pts}\;\Varid{f}\;(\Conid{Pure}\;\Varid{x})\;{}\<[43]%
\>[43]{}\Varid{n}\;{}\<[53]%
\>[53]{}\mathrel{=}\;\top{}\<[E]%
\\
\>[3]{}\mathit{terminates\text{-}in}\;\Varid{pts}\;\Varid{f}\;(\Conid{Op}\;\in\!\mathit{Head}\;\Varid{c}\;\Varid{k})\;{}\<[43]%
\>[43]{}\Conid{Zero}\;{}\<[53]%
\>[53]{}\mathrel{=}\;\bot{}\<[E]%
\\
\>[3]{}\mathit{terminates\text{-}in}\;\Varid{pts}\;\Varid{f}\;(\Conid{Op}\;\in\!\mathit{Head}\;\Varid{c}\;\Varid{k})\;{}\<[43]%
\>[43]{}(\Conid{Succ}\;\Varid{n})\;{}\<[53]%
\>[53]{}\mathrel{=}\;\mathit{terminates\text{-}in}\;\Varid{pts}\;\Varid{f}\;(\Varid{f}\;\Varid{c}\;\bind \;\Varid{k})\;\Varid{n}{}\<[E]%
\\
\>[3]{}\mathit{terminates\text{-}in}\;\Varid{pts}\;\Varid{f}\;(\Conid{Op}\;(\in\!\mathit{Tail}\;\Varid{i})\;\Varid{c}\;\Varid{k})\;{}\<[44]%
\>[44]{}\Varid{n}\;{}\<[53]%
\>[53]{}\mathrel{=}\;{}\<[E]%
\\
\>[3]{}\hsindent{2}{}\<[5]%
\>[5]{}\Varid{lookupPT}\;\Varid{pts}\;\Varid{i}\;\Varid{c}\;(\lambda\;\Varid{x}\;\to\;\mathit{terminates\text{-}in}\;\Varid{pts}\;\Varid{f}\;(\Varid{k}\;\Varid{x})\;\Varid{n}){}\<[E]%
\ColumnHook
\end{hscode}\resethooks
Given a program \ensuremath{\Conid{S}} that calls the recursive
function \ensuremath{\Varid{f}\;\mathbin{:}\;\Conid{I}\overset{\Varid{es}}{\looparrowright}\Conid{O}}, we check whether the computation requires no
more than a fixed number of steps to terminate.

Since \ensuremath{\Varid{dmatch}} always consumes a character before recurring,
we can bound the number of recursive calls with the length of the input string.
We formalize this argument in the lemma \ensuremath{\Varid{dmatchTerminates}}.
Note that \ensuremath{\Varid{dmatch'}} is defined using the \ensuremath{\Varid{hParser}} handler,
showing that we can mix denotational and predicate transformer semantics.
The proof goes by induction on this string.
Unfolding the recursive \ensuremath{\Varid{call}} gives \ensuremath{\Varid{integralTree}\;\Varid{r}\;\langle\$\rangle\;\Varid{dmatch'}\;(\Varid{d}\;\Varid{r}\;\Varid{/d}\;\Varid{x}\;\Varid{,}\;\Varid{xs})},
which we rewrite using the associativity monad law in a lemma called \ensuremath{\mathit{terminates\text{-}fmap}}.
\begin{hscode}\SaveRestoreHook
\column{B}{@{}>{\hspre}l<{\hspost}@{}}%
\column{3}{@{}>{\hspre}l<{\hspost}@{}}%
\column{5}{@{}>{\hspre}l<{\hspost}@{}}%
\column{7}{@{}>{\hspre}l<{\hspost}@{}}%
\column{35}{@{}>{\hspre}l<{\hspost}@{}}%
\column{E}{@{}>{\hspre}l<{\hspost}@{}}%
\>[3]{}\Varid{dmatchTerminates}\;\mathbin{:}\;(\Varid{r}\;\mathbin{:}\;\Conid{Regex})\;(\Varid{xs}\;\mathbin{:}\;\Conid{String})\;\to\;{}\<[E]%
\\
\>[3]{}\hsindent{2}{}\<[5]%
\>[5]{}\mathit{terminates\text{-}in}\;(\Varid{ptAll}\;::\;\Conid{Nil})\;(\Varid{dmatch'}\;\!\!)\;(\Varid{dmatch'}\;\!\!\;(\Varid{r}\;\Varid{,}\;\Varid{xs}))\;(\Varid{length}\;\Varid{xs}){}\<[E]%
\\
\>[3]{}\Varid{dmatchTerminates}\;\Varid{r}\;\Conid{Nil}\;\Keyword{with}\;\varepsilon?\;\Varid{r}{}\<[E]%
\\
\>[3]{}\Varid{dmatchTerminates}\;\Varid{r}\;\Conid{Nil}\;\mid \;\Varid{yes}\;\Varid{p}\;{}\<[35]%
\>[35]{}\mathrel{=}\;\Varid{tt}{}\<[E]%
\\
\>[3]{}\Varid{dmatchTerminates}\;\Varid{r}\;\Conid{Nil}\;\mid \;\Varid{no}\;\Varid{¬p}\;{}\<[35]%
\>[35]{}\mathrel{=}\;\Varid{tt}{}\<[E]%
\\
\>[3]{}\Varid{dmatchTerminates}\;\Varid{r}\;(\Varid{x}\;::\;\Varid{xs})\;\mathrel{=}\;\mathit{terminates\text{-}fmap}\;(\Varid{length}\;\Varid{xs})\;(\Varid{dmatch'}\;\!\!\;((\Varid{d}\;\Varid{r}\;\Varid{/d}\;\Varid{x})\;\Varid{,}\;\Varid{xs}))\;{}\<[E]%
\\
\>[3]{}\hsindent{2}{}\<[5]%
\>[5]{}(\Varid{dmatchTerminates}\;(\Varid{d}\;\Varid{r}\;\Varid{/d}\;\Varid{x})\;\Varid{xs}){}\<[E]%
\\
\>[3]{}\hsindent{2}{}\<[5]%
\>[5]{}\Keyword{where}{}\<[E]%
\\
\>[3]{}\hsindent{2}{}\<[5]%
\>[5]{}\mathit{terminates\text{-}fmap}\;\mathbin{:}\;\!\!\;\{\mskip1.5mu \Varid{f}\;\mathbin{:}\;\Conid{I}\overset{\Varid{es}}{\looparrowright}\Conid{O}\mskip1.5mu\}\;\{\mskip1.5mu \Varid{g}\;\mathbin{:}\;\Varid{a}\;\to\;\Varid{b}\mskip1.5mu\}\;(\Varid{n}\;\mathbin{:}\;\N)\;(\Conid{S}\;\mathbin{:}\;\Conid{Free}\;(\Conid{Rec}\;\Conid{I}\;\Conid{O}\;::\;\Varid{es})\;\Varid{a})\;\to\;{}\<[E]%
\\
\>[5]{}\hsindent{2}{}\<[7]%
\>[7]{}\mathit{terminates\text{-}in}\;\Varid{pts}\;\Varid{f}\;\Conid{S}\;\Varid{n}\;\to\;\mathit{terminates\text{-}in}\;\Varid{pts}\;\Varid{f}\;(\Varid{g}\;\langle\$\rangle\;\Conid{S})\;\Varid{n}{}\<[E]%
\ColumnHook
\end{hscode}\resethooks

Apart from termination, correctness consists of soundness and completeness: the
parse trees returned by \ensuremath{\Varid{dmatch}} should satisfy the specification given by
the original \ensuremath{\Conid{Match}} relation, and for any string that matches the regular
expression, \ensuremath{\Varid{dmatch}} should return a parse tree.  In the \ensuremath{\Varid{ptAll}} semantics, a
non-deterministic program \ensuremath{\Conid{S}} is refined by \ensuremath{\Conid{T}} if and only if the output values
of \ensuremath{\Conid{T}} are a subset of the output values of \ensuremath{\Conid{S}}; conversely \ensuremath{\Conid{S}} is refined by
\ensuremath{\Conid{T}} in the \ensuremath{\Varid{ptAny}} semantics if and only if the output values of \ensuremath{\Conid{S}} are a
subset of the output values of \ensuremath{\Conid{T}}. These properties allow us to express
program correctness in terms of refinement.

We can show soundness of \ensuremath{\Varid{dmatch}} by proving it refines \ensuremath{\Varid{match}}.  Transitivity
of the refinement relation then allows us to conclude that it also satisfies
the specification given by our original \ensuremath{\Conid{Match}} relation. The first step is to
show that the derivative operator is correct, i.e. \ensuremath{\Varid{d}\;\Varid{r}\;\Varid{/d}\;\Varid{x}} matches those
strings \ensuremath{\Varid{xs}} such that \ensuremath{\Varid{r}} matches \ensuremath{\Varid{x}\;::\;\Varid{xs}}.
\begin{hscode}\SaveRestoreHook
\column{B}{@{}>{\hspre}l<{\hspost}@{}}%
\column{3}{@{}>{\hspre}l<{\hspost}@{}}%
\column{E}{@{}>{\hspre}l<{\hspost}@{}}%
\>[3]{}\Varid{derivativeCorrect}\;\mathbin{:}\;\!\!\;\forall\;\Varid{r}\;\to\;\!\!\;\Conid{Match}\;(\Varid{d}\;\Varid{r}\;\Varid{/d}\;\Varid{x})\;\Varid{xs}\;\Varid{y}\;\to\;\Conid{Match}\;\Varid{r}\;(\Varid{x}\;::\;\Varid{xs})\;(\Varid{integralTree}\;\Varid{r}\;\Varid{y}){}\<[E]%
\ColumnHook
\end{hscode}\resethooks
The proof is straightforward by induction on the derivation of type \ensuremath{\Conid{Match}\;(\Varid{d}\;\Varid{r}\;\Varid{/d}\;\Varid{x})\;\Varid{xs}\;\Varid{y}}.

Using the preceding lemmas, we can prove the partial correctness of \ensuremath{\Varid{dmatch}}.
\begin{hscode}\SaveRestoreHook
\column{B}{@{}>{\hspre}l<{\hspost}@{}}%
\column{3}{@{}>{\hspre}l<{\hspost}@{}}%
\column{E}{@{}>{\hspre}l<{\hspost}@{}}%
\>[3]{}\Varid{dmatchSound}\;\mathbin{:}\;\forall\;\Varid{r}\;\Varid{xs}\;\to\;\llbracket\Varid{match}\;\!\!\;(\Varid{r}\;\Varid{,}\;\Varid{xs})\rrbracket_{\text{match}}\;\sqsubseteq\;\llbracket\Varid{dmatch'}\;\!\!\;(\Varid{r}\;\Varid{,}\;\Varid{xs})\rrbracket_{\text{match}}{}\<[E]%
\ColumnHook
\end{hscode}\resethooks
Since we need to perform the case distinctions of \ensuremath{\Varid{match}} and of \ensuremath{\Varid{dmatch}},
the proof is longer than that of \ensuremath{\Varid{matchSound}}.
Despite the length, most of it consists of this case distinction,
then giving a simple argument for each case.

Although we successfully proved \ensuremath{\Varid{dmatch}} is sound with respect to the \ensuremath{\Conid{Match}}
relation, it is not \emph{complete}: the function \ensuremath{\Varid{dmatch}} never makes a
non-deterministic choice. It will not return all possible parse trees that
satisfy the \ensuremath{\Conid{Match}} relation, only the first tree that it encounters.  We can,
however, prove that \ensuremath{\Varid{dmatch}} will find a parse tree if it exists.  To express
that \ensuremath{\Varid{dmatch}} returns any result at all, we use a trivially true postcondition;
by furthermore replacing the demonic choice of the \ensuremath{\Varid{ptAll}} semantics with the
angelic choice of \ensuremath{\Varid{ptAny}}, we require that \ensuremath{\Varid{dmatch}} \emph{must} return a
result:
\begin{hscode}\SaveRestoreHook
\column{B}{@{}>{\hspre}l<{\hspost}@{}}%
\column{3}{@{}>{\hspre}l<{\hspost}@{}}%
\column{5}{@{}>{\hspre}l<{\hspost}@{}}%
\column{E}{@{}>{\hspre}l<{\hspost}@{}}%
\>[3]{}\Varid{dmatchComplete}\;\mathbin{:}\;\forall\;\Varid{r}\;\Varid{xs}\;\Varid{y}\;\to\;\Conid{Match}\;\Varid{r}\;\Varid{xs}\;\Varid{y}\;\to\;{}\<[E]%
\\
\>[3]{}\hsindent{2}{}\<[5]%
\>[5]{}\llbracket\Varid{dmatch'}\;\!\!\;(\Varid{r}\;\Varid{,}\;\Varid{xs})\rrbracket_{\Varid{ptRec}\;\Varid{matchSpec}\;::\;\Varid{ptAny}\;::\;\Conid{Nil}}\;(\lambda\;\anonymous \;\to\;\top){}\<[E]%
\ColumnHook
\end{hscode}\resethooks
The proof is short, since \ensuremath{\Varid{dmatch}} can only \ensuremath{\Varid{fail}} when it encounters
an empty string and a regular expression that does not match the empty
string, which contradicts the assumption \ensuremath{\Conid{Match}\;\Varid{r}\;\Varid{xs}\;\Varid{y}}:
\begin{hscode}\SaveRestoreHook
\column{B}{@{}>{\hspre}l<{\hspost}@{}}%
\column{3}{@{}>{\hspre}l<{\hspost}@{}}%
\column{E}{@{}>{\hspre}l<{\hspost}@{}}%
\>[3]{}\Varid{dmatchComplete}\;\Varid{r}\;\Conid{Nil}\;\Varid{y}\;\Conid{H}\;\Keyword{with}\;\varepsilon?\;\Varid{r}{}\<[E]%
\\
\>[3]{}\Varid{...}\;\mid \;\Varid{yes}\;\Varid{p}\;\mathrel{=}\;\Varid{tt}{}\<[E]%
\\
\>[3]{}\Varid{...}\;\mid \;\Varid{no}\;\Varid{¬p}\;\mathrel{=}\;\Varid{¬p}\;(\anonymous \;\Varid{,}\;\Conid{H}){}\<[E]%
\\
\>[3]{}\Varid{dmatchComplete}\;\Varid{r}\;(\Varid{x}\;::\;\Varid{xs})\;\Varid{y}\;\Conid{H}\;\Varid{y'}\;\Conid{H'}\;\mathrel{=}\;\Varid{tt}{}\<[E]%
\ColumnHook
\end{hscode}\resethooks

In the proofs of \ensuremath{\Varid{dmatchSound}} and \ensuremath{\Varid{dmatchComplete}}, we demonstrate
the power of predicate transformer semantics for effects: by
separating syntax and semantics, we can easily describe different
aspects (soundness and completeness) of the one definition of
\ensuremath{\Varid{dmatch}}.  Since the soundness and completeness result we have proved
imply partial correctness, and partial correctness and termination
imply total correctness, we can conclude that \ensuremath{\Varid{dmatch}} is a totally
correct parser for regular languages.

\section{Discussion}

\subsection*{Related work}
The refinement calculus has traditionally been used to verify imperative programs~\cite{prog-from-spec}.
In this paper, however, we show how many of the ideas from the refinement calculus can also be used in the verification of functional programs~\cite{pt-semantics-for-effects}.
The \emph{Dijkstra monad}, introduced in the language F$\star$, also uses a predicate transformer semantics for verifying effectful programs
by collecting the proof obligations for verification~\cite{dijkstra-monad, dijkstra-monads-for-free, dijkstra-monads-for-all}.
This paper demonstrates how similar verification efforts can be undertaken directly in an interactive theorem prover such as Agda.
The separation of syntax and semantics in our approach allows for verification to be performed in several steps,
such as we did for \ensuremath{\Varid{dmatchTerminates}}, \ensuremath{\Varid{dmatchSound}} and \ensuremath{\Varid{dmatchComplete}}, adding new effects as we need them.

Our running example of the regular expression parser is inspired by the development of a regular expression parser by \citet{harper-regex}.
More recently, \citet{intrinsic-verification-regex} adapted the Functional Pearl to Agda.
A direct translation of \citeauthor{harper-regex}'s definitions is not possible:
they are rejected by Agda's termination checker because they are not structurally recursive.
\citeauthor{intrinsic-verification-regex} show how the defunctionalization of Harper's matcher, written in
continuation-passing style, is 
accepted by Agda's termination checker.

Formally verified parsers for a more general class of languages have been developed before:
\citet{total-parser-combinators, firsov-certification-context-free-grammars, simple-functional-cfg-parsing}, among others, have
previously shown how to verify parsers developed in a functional language.
In these developments, semantics are defined specialized to the domain of parsing,
while our semantics arise from combining a generic set of effect semantics.
Furthermore, we allow our parsers to be written using general recursion directly, whereas most existing approaches
deal with termination syntactically, either by incorporating delay and force operators in the grammar,
or explicitly passing around a proof of termination in the definition of the parser.
The modularity of our setup allows us to separate partial and total correctness cleanly.

There are various ways to represent a combination of effects such as used in parsers.
A traditional approach is to use \emph{monad transformers} to add each effect in turn, producing a complicated monad that incorporates all required operations~\cite{monad-transformers}.
More recently, \emph{graded monads} were introduced as a way to indicate more precisely the effects used in a specific computation~\cite{embedding-effect-systems,effects-and-monads}.
With some slight changes to the types of \ensuremath{\Conid{Pure}} and \ensuremath{\_\!\bind\!\_}, the \ensuremath{\Conid{Free}} monad can be viewed as graded over the free monoid \ensuremath{\Conid{List}\;\Conid{Sig}} generated by the type of effect signatures.
As this monad containing the computation is freely generated, it does not require us to assign any semantics to the effects ahead of time.

\subsection*{Open issues}
This paper builds upon our previous results \cite{pt-semantics-for-effects} by
demonstrating their use in non-trivial development. In the process, we show how to
\emph{combine} predicate transformer semantics and reason about
programs using a combination of effects.

Our approach relies on using coproducts to combine effect syntax. The
interaction between different effects means applying handlers in a different
order can result in different semantics.  We assign predicate transformer
semantics to a combination of effects all at once, specifying their interaction
explicitly---but we would still like to explore how to handle effects
one-by-one, allowing for greater flexibility when assigning semantics to
effectful programs~\cite{modular-algebraic-effects, effect-handlers-in-scope}.

\subsection*{Conclusions}
In conclusion, we have illustrated the approach to developing verified software
in a proof assistant using a predicate transformer semantics for effects for a
non-trivial example.  We believe this approach enables us to add new effects in
a modular fashion, while still being able to re-use any existing proofs.  Along
the way, we demonstrated how to combine different effects and define different
semantics for these effects, without impacting existing definitions.  As a
result, the verification effort---while conceptually more challenging at
times---remains fairly modular.

\paragraph{Acknowledgements}
T. Baanen has received funding from the NWO under the Vidi program (project No. 016.Vidi.189.037, Lean Forward).

\printbibliography

\appendix

\ifdefined\includeCFGs

\section{Parsing as effect} \label{sec:parser}
In the previous sections, we wrote parsers as nondeterministic functions.
For more complicated classes of languages than regular expressions, explicitly passing around the string to be parsed becomes cumbersome quickly.
The traditional solution is to switch from nondeterminism to stateful nondeterminism, where the state contains the unparsed portion of the string~\cite{hutton}.
The combination of nondeterminism and state can be represented by the \ensuremath{\Conid{ListOfSuccesses}} monad:
\begin{hscode}\SaveRestoreHook
\column{B}{@{}>{\hspre}l<{\hspost}@{}}%
\column{3}{@{}>{\hspre}l<{\hspost}@{}}%
\column{E}{@{}>{\hspre}l<{\hspost}@{}}%
\>[3]{}\Conid{ListOfSuccesses}\;\mathbin{:}\;\Conid{Set}\;\to\;\Conid{Set}{}\<[E]%
\\
\>[3]{}\Conid{ListOfSuccesses}\;\Varid{a}\;\mathrel{=}\;\Conid{String}\;\to\;\Conid{List}\;(\Varid{a}\;\times\;\Conid{String}){}\<[E]%
\ColumnHook
\end{hscode}\resethooks

Since our development makes use of algebraic effects,
we can introduce the effect of mutable state without having to change existing definitions.
We introduce this using the \ensuremath{\Conid{Parser}} effect, which has one command \ensuremath{\Conid{Symbol}}.
Calling \ensuremath{\Conid{Symbol}} will return the current symbol in the state (advancing the state by one) or fail if all symbols have been consumed.
\begin{hscode}\SaveRestoreHook
\column{B}{@{}>{\hspre}l<{\hspost}@{}}%
\column{3}{@{}>{\hspre}l<{\hspost}@{}}%
\column{5}{@{}>{\hspre}l<{\hspost}@{}}%
\column{E}{@{}>{\hspre}l<{\hspost}@{}}%
\>[3]{}\Keyword{data}\;\Conid{CParser}\;\mathbin{:}\;\Conid{Set}\;\Keyword{where}{}\<[E]%
\\
\>[3]{}\hsindent{2}{}\<[5]%
\>[5]{}\Conid{Symbol}\;\mathbin{:}\;\Conid{CParser}{}\<[E]%
\\
\>[3]{}\Conid{RParser}\;\mathbin{:}\;\Conid{CParser}\;\to\;\Conid{Set}{}\<[E]%
\\
\>[3]{}\Conid{RParser}\;\Conid{Symbol}\;\mathrel{=}\;\Conid{Char}{}\<[E]%
\\
\>[3]{}\Conid{Parser}\;\mathrel{=}\;\Varid{mkSig}\;\Conid{CParser}\;\Conid{RParser}{}\<[E]%
\\[\blanklineskip]%
\>[3]{}\Varid{symbol}\;\mathbin{:}\;\!\!\;\{\!\!\{\!\;\Varid{iP}\;\mathbin{:}\;\Conid{Parser}\;\in\;\Varid{es}\;\!\}\!\!\}\;\to\;\Conid{Free}\;\Varid{es}\;\Conid{Char}{}\<[E]%
\\
\>[3]{}\Varid{symbol}\;\{\!\!\{\!\;\Varid{iP}\;\!\}\!\!\}\;\mathrel{=}\;\Conid{Op}\;\Varid{iP}\;\Conid{Symbol}\;\Conid{Pure}{}\<[E]%
\ColumnHook
\end{hscode}\resethooks
We could add more commands such as \ensuremath{\Conid{EOF}} for detecting the end of the input, but we do not need them in the current development.
In the semantics we will define that parsing was successful if the input string has been completely consumed.

Note that \ensuremath{\Conid{Parser}} is not sufficient by itself to implement even simple parsers such as \ensuremath{\Varid{dmatch}}:
we need to be able to choose between parsing the next character or returning a value for the empty string.
This is why we usually combine \ensuremath{\Conid{Parser}} with nondeterminism and general recursion.

The denotational semantics of a parser in the \ensuremath{\Conid{Free}} monad take the form of a fold,
handling each command in the \ensuremath{\Conid{Parser}} monad.
\begin{hscode}\SaveRestoreHook
\column{B}{@{}>{\hspre}l<{\hspost}@{}}%
\column{3}{@{}>{\hspre}l<{\hspost}@{}}%
\column{5}{@{}>{\hspre}l<{\hspost}@{}}%
\column{E}{@{}>{\hspre}l<{\hspost}@{}}%
\>[3]{}\Varid{runParser}\;\mathbin{:}\;\!\!\;\Conid{Free}\;(\Conid{Nondet}\;::\;\Conid{Parser}\;::\;\Conid{Nil})\;\Varid{a}\;\to\;\Conid{ListOfSuccesses}\;\Varid{a}{}\<[E]%
\\
\>[3]{}\Varid{runParser}\;(\Conid{Pure}\;\Varid{x})\;\Conid{Nil}\;\mathrel{=}\;(\Varid{x}\;\Varid{,}\;\Conid{Nil})\;::\;\Conid{Nil}{}\<[E]%
\\
\>[3]{}\Varid{runParser}\;(\Conid{Pure}\;\Varid{x})\;(\anonymous \;::\;\anonymous )\;\mathrel{=}\;\Conid{Nil}{}\<[E]%
\\
\>[3]{}\Varid{runParser}\;(\Conid{Op}\;\in\!\mathit{Head}\;\Conid{Fail}\;\Varid{k})\;\Varid{xs}\;\mathrel{=}\;\Conid{Nil}{}\<[E]%
\\
\>[3]{}\Varid{runParser}\;(\Conid{Op}\;\in\!\mathit{Head}\;\Conid{Choice}\;\Varid{k})\;\Varid{xs}\;\mathrel{=}\;{}\<[E]%
\\
\>[3]{}\hsindent{2}{}\<[5]%
\>[5]{}\Varid{runParser}\;(\Varid{k}\;\Conid{True})\;\Varid{xs}\;\plus \;\Varid{runParser}\;(\Varid{k}\;\Conid{False})\;\Varid{xs}{}\<[E]%
\\
\>[3]{}\Varid{runParser}\;(\Conid{Op}\;(\in\!\mathit{Tail}\;\in\!\mathit{Head})\;\Conid{Symbol}\;\Varid{k})\;\Conid{Nil}\;\mathrel{=}\;\Conid{Nil}{}\<[E]%
\\
\>[3]{}\Varid{runParser}\;(\Conid{Op}\;(\in\!\mathit{Tail}\;\in\!\mathit{Head})\;\Conid{Symbol}\;\Varid{k})\;(\Varid{x}\;::\;\Varid{xs})\;\mathrel{=}\;\Varid{runParser}\;(\Varid{k}\;\Varid{x})\;\Varid{xs}{}\<[E]%
\ColumnHook
\end{hscode}\resethooks

In this article, we are more interested in the predicate transformer semantics of \ensuremath{\Conid{Parser}}.
Since the semantics of \ensuremath{\Conid{Parser}} refer to a state, the predicates depend on this state.
We can incorporate a mutable state of type \ensuremath{\Varid{s}} in predicate transformer semantics
by replacing the propositions in \ensuremath{\Conid{Set}} with predicates over the state in \ensuremath{\Varid{s}\;\to\;\Conid{Set}}.
We define the resulting type of stateful predicate transformers for an effect with signature \ensuremath{\Varid{e}} to be \ensuremath{\text{\itshape PT}^S\;\Varid{s}\;\Varid{e}}, as follows:
\begin{hscode}\SaveRestoreHook
\column{B}{@{}>{\hspre}l<{\hspost}@{}}%
\column{3}{@{}>{\hspre}l<{\hspost}@{}}%
\column{5}{@{}>{\hspre}l<{\hspost}@{}}%
\column{7}{@{}>{\hspre}l<{\hspost}@{}}%
\column{E}{@{}>{\hspre}l<{\hspost}@{}}%
\>[3]{}\Keyword{record}\;\text{\itshape PT}^S\;(\Varid{s}\;\mathbin{:}\;\Conid{Set})\;(\Varid{e}\;\mathbin{:}\;\Conid{Sig})\;\mathbin{:}\;\Conid{Set}\;\Keyword{where}{}\<[E]%
\\
\>[3]{}\hsindent{2}{}\<[5]%
\>[5]{}\Keyword{constructor}\;\Varid{mkPTS}{}\<[E]%
\\
\>[3]{}\hsindent{2}{}\<[5]%
\>[5]{}\Keyword{field}{}\<[E]%
\\
\>[5]{}\hsindent{2}{}\<[7]%
\>[7]{}\Varid{pt}\;\mathbin{:}\;(\Varid{c}\;\mathbin{:}\;\Conid{C}\;\Varid{e})\;\to\;(\Conid{R}\;\Varid{e}\;\Varid{c}\;\to\;\Varid{s}\;\to\;\Conid{Set})\;\to\;\Varid{s}\;\to\;\Conid{Set}{}\<[E]%
\\
\>[5]{}\hsindent{2}{}\<[7]%
\>[7]{}\Varid{mono}\;\mathbin{:}\;\forall\;\Varid{c}\;\Conid{P}\;\Conid{P'}\;\to\;(\forall\;\Varid{x}\;\Varid{t}\;\to\;\Conid{P}\;\Varid{x}\;\Varid{t}\;\to\;\Conid{P'}\;\Varid{x}\;\Varid{t})\;\to\;\Varid{pt}\;\Varid{c}\;\Conid{P}\;\subseteq\;\Varid{pt}\;\Varid{c}\;\Conid{P'}{}\<[E]%
\ColumnHook
\end{hscode}\resethooks
If we define \ensuremath{\text{\itshape PTs}^S} and \ensuremath{\Varid{lookupPTS}} analogously to \ensuremath{\Conid{PTs}} and \ensuremath{\Varid{lookupPT}}, 
we have found a predicate transformer semantics that incorporates the current state:
\begin{hscode}\SaveRestoreHook
\column{B}{@{}>{\hspre}l<{\hspost}@{}}%
\column{3}{@{}>{\hspre}l<{\hspost}@{}}%
\column{E}{@{}>{\hspre}l<{\hspost}@{}}%
\>[3]{}\llbracket \_ \rrbracket^S\;\mathbin{:}\;\!\!\;(\Varid{pts}\;\mathbin{:}\;\text{\itshape PTs}^S\;\Varid{s}\;\Varid{es})\;\to\;\Conid{Free}\;\Varid{es}\;\Varid{a}\;\to\;(\Varid{a}\;\to\;\Varid{s}\;\to\;\Conid{Set})\;\to\;\Varid{s}\;\to\;\Conid{Set}{}\<[E]%
\\
\>[3]{}\llbracket\Conid{Pure}\;\Varid{x}\rrbracket^S_{\Varid{pts}}\;\Conid{P}\;\mathrel{=}\;\Conid{P}\;\Varid{x}{}\<[E]%
\\
\>[3]{}\llbracket\Conid{Op}\;\Varid{i}\;\Varid{c}\;\Varid{k}\rrbracket^S_{\Varid{pts}}\;\Conid{P}\;\mathrel{=}\;\Varid{lookupPTS}\;\Varid{pts}\;\Varid{i}\;\Varid{c}\;(\lambda\;\Varid{x}\;\to\;\llbracket\Varid{k}\;\Varid{x}\rrbracket^S_{\Varid{pts}}\;\Conid{P}){}\<[E]%
\ColumnHook
\end{hscode}\resethooks

In this definition for \ensuremath{\llbracket \_ \rrbracket^S}, we assume that all effects share access to one mutable variable of type \ensuremath{\Varid{s}}.
We can allow for more variables by setting \ensuremath{\Varid{s}} to be a product type over the effects.
With a suitable modification of the predicate transformers,
we could set it up so that each effect can only modify its own associated variable.
Thus, the previous definition is not limited in generality by writing it only for one variable.

To give the predicate transformer semantics of the \ensuremath{\Conid{Parser}} effect,
we need to choose the meaning of failure, for the case where the next character is needed
and all characters have already been consumed.
Since we want all results returned by the parser to be correct,
we use demonic choice and the \ensuremath{\Varid{ptAll}} predicate transformer
as the semantics for \ensuremath{\Conid{Nondet}}.
Using \ensuremath{\Varid{ptAll}}'s semantics for the \ensuremath{\Conid{Fail}} command gives the following semantics for the \ensuremath{\Conid{Parser}} effect.
\begin{hscode}\SaveRestoreHook
\column{B}{@{}>{\hspre}l<{\hspost}@{}}%
\column{3}{@{}>{\hspre}l<{\hspost}@{}}%
\column{E}{@{}>{\hspre}l<{\hspost}@{}}%
\>[3]{}\Varid{ptParse}\;\mathbin{:}\;\text{\itshape PT}^S\;\Conid{String}\;\Conid{Parser}{}\<[E]%
\\
\>[3]{}\Conid{PTS.pt}\;\Varid{ptParse}\;\Conid{Symbol}\;\Conid{P}\;\Conid{Nil}\;\mathrel{=}\;\top{}\<[E]%
\\
\>[3]{}\Conid{PTS.pt}\;\Varid{ptParse}\;\Conid{Symbol}\;\Conid{P}\;(\Varid{x}\;::\;\Varid{xs})\;\mathrel{=}\;\Conid{P}\;\Varid{x}\;\Varid{xs}{}\<[E]%
\ColumnHook
\end{hscode}\resethooks

With the predicate transformer semantics of \ensuremath{\Conid{Parser}},
we can define the language accepted by a parser in the \ensuremath{\Conid{Free}} monad as a predicate over strings:
a string \ensuremath{\Varid{xs}} is in the language of a parser \ensuremath{\Conid{S}} if the postcondition ``all characters have been consumed'' is satisfied.
\begin{hscode}\SaveRestoreHook
\column{B}{@{}>{\hspre}l<{\hspost}@{}}%
\column{3}{@{}>{\hspre}l<{\hspost}@{}}%
\column{E}{@{}>{\hspre}l<{\hspost}@{}}%
\>[3]{}\Varid{empty?}\;\mathbin{:}\;\!\!\;\!\!\;\Conid{List}\;\Varid{a}\;\to\;\Conid{Set}{}\<[E]%
\\
\>[3]{}\Varid{empty?}\;\Conid{Nil}\;\mathrel{=}\;\top{}\<[E]%
\\
\>[3]{}\Varid{empty?}\;(\anonymous \;::\;\anonymous )\;\mathrel{=}\;\bot{}\<[E]%
\\[\blanklineskip]%
\>[3]{}\_\! \in \![\_]\;\mathbin{:}\;\!\!\;\Conid{String}\;\to\;\Conid{Free}\;(\Conid{Nondet}\;::\;\Conid{Parser}\;::\;\Conid{Nil})\;\Varid{a}\;\to\;\Conid{Set}{}\<[E]%
\\
\>[3]{}\Varid{xs}\;\in[\;\Conid{S}\;\mskip1.5mu]\;\mathrel{=}\;\llbracket\Conid{S}\rrbracket^S_{\Varid{ptAll}\;::\;\Varid{ptParse}\;::\;\Conid{Nil}}\;(\lambda\;\anonymous \;\to\;\Varid{empty?})\;\Varid{xs}{}\<[E]%
\ColumnHook
\end{hscode}\resethooks

\section{Parsing context-free languages} \label{sec:fromProductions}
In Section \ref{sec:dmatch}, we developed and formally verified a parser for regular languages.
The class of regular languages is small, and does not include most programming languages.
A class of languages that is more expressive than the regular languages,
while remaining tractable in parsing is that of the \emph{context-free language}.
The expressiveness of context-free languages is enough to cover most programming languages used in practice~\cite{dragon-book}.
We will represent context-free languages in Agda by giving a grammar in the style of \citet{dependent-grammar},
in a similar way as we represent a regular language using an element of the \ensuremath{\Conid{Regex}} type.
Following their development, we parametrize our definitions over a collection of non-terminal symbols.
\begin{hscode}\SaveRestoreHook
\column{B}{@{}>{\hspre}l<{\hspost}@{}}%
\column{3}{@{}>{\hspre}l<{\hspost}@{}}%
\column{5}{@{}>{\hspre}l<{\hspost}@{}}%
\column{E}{@{}>{\hspre}l<{\hspost}@{}}%
\>[B]{}\Keyword{record}\;\Conid{GrammarSymbols}\;\mathbin{:}\;\Conid{Set}\;\Keyword{where}{}\<[E]%
\\
\>[B]{}\hsindent{3}{}\<[3]%
\>[3]{}\Keyword{field}{}\<[E]%
\\
\>[3]{}\hsindent{2}{}\<[5]%
\>[5]{}\Conid{Nonterm}\;\mathbin{:}\;\Conid{Set}{}\<[E]%
\\
\>[3]{}\hsindent{2}{}\<[5]%
\>[5]{}\llbracket\!\_\!\rrbracket\;\mathbin{:}\;\Conid{Nonterm}\;\to\;\Conid{Set}{}\<[E]%
\\
\>[3]{}\hsindent{2}{}\<[5]%
\>[5]{}\_\!\mathbin{\smash{\overset{\raisebox{-0.2em}{\tiny ?}}{=}}}\!\_\;\mathbin{:}\;\Conid{Decidable}\;\{\mskip1.5mu \Conid{A}\;\mathrel{=}\;\Conid{Nonterm}\mskip1.5mu\}\;\_\!==\!\_{}\<[E]%
\ColumnHook
\end{hscode}\resethooks
The elements of the type \ensuremath{\Conid{Char}} are the \emph{terminal} symbols.
The elements of the type \ensuremath{\Conid{Nonterm}} are the \emph{non-terminal} symbols, representing the language constructs.
As for \ensuremath{\Conid{Char}}, we also need to be able to decide the equality of non-terminals.
The (disjoint) union of \ensuremath{\Conid{Char}} and \ensuremath{\Conid{Nonterm}} gives all the symbols that we can use in defining the grammar.
\begin{hscode}\SaveRestoreHook
\column{B}{@{}>{\hspre}l<{\hspost}@{}}%
\column{3}{@{}>{\hspre}l<{\hspost}@{}}%
\column{E}{@{}>{\hspre}l<{\hspost}@{}}%
\>[3]{}\Conid{Symbol}\;\mathrel{=}\;\Conid{Either}\;\Conid{Char}\;\Conid{Nonterm}{}\<[E]%
\\
\>[3]{}\Conid{Symbols}\;\mathrel{=}\;\Conid{List}\;\Conid{Symbol}{}\<[E]%
\ColumnHook
\end{hscode}\resethooks
For each non-terminal \ensuremath{\Conid{A}}, our goal is to parse a string into a value of type \ensuremath{\llbracket\;\Conid{A}\;\rrbracket},
based on a set of production rules.
A production rule $A \to xs$ gives a way to expand the non-terminal \ensuremath{\Conid{A}} into a list of symbols \ensuremath{\Varid{xs}},
such that successfully matching each symbol of \ensuremath{\Varid{xs}} with parts of a string
gives a match of the string with \ensuremath{\Conid{A}}.
Since matching a non-terminal symbol \ensuremath{\Conid{B}} with a (part of a) string results in a value of type \ensuremath{\llbracket\;\Conid{B}\;\rrbracket},
a production rule for \ensuremath{\Conid{A}} is associated with a \emph{semantic function} that takes all values arising from submatches
and returns a value of type \ensuremath{\llbracket\;\Conid{A}\;\rrbracket},
as expressed by the following type:
\begin{hscode}\SaveRestoreHook
\column{B}{@{}>{\hspre}l<{\hspost}@{}}%
\column{3}{@{}>{\hspre}l<{\hspost}@{}}%
\column{12}{@{}>{\hspre}l<{\hspost}@{}}%
\column{19}{@{}>{\hspre}l<{\hspost}@{}}%
\column{E}{@{}>{\hspre}l<{\hspost}@{}}%
\>[3]{}\llbracket\!\_\!\|\!\_\!\rrbracket\;\mathbin{:}\;\Conid{Symbols}\;\to\;\Conid{Nonterm}\;\to\;\Conid{Set}{}\<[E]%
\\
\>[3]{}\llbracket\;\Conid{Nil}\;{}\<[19]%
\>[19]{}\|\;\Conid{A}\;\rrbracket\;\mathrel{=}\;\llbracket\;\Conid{A}\;\rrbracket{}\<[E]%
\\
\>[3]{}\llbracket\;\Conid{Inl}\;\Varid{x}\;{}\<[12]%
\>[12]{}::\;\Varid{xs}\;{}\<[19]%
\>[19]{}\|\;\Conid{A}\;\rrbracket\;\mathrel{=}\;\llbracket\;\Varid{xs}\;\|\;\Conid{A}\;\rrbracket{}\<[E]%
\\
\>[3]{}\llbracket\;\Conid{Inr}\;\Conid{B}\;{}\<[12]%
\>[12]{}::\;\Varid{xs}\;{}\<[19]%
\>[19]{}\|\;\Conid{A}\;\rrbracket\;\mathrel{=}\;\llbracket\;\Conid{B}\;\rrbracket\;\to\;\llbracket\;\Varid{xs}\;\|\;\Conid{A}\;\rrbracket{}\<[E]%
\ColumnHook
\end{hscode}\resethooks
Now we can define the type of production rules. A rule of the form $A \to B c D$ is represented as \ensuremath{\Varid{prod}\;\Conid{A}\;(\Conid{Inr}\;\Conid{B}\;::\;\Conid{Inl}\;\Varid{c}\;::\;\Conid{Inr}\;\Conid{D}\;::\;\Conid{Nil})\;\Varid{f}} for some \ensuremath{\Varid{f}}.
\begin{hscode}\SaveRestoreHook
\column{B}{@{}>{\hspre}l<{\hspost}@{}}%
\column{3}{@{}>{\hspre}l<{\hspost}@{}}%
\column{5}{@{}>{\hspre}l<{\hspost}@{}}%
\column{7}{@{}>{\hspre}l<{\hspost}@{}}%
\column{E}{@{}>{\hspre}l<{\hspost}@{}}%
\>[3]{}\Keyword{record}\;\Conid{Prod}\;\mathbin{:}\;\Conid{Set}\;\Keyword{where}{}\<[E]%
\\
\>[3]{}\hsindent{2}{}\<[5]%
\>[5]{}\Keyword{constructor}\;\Varid{prod}{}\<[E]%
\\
\>[3]{}\hsindent{2}{}\<[5]%
\>[5]{}\Keyword{field}{}\<[E]%
\\
\>[5]{}\hsindent{2}{}\<[7]%
\>[7]{}\Varid{lhs}\;\mathbin{:}\;\Conid{Nonterm}{}\<[E]%
\\
\>[5]{}\hsindent{2}{}\<[7]%
\>[7]{}\Varid{rhs}\;\mathbin{:}\;\Conid{Symbols}{}\<[E]%
\\
\>[5]{}\hsindent{2}{}\<[7]%
\>[7]{}\Varid{sem}\;\mathbin{:}\;\llbracket\;\Varid{rhs}\;\|\;\Varid{lhs}\;\rrbracket{}\<[E]%
\ColumnHook
\end{hscode}\resethooks
We use the abbreviation \ensuremath{\Conid{Prods}} to represent a list of productions,
and a grammar will consist of the list of all relevant productions.

We want to show that a generally recursive function making use of the effects \ensuremath{\Conid{Parser}} and \ensuremath{\Conid{Nondet}} can parse any context-free grammar.
To show this claim, we implement a function \ensuremath{\Varid{fromProds}} that constructs a parser for any context-free grammar given as a list of \ensuremath{\Conid{Prod}}s,
then formally verify the correctness of \ensuremath{\Varid{fromProds}}.
Our implementation mirrors the definition of the \ensuremath{\Varid{generateParser}} function by \citeauthor{dependent-grammar},
differing in the naming and in the system that the parser is written in:
our implementation uses the \ensuremath{\Conid{Free}} monad and algebraic effects, while \citeauthor{dependent-grammar} use a monad \ensuremath{\Conid{Parser}} that is based on parser combinators.

We start by defining two auxiliary types, used as abbreviations in our code.
\begin{hscode}\SaveRestoreHook
\column{B}{@{}>{\hspre}l<{\hspost}@{}}%
\column{3}{@{}>{\hspre}l<{\hspost}@{}}%
\column{5}{@{}>{\hspre}l<{\hspost}@{}}%
\column{7}{@{}>{\hspre}l<{\hspost}@{}}%
\column{E}{@{}>{\hspre}l<{\hspost}@{}}%
\>[3]{}\Conid{FreeParser}\;\mathrel{=}\;\Conid{Free}\;(\Varid{mkSig}\;\Conid{Nonterm}\;\llbracket\!\_\!\rrbracket\;::\;\Conid{Nondet}\;::\;\Conid{Parser}\;::\;\Conid{Nil}){}\<[E]%
\\[\blanklineskip]%
\>[3]{}\Keyword{record}\;\Conid{ProdRHS}\;(\Conid{A}\;\mathbin{:}\;\Conid{Nonterm})\;\mathbin{:}\;\Conid{Set}\;\Keyword{where}{}\<[E]%
\\
\>[3]{}\hsindent{2}{}\<[5]%
\>[5]{}\Keyword{constructor}\;\Varid{prodrhs}{}\<[E]%
\\
\>[3]{}\hsindent{2}{}\<[5]%
\>[5]{}\Keyword{field}{}\<[E]%
\\
\>[5]{}\hsindent{2}{}\<[7]%
\>[7]{}\Varid{rhs}\;\mathbin{:}\;\Conid{Symbols}{}\<[E]%
\\
\>[5]{}\hsindent{2}{}\<[7]%
\>[7]{}\Varid{sem}\;\mathbin{:}\;\llbracket\;\Varid{rhs}\;\|\;\Conid{A}\;\rrbracket{}\<[E]%
\ColumnHook
\end{hscode}\resethooks

The core algorithm for parsing a context-free grammar consists of the following functions,
calling each other in mutual recursion:
\begin{hscode}\SaveRestoreHook
\column{B}{@{}>{\hspre}l<{\hspost}@{}}%
\column{3}{@{}>{\hspre}l<{\hspost}@{}}%
\column{16}{@{}>{\hspre}l<{\hspost}@{}}%
\column{E}{@{}>{\hspre}l<{\hspost}@{}}%
\>[3]{}\Varid{fromProds}\;{}\<[16]%
\>[16]{}\mathbin{:}\;(\Conid{A}\;\mathbin{:}\;\Conid{Nonterm})\;\to\;\Conid{FreeParser}\;\llbracket\;\Conid{A}\;\rrbracket{}\<[E]%
\\
\>[3]{}\Varid{filterLHS}\;{}\<[16]%
\>[16]{}\mathbin{:}\;(\Conid{A}\;\mathbin{:}\;\Conid{Nonterm})\;\to\;\Conid{Prods}\;\to\;\Conid{List}\;(\Conid{ProdRHS}\;\Conid{A}){}\<[E]%
\\
\>[3]{}\Varid{fromProd}\;{}\<[16]%
\>[16]{}\mathbin{:}\;\!\!\;\Conid{ProdRHS}\;\Conid{A}\;\to\;\Conid{FreeParser}\;\llbracket\;\Conid{A}\;\rrbracket{}\<[E]%
\\
\>[3]{}\Varid{buildParser}\;{}\<[16]%
\>[16]{}\mathbin{:}\;\!\!\;(\Varid{xs}\;\mathbin{:}\;\Conid{Symbols})\;\to\;\Conid{FreeParser}\;(\llbracket\;\Varid{xs}\;\|\;\Conid{A}\;\rrbracket\;\to\;\llbracket\;\Conid{A}\;\rrbracket){}\<[E]%
\\
\>[3]{}\Varid{exact}\;{}\<[16]%
\>[16]{}\mathbin{:}\;\!\!\;\Varid{a}\;\to\;\Conid{Char}\;\to\;\Conid{FreeParser}\;\Varid{a}{}\<[E]%
\ColumnHook
\end{hscode}\resethooks
The main function is \ensuremath{\Varid{fromProds}}: given a non-terminal,
it selects the productions with this non-terminal on the left hand side using \ensuremath{\Varid{filterLHS}},
and makes a nondeterministic choice between them.
\begin{hscode}\SaveRestoreHook
\column{B}{@{}>{\hspre}l<{\hspost}@{}}%
\column{3}{@{}>{\hspre}l<{\hspost}@{}}%
\column{19}{@{}>{\hspre}l<{\hspost}@{}}%
\column{E}{@{}>{\hspre}l<{\hspost}@{}}%
\>[3]{}\Varid{filterLHS}\;\Conid{A}\;\Conid{Nil}\;\mathrel{=}\;\Conid{Nil}{}\<[E]%
\\
\>[3]{}\Varid{filterLHS}\;\Conid{A}\;(\Varid{prod}\;\Varid{lhs}\;\Varid{rhs}\;\Varid{sem}\;::\;\Varid{ps})\;\Keyword{with}\;\Conid{A}\;\mathbin{\smash{\overset{\raisebox{-0.2em}{\tiny ?}}{=}}}\;\Varid{lhs}{}\<[E]%
\\
\>[3]{}\Varid{...}\;\mid \;\Varid{yes}\;\Varid{refl}\;{}\<[19]%
\>[19]{}\mathrel{=}\;\Varid{prodrhs}\;\Varid{rhs}\;\Varid{sem}\;::\;\Varid{filterLHS}\;\Conid{A}\;\Varid{ps}{}\<[E]%
\\
\>[3]{}\Varid{...}\;\mid \;\Varid{no}\;\anonymous \;{}\<[19]%
\>[19]{}\mathrel{=}\;\Varid{filterLHS}\;\Conid{A}\;\Varid{ps}{}\<[E]%
\\[\blanklineskip]%
\>[3]{}\Varid{fromProds}\;\Conid{A}\;\mathrel{=}\;\Varid{choices}\;\!\!\;(\Varid{map}\;\Varid{fromProd}\;(\Varid{filterLHS}\;\Conid{A}\;\Varid{prods})){}\<[E]%
\ColumnHook
\end{hscode}\resethooks
The \ensuremath{\Varid{choices}} operator takes a list of computations and nondeterministically chooses one of them to execute.

The function \ensuremath{\Varid{fromProd}} takes a single production and tries to parse the input string using this production.
It then uses the semantic function of the production to give the resulting value.
\begin{hscode}\SaveRestoreHook
\column{B}{@{}>{\hspre}l<{\hspost}@{}}%
\column{3}{@{}>{\hspre}l<{\hspost}@{}}%
\column{E}{@{}>{\hspre}l<{\hspost}@{}}%
\>[3]{}\Varid{fromProd}\;(\Varid{prodrhs}\;\Varid{rhs}\;\Varid{sem})\;\mathrel{=}\;\Varid{buildParser}\;\Varid{rhs}\;\bind \;\lambda\;\Varid{f}\;\to\;\Conid{Pure}\;(\Varid{f}\;\Varid{sem}){}\<[E]%
\ColumnHook
\end{hscode}\resethooks
The function \ensuremath{\Varid{buildParser}} iterates over the \ensuremath{\Conid{Symbols}}, calling \ensuremath{\Varid{exact}} for each literal character symbol,
and making a recursive \ensuremath{\Varid{call}} to \ensuremath{\Varid{fromProds}} for each non-terminal symbol.
\begin{hscode}\SaveRestoreHook
\column{B}{@{}>{\hspre}l<{\hspost}@{}}%
\column{3}{@{}>{\hspre}l<{\hspost}@{}}%
\column{5}{@{}>{\hspre}l<{\hspost}@{}}%
\column{23}{@{}>{\hspre}l<{\hspost}@{}}%
\column{E}{@{}>{\hspre}l<{\hspost}@{}}%
\>[3]{}\Varid{buildParser}\;\Conid{Nil}\;\mathrel{=}\;\Conid{Pure}\;\Varid{id}{}\<[E]%
\\
\>[3]{}\Varid{buildParser}\;(\Conid{Inl}\;\Varid{x}\;{}\<[23]%
\>[23]{}::\;\Varid{xs})\;\mathrel{=}\;\Varid{exact}\;\Varid{tt}\;\Varid{x}\;\bind \;\lambda\;\anonymous \;\to\;\Varid{buildParser}\;\Varid{xs}{}\<[E]%
\\
\>[3]{}\Varid{buildParser}\;(\Conid{Inr}\;\Conid{B}\;{}\<[23]%
\>[23]{}::\;\Varid{xs})\;\mathrel{=}\;\textrm{\bfseries do}\;{}\<[E]%
\\
\>[3]{}\hsindent{2}{}\<[5]%
\>[5]{}\Varid{x}\;\leftarrow \;\Varid{call}\;\!\!\;\Conid{B}\;{}\<[E]%
\\
\>[3]{}\hsindent{2}{}\<[5]%
\>[5]{}\Varid{o}\;\leftarrow \;\Varid{buildParser}\;\Varid{xs}\;{}\<[E]%
\\
\>[3]{}\hsindent{2}{}\<[5]%
\>[5]{}\Conid{Pure}\;\lambda\;\Varid{f}\;\to\;\Varid{o}\;(\Varid{f}\;\Varid{x}){}\<[E]%
\ColumnHook
\end{hscode}\resethooks
Finally, \ensuremath{\Varid{exact}} uses the \ensuremath{\Varid{symbol}} command to check that the next character in the string is as expected,
and \ensuremath{\Varid{fail}}s if this is not the case.
\begin{hscode}\SaveRestoreHook
\column{B}{@{}>{\hspre}l<{\hspost}@{}}%
\column{3}{@{}>{\hspre}l<{\hspost}@{}}%
\column{E}{@{}>{\hspre}l<{\hspost}@{}}%
\>[3]{}\Varid{exact}\;\Varid{x}\;\Varid{t}\;\mathrel{=}\;\Varid{symbol}\;\!\!\;\bind \;\lambda\;\Varid{t'}\;\to\;\textrm{\bfseries if}\;\Varid{t}\;\mathbin{\smash{\overset{\raisebox{-0.2em}{\tiny ?}}{=}}}\;\Varid{t'}\;\textrm{\bfseries then}\;\Conid{Pure}\;\Varid{x}\;\textrm{\bfseries else}\;\Varid{fail}\;\!\!{}\<[E]%
\ColumnHook
\end{hscode}\resethooks

\section{Partial correctness of the parser} \label{sec:partialCorrectness}
Partial correctness of the parser is relatively simple to show,
as soon as we have a specification.
Since we want to prove that \ensuremath{\Varid{fromProds}} correctly parses any given context free grammar given as an element of \ensuremath{\Conid{Prods}},
the specification consists of a relation between many sets:
the production rules, an input string, a non-terminal, the output of the parser, and the remaining unparsed string.
Due to the many arguments, the notation is unfortunately somewhat unwieldy.
To make it a bit easier to read, we define two relations in mutual recursion,
one for all productions of a non-terminal,
and for matching a string with a single production rule.

\begin{hscode}\SaveRestoreHook
\column{B}{@{}>{\hspre}l<{\hspost}@{}}%
\column{3}{@{}>{\hspre}l<{\hspost}@{}}%
\column{5}{@{}>{\hspre}l<{\hspost}@{}}%
\column{16}{@{}>{\hspre}l<{\hspost}@{}}%
\column{E}{@{}>{\hspre}l<{\hspost}@{}}%
\>[3]{}\Keyword{data}\;\_\!\vdash\!\_\!\in \llbracket\!\_\!\rrbracket \Rightarrow\!\_\!\Varid{,}\!\_\;\Varid{prods}\;\Keyword{where}{}\<[E]%
\\
\>[3]{}\hsindent{2}{}\<[5]%
\>[5]{}\Conid{Produce}\;\mathbin{:}\;{}\<[16]%
\>[16]{}\!\!\;\Varid{prod}\;\Varid{lhs}\;\Varid{rhs}\;\Varid{sem}\;\in\;\Varid{prods}\;\to\;{}\<[E]%
\\
\>[16]{}\Varid{prods}\;\vdash\;\Varid{xs}\;\sim\;\Varid{rhs}\;\Rightarrow \;\Varid{f}\;\Varid{,}\;\Varid{ys}\;\to\;{}\<[E]%
\\
\>[16]{}\Varid{prods}\;\vdash\;\Varid{xs}\;\in \llbracket\;\Varid{lhs}\;\rrbracket \Rightarrow\;\Varid{f}\;\Varid{sem}\;\Varid{,}\;\Varid{ys}{}\<[E]%
\\
\>[3]{}\Keyword{data}\;\_\!\vdash\!\_\!\sim\!\_\!\Rightarrow \!\_\!\Varid{,}\!\_\;\Varid{prods}\;\Keyword{where}{}\<[E]%
\\
\>[3]{}\hsindent{2}{}\<[5]%
\>[5]{}\Conid{Done}\;\mathbin{:}\;{}\<[16]%
\>[16]{}\!\!\;\Varid{prods}\;\vdash\;\Varid{xs}\;\sim\;\Conid{Nil}\;\Rightarrow \;\Varid{id}\;\Varid{,}\;\Varid{xs}{}\<[E]%
\\
\>[3]{}\hsindent{2}{}\<[5]%
\>[5]{}\Conid{Next}\;\mathbin{:}\;{}\<[16]%
\>[16]{}\!\!\;\Varid{prods}\;\vdash\;\Varid{xs}\;\sim\;\Varid{ps}\;\Rightarrow \;\Varid{o}\;\Varid{,}\;\Varid{ys}\;\to\;{}\<[E]%
\\
\>[16]{}\Varid{prods}\;\vdash\;(\Varid{x}\;::\;\Varid{xs})\;\sim\;(\Conid{Inl}\;\Varid{x}\;::\;\Varid{ps})\;\Rightarrow \;\Varid{o}\;\Varid{,}\;\Varid{ys}{}\<[E]%
\\
\>[3]{}\hsindent{2}{}\<[5]%
\>[5]{}\Conid{Call}\;\mathbin{:}\;{}\<[16]%
\>[16]{}\!\!\;\Varid{prods}\;\vdash\;\Varid{xs}\;\in \llbracket\;\Conid{A}\;\rrbracket \Rightarrow\;\Varid{o}\;\Varid{,}\;\Varid{ys}\;\to\;{}\<[E]%
\\
\>[16]{}\Varid{prods}\;\vdash\;\Varid{ys}\;\sim\;\Varid{ps}\;\Rightarrow \;\Varid{f}\;\Varid{,}\;\Varid{zs}\;\to\;{}\<[E]%
\\
\>[16]{}\Varid{prods}\;\vdash\;\Varid{xs}\;\sim\;(\Conid{Inr}\;\Conid{A}\;::\;\Varid{ps})\;\Rightarrow \;(\lambda\;\Varid{g}\;\to\;\Varid{f}\;(\Varid{g}\;\Varid{o}))\;\Varid{,}\;\Varid{zs}{}\<[E]%
\ColumnHook
\end{hscode}\resethooks
With these relations, we can define the specification \ensuremath{\Varid{parserSpec}} to be equal to \ensuremath{\_\!\vdash\!\_\!\in \llbracket\!\_\!\rrbracket \Rightarrow\!\_\!\Varid{,}\!\_} (up to reordering some arguments),
and show that \ensuremath{\Varid{fromProds}} refines this specification.
To state that the refinement relation holds, we first need to determine the semantics of the effects.
We choose \ensuremath{\Varid{ptAll}} as the semantics of nondeterminism, since we want to ensure all output of the parser is correct.
\begin{hscode}\SaveRestoreHook
\column{B}{@{}>{\hspre}l<{\hspost}@{}}%
\column{3}{@{}>{\hspre}l<{\hspost}@{}}%
\column{5}{@{}>{\hspre}l<{\hspost}@{}}%
\column{E}{@{}>{\hspre}l<{\hspost}@{}}%
\>[3]{}\Varid{pts}\;\Varid{prods}\;\mathrel{=}\;\Varid{ptRec}\;(\Varid{parserSpec}\;\Varid{prods})\;::\;\Varid{ptAll}\;::\;\Varid{ptParse}\;::\;\Conid{Nil}{}\<[E]%
\\[\blanklineskip]%
\>[3]{}\llbracket\Conid{S}\rrbracket_{\text{fromProd}}\ \Varid{prods}\;\mathrel{=}\;\llbracket\Conid{S}\rrbracket^S_{\Varid{pts}\;\Varid{prods}}{}\<[E]%
\\[\blanklineskip]%
\>[3]{}\Varid{partialCorrectness}\;\mathbin{:}\;(\Varid{prods}\;\mathbin{:}\;\Conid{Prods})\;(\Conid{A}\;\mathbin{:}\;\Conid{Nonterm})\;\to\;{}\<[E]%
\\
\>[3]{}\hsindent{2}{}\<[5]%
\>[5]{}\llbracket\top,(\Varid{parserSpec}\;\Varid{prods}\;\Conid{A})\rrbracket_{\text{spec}}\;\sqsubseteq\;\llbracket\Varid{fromProds}\;\Varid{prods}\;\Conid{A}\rrbracket_{\text{fromProd}}\ \Varid{prods}{}\<[E]%
\ColumnHook
\end{hscode}\resethooks

Let us fix the production rules \ensuremath{\Varid{prods}}.
How do we prove the partial correctness of a parser for \ensuremath{\Varid{prods}}?
Since the structure of \ensuremath{\Varid{fromProds}} is of a nondeterministic choice between productions to be parsed,
and we want to show that all alternatives for a choice result in success,
we will first give a lemma expressing the correctness of each alternative.
Correctness in this case is expressed by the semantics of a single production rule,
i.e. the \ensuremath{\_\!\vdash\!\_\!\sim\!\_\!\Rightarrow \!\_\!\Varid{,}\!\_} relation.
Thus, we want to prove the following lemma:
\begin{hscode}\SaveRestoreHook
\column{B}{@{}>{\hspre}l<{\hspost}@{}}%
\column{5}{@{}>{\hspre}l<{\hspost}@{}}%
\column{7}{@{}>{\hspre}l<{\hspost}@{}}%
\column{E}{@{}>{\hspre}l<{\hspost}@{}}%
\>[5]{}\Varid{parseStep}\;\mathbin{:}\;\forall\;\Conid{A}\;\Varid{xs}\;\Conid{P}\;\Varid{str}\;\to\;{}\<[E]%
\\
\>[5]{}\hsindent{2}{}\<[7]%
\>[7]{}(\forall\;\Varid{o}\;\Varid{str'}\;\to\;\Varid{prods}\;\vdash\;\Varid{str}\;\sim\;\Varid{xs}\;\Rightarrow \;\Varid{o}\;\Varid{,}\;\Varid{str'}\;\to\;\Conid{P}\;\Varid{o}\;\Varid{str'})\;\to\;{}\<[E]%
\\
\>[5]{}\hsindent{2}{}\<[7]%
\>[7]{}\llbracket\Varid{buildParser}\;\Varid{prods}\;\Varid{xs}\rrbracket_{\text{fromProd}}\ \Varid{prods}\;\Conid{P}\;\Varid{str}{}\<[E]%
\ColumnHook
\end{hscode}\resethooks
The lemma can be proved by reproducing the case distinctions used to define \ensuremath{\Varid{buildParser}};
there is no complication apart from having to use the \ensuremath{\Varid{wpToBind}} lemma to deal with the \ensuremath{\_\!\bind\!\_} operator in a few places.
\begin{hscode}\SaveRestoreHook
\column{B}{@{}>{\hspre}l<{\hspost}@{}}%
\column{5}{@{}>{\hspre}l<{\hspost}@{}}%
\column{7}{@{}>{\hspre}l<{\hspost}@{}}%
\column{9}{@{}>{\hspre}l<{\hspost}@{}}%
\column{E}{@{}>{\hspre}l<{\hspost}@{}}%
\>[5]{}\Varid{parseStep}\;\Conid{A}\;\Conid{Nil}\;\Conid{P}\;\Varid{t}\;\Conid{H}\;\mathrel{=}\;\Conid{H}\;\Varid{id}\;\Varid{t}\;\Conid{Done}{}\<[E]%
\\
\>[5]{}\Varid{parseStep}\;\Conid{A}\;(\Conid{Inl}\;\Varid{x}\;::\;\Varid{xs})\;\Conid{P}\;\Conid{Nil}\;\Conid{H}\;\mathrel{=}\;\Varid{tt}{}\<[E]%
\\
\>[5]{}\Varid{parseStep}\;\Conid{A}\;(\Conid{Inl}\;\Varid{x}\;::\;\Varid{xs})\;\Conid{P}\;(\Varid{x'}\;::\;\Varid{t})\;\Conid{H}\;\Keyword{with}\;\Varid{x}\;\mathbin{\smash{\overset{\raisebox{-0.2em}{\tiny ?}}{=}}}\;\Varid{x'}{}\<[E]%
\\
\>[5]{}\Varid{...}\;\mid \;\Varid{yes}\;\Varid{refl}\;\mathrel{=}\;\Varid{parseStep}\;\Conid{A}\;\Varid{xs}\;\Conid{P}\;\Varid{t}\;\lambda\;\Varid{o}\;\Varid{t'}\;\Conid{H'}\;\to\;\Conid{H}\;\Varid{o}\;\Varid{t'}\;(\Conid{Next}\;\Conid{H'}){}\<[E]%
\\
\>[5]{}\Varid{...}\;\mid \;\Varid{no}\;\Varid{¬p}\;\mathrel{=}\;\Varid{tt}{}\<[E]%
\\
\>[5]{}\Varid{parseStep}\;\Conid{A}\;(\Conid{Inr}\;\Conid{B}\;::\;\Varid{xs})\;\Conid{P}\;\Varid{t}\;\Conid{H}\;\Varid{o}\;\Varid{t'}\;\Conid{Ho}\;\mathrel{=}\;{}\<[E]%
\\
\>[5]{}\hsindent{2}{}\<[7]%
\>[7]{}\Varid{wpToBind}\;(\Varid{buildParser}\;\Varid{prods}\;\Varid{xs})\;\anonymous \;\anonymous \;{}\<[E]%
\\
\>[7]{}\hsindent{2}{}\<[9]%
\>[9]{}(\Varid{parseStep}\;\Conid{A}\;\Varid{xs}\;\anonymous \;\Varid{t'}\;\lambda\;\Varid{o'}\;\Varid{str'}\;\Conid{Ho'}\;\to\;\Conid{H}\;\anonymous \;\anonymous \;(\Conid{Call}\;\Conid{Ho}\;\Conid{Ho'})){}\<[E]%
\ColumnHook
\end{hscode}\resethooks

To combine the \ensuremath{\Varid{parseStep}} for each of the productions that the nondeterministic choice is made between,
it is tempting to define another lemma \ensuremath{\Varid{filterStep}} by induction on the list of productions.
But we must be careful that the productions that are used in the \ensuremath{\Varid{parseStep}} are the full list \ensuremath{\Varid{prods}},
not the sublist \ensuremath{\Varid{prods'}} used in the induction step.
Additionally, we must also make sure that \ensuremath{\Varid{prods'}} is indeed a sublist,
since using an incorrect production rule in the \ensuremath{\Varid{parseStep}} will result in an invalid result.
Thus, we parametrize \ensuremath{\Varid{filterStep}} by a list \ensuremath{\Varid{prods'}} and a proof that it is a sublist of \ensuremath{\Varid{prods}}.
Again, the proof uses the same distinction as \ensuremath{\Varid{fromProds}} does,
and uses the \ensuremath{\Varid{wpToBind}} lemma to deal with the \ensuremath{\_\!\bind\!\_} operator.
\begin{hscode}\SaveRestoreHook
\column{B}{@{}>{\hspre}l<{\hspost}@{}}%
\column{5}{@{}>{\hspre}l<{\hspost}@{}}%
\column{7}{@{}>{\hspre}l<{\hspost}@{}}%
\column{E}{@{}>{\hspre}l<{\hspost}@{}}%
\>[5]{}\Varid{filterStep}\;\mathbin{:}\;\forall\;\Varid{prods'}\;\Conid{A}\;\to\;(\!\!\;\Varid{p}\;\in\;\Varid{prods'}\;\to\;\Varid{p}\;\in\;\Varid{prods})\;\to\;{}\<[E]%
\\
\>[5]{}\hsindent{2}{}\<[7]%
\>[7]{}\llbracket\top,(\Varid{parserSpec}\;\Varid{prods}\;\Conid{A})\rrbracket_{\text{spec}}\;\sqsubseteq\;\llbracket\Varid{choices}\;\!\!\;(\Varid{map}\;(\Varid{fromProd}\;\Varid{prods})\;(\Varid{filterLHS}\;\Varid{prods}\;\Conid{A}\;\Varid{prods'}))\rrbracket_{\text{fromProd}}\ \Varid{prods}{}\<[E]%
\\
\>[5]{}\Varid{filterStep}\;\Conid{Nil}\;\Conid{A}\;\Varid{subset}\;\Conid{P}\;\Varid{xs}\;\Conid{H}\;\mathrel{=}\;\Varid{tt}{}\<[E]%
\\
\>[5]{}\Varid{filterStep}\;(\Varid{prod}\;\Varid{lhs}\;\Varid{rhs}\;\Varid{sem}\;::\;\Varid{prods'})\;\Conid{A}\;\Varid{subset}\;\Conid{P}\;\Varid{xs}\;\Conid{H}\;\Keyword{with}\;\Conid{A}\;\mathbin{\smash{\overset{\raisebox{-0.2em}{\tiny ?}}{=}}}\;\Varid{lhs}{}\<[E]%
\\
\>[5]{}\Varid{filterStep}\;(\Varid{prod}\;\Varid{.A}\;\Varid{rhs}\;\Varid{sem}\;::\;\Varid{prods'})\;\Conid{A}\;\Varid{subset}\;\Conid{P}\;\Varid{xs}\;(\anonymous \;\Varid{,}\;\Conid{H})\;\mid \;\Varid{yes}\;\Varid{refl}\;{}\<[E]%
\\
\>[5]{}\hsindent{2}{}\<[7]%
\>[7]{}\mathrel{=}\;\Varid{wpToBind}\;(\Varid{buildParser}\;\Varid{prods}\;\Varid{rhs})\;\anonymous \;\anonymous \;{}\<[E]%
\\
\>[5]{}\hsindent{2}{}\<[7]%
\>[7]{}(\Varid{parseStep}\;\Conid{A}\;\Varid{rhs}\;\anonymous \;\Varid{xs}\;\lambda\;\Varid{o}\;\Varid{t'}\;\Conid{H'}\;\to\;\Conid{H}\;\anonymous \;\anonymous \;(\Conid{Produce}\;(\Varid{subset}\;\in\!\mathit{Head})\;\Conid{H'}))\;{}\<[E]%
\\
\>[5]{}\hsindent{2}{}\<[7]%
\>[7]{}\Varid{,}\;\Varid{filterStep}\;\Varid{prods'}\;\Conid{A}\;(\Varid{subset}\;\circ\;\in\!\mathit{Tail})\;\Conid{P}\;\Varid{xs}\;(\anonymous \;\Varid{,}\;\Conid{H}){}\<[E]%
\\
\>[5]{}\Varid{...}\;\mid \;\Varid{no}\;\Varid{¬p}\;\mathrel{=}\;\Varid{filterStep}\;\Varid{prods'}\;\Conid{A}\;(\Varid{subset}\;\circ\;\in\!\mathit{Tail})\;\Conid{P}\;\Varid{xs}\;\Conid{H}{}\<[E]%
\ColumnHook
\end{hscode}\resethooks

With these lemmas, \ensuremath{\Varid{partialCorrectness}} just consists of applying \ensuremath{\Varid{filterStep}} to the subset of \ensuremath{\Varid{prods}} consisting of \ensuremath{\Varid{prods}} itself.

\section{Termination of the parser} \label{sec:fromProds-terminates}
To show termination we need a somewhat more subtle argument:
since we are able to call the same non-terminal repeatedly,
termination cannot be shown simply by inspecting each alternative in the definition.
Consider the grammar given by $E \rightarrow a E; E \rightarrow b$,
where we see that the string that matches $E$ in the recursive case is shorter than the original string,
but the definition itself can be expanded to unbounded length.
By taking into account the current state, i.e. the string to be parsed, in the variant,
we can show that a decreasing string length leads to termination.

But not all grammars feature this decreasing string length in the recursive case,
with the most pathological case being those of the form $E \to E$.
The issues do not only occur in edge cases: the grammar $E \to E + E; E \to 1$ representing very simple expressions
will already result in non-termination for \ensuremath{\Varid{fromProds}}
as it will go in recursion on the first non-terminal without advancing the input string.
Since the position in the string and current non-terminal together fully determine the state of \ensuremath{\Varid{fromParsers}},
it will not terminate.
We need to ensure that the grammars passed to the parser do not allow for such loops.

Intuitively, the condition on the grammars should be that they are not \emph{left-recursive},
since in that case, the parser should always advance its position in the string before it encounters the same non-terminal.
This means that the number of recursive calls to \ensuremath{\Varid{fromProds}} is bounded
by the length of the string times the number of different non-terminals occurring in the production rules.
The type we will use to describe the predicate ``there is no left recursion'' is constructively somewhat stronger:
we define a left-recursion chain from $A$ to $B$ to be
a sequence of non-terminals $A, \dots, A_i, A_{i+1}, \dots, B$,
such that for each adjacent pair $A_i, A_{i+1}$ in the chain, there is a production of the form $A_{i+1} \to B_1 B_2 \dots B_n A_i \dots$, where $B_1 \dots B_n$ are all non-terminals.
In other words, we can advance the parser to $A$ starting in $B$ without consuming a character.
Disallowing (unbounded) left recursion is not a limitation for our parsers:
\citet{dependent-grammar} have shown that the \emph{left-corner transform}
can transform left-recursive grammars into an equivalent grammar without left recursion.
Moreover, they have implemented this transform, including formal verification, in Agda.
In this work, we assume that the left-corner transform has already been applied if needed,
so that there is an upper bound on the length of left-recursive chains in the grammar.

We formalize one link of this left-recursive chain in the type \ensuremath{\Conid{LRec}},
while a list of such links forms the \ensuremath{\Conid{Chain}} data type.
\begin{hscode}\SaveRestoreHook
\column{B}{@{}>{\hspre}l<{\hspost}@{}}%
\column{3}{@{}>{\hspre}l<{\hspost}@{}}%
\column{5}{@{}>{\hspre}l<{\hspost}@{}}%
\column{7}{@{}>{\hspre}l<{\hspost}@{}}%
\column{E}{@{}>{\hspre}l<{\hspost}@{}}%
\>[3]{}\Keyword{record}\;\Conid{LRec}\;(\Varid{prods}\;\mathbin{:}\;\Conid{Prods})\;(\Conid{A}\;\Conid{B}\;\mathbin{:}\;\Conid{Nonterm})\;\mathbin{:}\;\Conid{Set}\;\Keyword{where}{}\<[E]%
\\
\>[3]{}\hsindent{2}{}\<[5]%
\>[5]{}\Keyword{field}{}\<[E]%
\\
\>[5]{}\hsindent{2}{}\<[7]%
\>[7]{}\Varid{rec}\;\mathbin{:}\;\Varid{prod}\;\Conid{A}\;(\Varid{map}\;\Conid{Inr}\;\Varid{xs}\;\plus \;(\Conid{Inr}\;\Conid{B}\;::\;\Varid{ys}))\;\Varid{sem}\;\in\;\Varid{prods}{}\<[E]%
\ColumnHook
\end{hscode}\resethooks
(We leave \ensuremath{\Varid{xs}}, \ensuremath{\Varid{ys}} and \ensuremath{\Varid{sem}} as implicit fields of \ensuremath{\Conid{LRec}}, since they are fixed by the type of \ensuremath{\Varid{rec}}.)
\begin{hscode}\SaveRestoreHook
\column{B}{@{}>{\hspre}l<{\hspost}@{}}%
\column{3}{@{}>{\hspre}l<{\hspost}@{}}%
\column{5}{@{}>{\hspre}l<{\hspost}@{}}%
\column{E}{@{}>{\hspre}l<{\hspost}@{}}%
\>[3]{}\Keyword{data}\;\Conid{Chain}\;(\Varid{prods}\;\mathbin{:}\;\Conid{Prods})\;\mathbin{:}\;\Conid{Nonterm}\;\to\;\Conid{Nonterm}\;\to\;\Conid{Set}\;\Keyword{where}{}\<[E]%
\\
\>[3]{}\hsindent{2}{}\<[5]%
\>[5]{}\Conid{Nil}\;\mathbin{:}\;\!\!\;\Conid{Chain}\;\Varid{prods}\;\Conid{A}\;\Conid{A}{}\<[E]%
\\
\>[3]{}\hsindent{2}{}\<[5]%
\>[5]{}\_\!::\!\_\;\mathbin{:}\;\!\!\;\Conid{LRec}\;\Varid{prods}\;\Conid{B}\;\Conid{A}\;\to\;\Conid{Chain}\;\Varid{prods}\;\Conid{A}\;\Conid{C}\;\to\;\Conid{Chain}\;\Varid{prods}\;\Conid{B}\;\Conid{C}{}\<[E]%
\ColumnHook
\end{hscode}\resethooks
Now we say that a set of productions has no left recursion if all such chains have an upper bound on their length.
\begin{hscode}\SaveRestoreHook
\column{B}{@{}>{\hspre}l<{\hspost}@{}}%
\column{3}{@{}>{\hspre}l<{\hspost}@{}}%
\column{E}{@{}>{\hspre}l<{\hspost}@{}}%
\>[3]{}\Varid{chainLength}\;\mathbin{:}\;\!\!\;\Conid{Chain}\;\Varid{prods}\;\Conid{A}\;\Conid{B}\;\to\;\N{}\<[E]%
\\
\>[3]{}\Varid{chainLength}\;\Conid{Nil}\;\mathrel{=}\;\Varid{0}{}\<[E]%
\\
\>[3]{}\Varid{chainLength}\;(\Varid{c}\;::\;\Varid{cs})\;\mathrel{=}\;\Conid{Succ}\;(\Varid{chainLength}\;\Varid{cs}){}\<[E]%
\\[\blanklineskip]%
\>[3]{}\Varid{leftRecBound}\;\mathbin{:}\;\Conid{Prods}\;\to\;\N\;\to\;\Conid{Set}{}\<[E]%
\\
\>[3]{}\Varid{leftRecBound}\;\Varid{prods}\;\Varid{n}\;\mathrel{=}\;\!\!\;(\Varid{cs}\;\mathbin{:}\;\Conid{Chain}\;\Varid{prods}\;\Conid{A}\;\Conid{B})\;\to\;\Varid{chainLength}\;\Varid{cs}\;\mathbin{<}\;\Varid{n}{}\<[E]%
\ColumnHook
\end{hscode}\resethooks
If we have this bound on left recursion, we are able to prove termination,
since each call to \ensuremath{\Varid{fromProds}} will be made either after we have consumed an extra character,
or it is a left-recursive step, of which there is an upper bound on the sequence.

This informal proof fits better with a different notion of termination than in the petrol-driven semantics.
The petrol-driven semantics are based on a syntactic argument:
we know a computation terminates because expanding the call tree will eventually result in no more \ensuremath{\Varid{call}}s.
Here, we want to capture the notion that a recursive definition terminates
if all recursive calls are made to a smaller argument, according to a well-founded relation.
\begin{Def}[\cite{aczel-acc}]
In intuitionistic type theory, we say that a relation \ensuremath{\_\!\mathbin{\prec}\!\_} on a type \ensuremath{\Varid{a}} is well-founded if all elements \ensuremath{\Varid{x}\;\mathbin{:}\;\Varid{a}} are \emph{accessible},
which is defined by (well-founded) recursion to be the case if all elements in the downset of \ensuremath{\Varid{x}} are accessible.
\begin{hscode}\SaveRestoreHook
\column{B}{@{}>{\hspre}l<{\hspost}@{}}%
\column{3}{@{}>{\hspre}l<{\hspost}@{}}%
\column{5}{@{}>{\hspre}l<{\hspost}@{}}%
\column{E}{@{}>{\hspre}l<{\hspost}@{}}%
\>[3]{}\Keyword{data}\;\Conid{Acc}\;\!\!\;(\_\!\mathbin{\prec}\!\_\;\mathbin{:}\;\Varid{a}\;\to\;\Varid{a}\;\to\;\Conid{Set})\;\mathbin{:}\;\Varid{a}\;\to\;\Conid{Set}\;\Keyword{where}{}\<[E]%
\\
\>[3]{}\hsindent{2}{}\<[5]%
\>[5]{}\Varid{acc}\;\mathbin{:}\;\!\!\;(\forall\;\Varid{y}\;\to\;\Varid{y}\;\mathbin{\prec}\;\Varid{x}\;\to\;\Conid{Acc}\;\_\!\mathbin{\prec}\!\_\;\Varid{y})\;\to\;\Conid{Acc}\;\_\!\mathbin{\prec}\!\_\;\Varid{x}{}\<[E]%
\ColumnHook
\end{hscode}\resethooks
\end{Def}
To see that this is equivalent to the definition of well-foundedness in set theory,
recall that a relation \ensuremath{\_\!\mathbin{\prec}\!\_} on a set \ensuremath{\Varid{a}} is well-founded if and only if there is a monotone function from \ensuremath{\Varid{a}} to a well-founded order.
Since all inductive data types are well-founded,
and the termination checker ensures that the argument to \ensuremath{\Varid{acc}} is a monotone function,
there is a function from \ensuremath{\Varid{x}\;\mathbin{:}\;\Varid{a}} to \ensuremath{\Conid{Acc}\;\_\!\mathbin{\prec}\!\_\;\Varid{x}} if and only if \ensuremath{\_\!\mathbin{\prec}\!\_} is a well-founded relation in the set-theoretic sense.

The condition that all calls are made to a smaller argument is related to the notion of a loop \emph{variant}
in imperative languages.
While an invariant is a predicate that is true at the start and end of each looping step,
the variant is a relation that holds between successive looping steps.
\begin{Def}
Given a recursive definition \ensuremath{\Varid{f}\;\mathbin{:}\;\Conid{I}\overset{\Varid{es}}{\looparrowright}\Conid{O}},
a relation \ensuremath{\_\!\mathbin{\prec}\!\_} on \ensuremath{\Conid{C}} is a recursive \emph{variant} if for each argument \ensuremath{\Varid{c}},
and each recursive call made to \ensuremath{\Varid{c'}} in the evaluation of \ensuremath{\Varid{f}\;\Varid{c}},
we have \ensuremath{\Varid{c'}\;\mathbin{\prec}\;\Varid{c}}.
Formally:
\begin{hscode}\SaveRestoreHook
\column{B}{@{}>{\hspre}l<{\hspost}@{}}%
\column{3}{@{}>{\hspre}l<{\hspost}@{}}%
\column{5}{@{}>{\hspre}l<{\hspost}@{}}%
\column{7}{@{}>{\hspre}l<{\hspost}@{}}%
\column{E}{@{}>{\hspre}l<{\hspost}@{}}%
\>[3]{}\Varid{variant'}\;\mathbin{:}\;\!\!\;(\Varid{pts}\;\mathbin{:}\;\text{\itshape PTs}^S\;\Varid{s}\;(\Varid{mkSig}\;\Conid{C}\;\Conid{R}\;::\;\Varid{es}))\;(\Varid{f}\;\mathbin{:}\;\Conid{C}\overset{\Varid{es}}{\looparrowright}\Conid{R})\;{}\<[E]%
\\
\>[3]{}\hsindent{2}{}\<[5]%
\>[5]{}(\_\!\mathbin{\prec}\!\_\;\mathbin{:}\;(\Conid{C}\;\times\;\Varid{s})\;\to\;(\Conid{C}\;\times\;\Varid{s})\;\to\;\Conid{Set})\;{}\<[E]%
\\
\>[3]{}\hsindent{2}{}\<[5]%
\>[5]{}(\Varid{c}\;\mathbin{:}\;\Conid{C})\;(\Varid{t}\;\mathbin{:}\;\Varid{s})\;(\Conid{S}\;\mathbin{:}\;\Conid{Free}\;(\Varid{mkSig}\;\Conid{C}\;\Conid{R}\;::\;\Varid{es})\;\Varid{a})\;\to\;\Varid{s}\;\to\;\Conid{Set}{}\<[E]%
\\
\>[3]{}\Varid{variant'}\;\Varid{pts}\;\Varid{f}\;\_\!\mathbin{\prec}\!\_\;\Varid{c}\;\Varid{t}\;(\Conid{Pure}\;\Varid{x})\;\Varid{t'}\;\mathrel{=}\;\top{}\<[E]%
\\
\>[3]{}\Varid{variant'}\;\Varid{pts}\;\Varid{f}\;\_\!\mathbin{\prec}\!\_\;\Varid{c}\;\Varid{t}\;(\Conid{Op}\;\in\!\mathit{Head}\;\Varid{c'}\;\Varid{k})\;\Varid{t'}\;{}\<[E]%
\\
\>[3]{}\hsindent{2}{}\<[5]%
\>[5]{}\mathrel{=}\;((\Varid{c'}\;\Varid{,}\;\Varid{t'})\;\mathbin{\prec}\;(\Varid{c}\;\Varid{,}\;\Varid{t}))\;\times\;\Varid{lookupPTS}\;\Varid{pts}\;\in\!\mathit{Head}\;\Varid{c'}\;{}\<[E]%
\\
\>[5]{}\hsindent{2}{}\<[7]%
\>[7]{}(\lambda\;\Varid{x}\;\to\;\Varid{variant'}\;\Varid{pts}\;\Varid{f}\;\_\!\mathbin{\prec}\!\_\;\Varid{c}\;\Varid{t}\;(\Varid{k}\;\Varid{x}))\;\Varid{t'}{}\<[E]%
\\
\>[3]{}\Varid{variant'}\;\Varid{pts}\;\Varid{f}\;\_\!\mathbin{\prec}\!\_\;\Varid{c}\;\Varid{t}\;(\Conid{Op}\;(\in\!\mathit{Tail}\;\Varid{i})\;\Varid{c'}\;\Varid{k})\;\Varid{t'}\;{}\<[E]%
\\
\>[3]{}\hsindent{2}{}\<[5]%
\>[5]{}\mathrel{=}\;\Varid{lookupPTS}\;\Varid{pts}\;(\in\!\mathit{Tail}\;\Varid{i})\;\Varid{c'}\;(\lambda\;\Varid{x}\;\to\;\Varid{variant'}\;\Varid{pts}\;\Varid{f}\;\_\!\mathbin{\prec}\!\_\;\Varid{c}\;\Varid{t}\;(\Varid{k}\;\Varid{x}))\;\Varid{t'}{}\<[E]%
\\[\blanklineskip]%
\>[3]{}\Varid{variant}\;\mathbin{:}\;\!\!\;(\Varid{pts}\;\mathbin{:}\;\text{\itshape PTs}^S\;\Varid{s}\;(\Varid{mkSig}\;\Conid{C}\;\Conid{R}\;::\;\Varid{es}))\;(\Varid{f}\;\mathbin{:}\;\Conid{C}\overset{\Varid{es}}{\looparrowright}\Conid{R})\;\to\;{}\<[E]%
\\
\>[3]{}\hsindent{2}{}\<[5]%
\>[5]{}(\_\!\mathbin{\prec}\!\_\;\mathbin{:}\;(\Conid{C}\;\times\;\Varid{s})\;\to\;(\Conid{C}\;\times\;\Varid{s})\;\to\;\Conid{Set})\;\to\;\Conid{Set}{}\<[E]%
\\
\>[3]{}\Varid{variant}\;\!\!\;\!\!\;\!\!\;\!\!\;\Varid{pts}\;\Varid{f}\;\_\!\mathbin{\prec}\!\_\;\mathrel{=}\;\forall\;\Varid{c}\;\Varid{t}\;\to\;\Varid{variant'}\;\Varid{pts}\;\Varid{f}\;\_\!\mathbin{\prec}\!\_\;\Varid{c}\;\Varid{t}\;(\Varid{f}\;\Varid{c})\;\Varid{t}{}\<[E]%
\ColumnHook
\end{hscode}\resethooks
\end{Def}
Note that \ensuremath{\Varid{variant}} depends on the semantics \ensuremath{\Varid{pts}} we give to the recursive function \ensuremath{\Varid{f}}.
We cannot derive the semantics in \ensuremath{\Varid{variant}} from the structure of \ensuremath{\Varid{f}} as we do for the petrol-driven semantics,
since we do not yet know whether \ensuremath{\Varid{f}} terminates.
Using \ensuremath{\Varid{variant}}, we can define another termination condition on \ensuremath{\Varid{f}}:
there is a well-founded variant for \ensuremath{\Varid{f}}.
\begin{hscode}\SaveRestoreHook
\column{B}{@{}>{\hspre}l<{\hspost}@{}}%
\column{3}{@{}>{\hspre}l<{\hspost}@{}}%
\column{5}{@{}>{\hspre}l<{\hspost}@{}}%
\column{7}{@{}>{\hspre}l<{\hspost}@{}}%
\column{E}{@{}>{\hspre}l<{\hspost}@{}}%
\>[3]{}\Keyword{record}\;\Conid{Termination}\;\!\!\;(\Varid{pts}\;\mathbin{:}\;\text{\itshape PTs}^S\;\Varid{s}\;(\Varid{mkSig}\;\Conid{C}\;\Conid{R}\;::\;\Varid{es}))\;(\Varid{f}\;\mathbin{:}\;\Conid{C}\overset{\Varid{es}}{\looparrowright}\Conid{R})\;\mathbin{:}\;\Conid{Set}\;\Keyword{where}{}\<[E]%
\\
\>[3]{}\hsindent{2}{}\<[5]%
\>[5]{}\Keyword{field}{}\<[E]%
\\
\>[5]{}\hsindent{2}{}\<[7]%
\>[7]{}\_\!\mathbin{\prec}\!\_\;\mathbin{:}\;(\Conid{C}\;\times\;\Varid{s})\;\to\;(\Conid{C}\;\times\;\Varid{s})\;\to\;\Conid{Set}{}\<[E]%
\\
\>[5]{}\hsindent{2}{}\<[7]%
\>[7]{}\Varid{w-f}\;\mathbin{:}\;\forall\;\Varid{c}\;\Varid{t}\;\to\;\Conid{Acc}\;\_\!\mathbin{\prec}\!\_\;(\Varid{c}\;\Varid{,}\;\Varid{t}){}\<[E]%
\\
\>[5]{}\hsindent{2}{}\<[7]%
\>[7]{}\Varid{var}\;\mathbin{:}\;\Varid{variant}\;\Varid{pts}\;\Varid{f}\;\_\!\mathbin{\prec}\!\_{}\<[E]%
\ColumnHook
\end{hscode}\resethooks
A generally recursive function that terminates in the petrol-driven semantics also has a well-founded variant,
given by the well-order \ensuremath{\_\!\mathbin{<}\!\_} on the amount of fuel consumed by each call.
The converse also holds: if we have a descending chain of calls \ensuremath{\Varid{cs}} after calling \ensuremath{\Varid{f}} with argument \ensuremath{\Varid{c}},
we can use induction on the type \ensuremath{\Conid{Acc}\;\_\!\mathbin{\prec}\!\_\;\Varid{c}} to bound the length of \ensuremath{\Varid{cs}}.
This bound gives the amount of fuel consumed by evaluating a call to \ensuremath{\Varid{f}} on \ensuremath{\Varid{c}}.

In our case, the relation \ensuremath{\Conid{RecOrder}} will work as a recursive variant for \ensuremath{\Varid{fromProds}}:
\begin{hscode}\SaveRestoreHook
\column{B}{@{}>{\hspre}l<{\hspost}@{}}%
\column{3}{@{}>{\hspre}l<{\hspost}@{}}%
\column{5}{@{}>{\hspre}l<{\hspost}@{}}%
\column{7}{@{}>{\hspre}l<{\hspost}@{}}%
\column{E}{@{}>{\hspre}l<{\hspost}@{}}%
\>[3]{}\Keyword{data}\;\Conid{RecOrder}\;(\Varid{prods}\;\mathbin{:}\;\Conid{Prods})\;\mathbin{:}\;(\Varid{x}\;\Varid{y}\;\mathbin{:}\;\Conid{Nonterm}\;\times\;\Conid{String})\;\to\;\Conid{Set}\;\Keyword{where}{}\<[E]%
\\
\>[3]{}\hsindent{2}{}\<[5]%
\>[5]{}\Conid{Left}\;\mathbin{:}\;\!\!\;\Varid{length}\;\Varid{str}\;\mathbin{<}\;\Varid{length}\;\Varid{str'}\;\to\;{}\<[E]%
\\
\>[5]{}\hsindent{2}{}\<[7]%
\>[7]{}\Conid{RecOrder}\;\Varid{prods}\;(\Conid{A}\;\Varid{,}\;\Varid{str})\;(\Conid{B}\;\Varid{,}\;\Varid{str'}){}\<[E]%
\\
\>[3]{}\hsindent{2}{}\<[5]%
\>[5]{}\Conid{Right}\;\mathbin{:}\;\!\!\;\Varid{length}\;\Varid{str}\;\mathbin{\le}\;\Varid{length}\;\Varid{str'}\;\to\;{}\<[E]%
\\
\>[5]{}\hsindent{2}{}\<[7]%
\>[7]{}\Conid{LRec}\;\Varid{prods}\;\Conid{A}\;\Conid{B}\;\to\;\Conid{RecOrder}\;\Varid{prods}\;(\Conid{A}\;\Varid{,}\;\Varid{str})\;(\Conid{B}\;\Varid{,}\;\Varid{str'}){}\<[E]%
\ColumnHook
\end{hscode}\resethooks
With the definition of \ensuremath{\Conid{RecOrder}}, we can complete the correctness proof of \ensuremath{\Varid{fromProds}},
by giving an element of the corresponding \ensuremath{\Conid{Termination}} type.
We assume that the length of recursion is bounded by \ensuremath{\Varid{bound}\;\mathbin{:}\;\N}.
\begin{hscode}\SaveRestoreHook
\column{B}{@{}>{\hspre}l<{\hspost}@{}}%
\column{3}{@{}>{\hspre}l<{\hspost}@{}}%
\column{5}{@{}>{\hspre}l<{\hspost}@{}}%
\column{E}{@{}>{\hspre}l<{\hspost}@{}}%
\>[3]{}\Varid{fromProdsTerminates}\;\mathbin{:}\;\forall\;\Varid{prods}\;\Varid{bound}\;\to\;\Varid{leftRecBound}\;\Varid{prods}\;\Varid{bound}\;\to\;{}\<[E]%
\\
\>[3]{}\hsindent{2}{}\<[5]%
\>[5]{}\Conid{Termination}\;(\Varid{pts}\;\Varid{prods})\;(\Varid{fromProds}\;\Varid{prods}){}\<[E]%
\\
\>[3]{}\Conid{Termination.}\_\!\mathbin{\prec}\!\_\;(\Varid{fromProdsTerminates}\;\Varid{prods}\;\Varid{bound}\;\Conid{H})\;\mathrel{=}\;\Conid{RecOrder}\;\Varid{prods}{}\<[E]%
\ColumnHook
\end{hscode}\resethooks
To show that the relation \ensuremath{\Conid{RecOrder}} is well-founded,
we need to show that there is no infinite descending chain starting from some non-terminal \ensuremath{\Conid{A}} and string \ensuremath{\Varid{str}}.
The proof is based on iteration on two natural numbers \ensuremath{\Varid{n}} and \ensuremath{\Varid{k}},
which form an upper bound on the number of allowed left-recursive calls in sequence and unconsumed characters in the string respectively.
Note that the number \ensuremath{\Varid{bound}} is an upper bound for \ensuremath{\Varid{n}} and the length of the input string is an upper bound for \ensuremath{\Varid{k}}.
Since each non-terminal in the production will decrease \ensuremath{\Varid{n}} and each terminal will decrease \ensuremath{\Varid{k}},
we eventually reach the base case \ensuremath{\Varid{0}} for either.
If \ensuremath{\Varid{n}} is zero, we have made more than \ensuremath{\Varid{bound}} left-recursive calls, contradicting the assumption that we have bounded left recursion.
If \ensuremath{\Varid{k}} is zero, we have consumed more than \ensuremath{\Varid{length}\;\Varid{str}} characters of \ensuremath{\Varid{str}}, also a contradiction.
\begin{hscode}\SaveRestoreHook
\column{B}{@{}>{\hspre}l<{\hspost}@{}}%
\column{3}{@{}>{\hspre}l<{\hspost}@{}}%
\column{5}{@{}>{\hspre}l<{\hspost}@{}}%
\column{7}{@{}>{\hspre}l<{\hspost}@{}}%
\column{E}{@{}>{\hspre}l<{\hspost}@{}}%
\>[3]{}\Conid{Termination.w-f}\;(\Varid{fromProdsTerminates}\;\Varid{prods}\;\Varid{bound}\;\Conid{H})\;\Conid{A}\;\Varid{str}\;{}\<[E]%
\\
\>[3]{}\hsindent{2}{}\<[5]%
\>[5]{}\mathrel{=}\;\Varid{acc}\;(\Varid{go}\;\Conid{A}\;\Varid{str}\;(\Varid{length}\;\Varid{str})\;\text{\itshape $\le$-refl}\;\Varid{bound}\;\Conid{Nil}\;\text{\itshape $\le$-refl}){}\<[E]%
\\
\>[3]{}\hsindent{2}{}\<[5]%
\>[5]{}\Keyword{where}{}\<[E]%
\\
\>[3]{}\hsindent{2}{}\<[5]%
\>[5]{}\Varid{go}\;\mathbin{:}\;\!\!\;\forall\;\Conid{A}\;\Varid{str}\;\to\;{}\<[E]%
\\
\>[5]{}\hsindent{2}{}\<[7]%
\>[7]{}(\Varid{k}\;\mathbin{:}\;\N)\;\to\;\Varid{length}\;\Varid{str}\;\mathbin{\le}\;\Varid{k}\;\to\;{}\<[E]%
\\
\>[5]{}\hsindent{2}{}\<[7]%
\>[7]{}(\Varid{n}\;\mathbin{:}\;\N)\;(\Varid{cs}\;\mathbin{:}\;\Conid{Chain}\;\Varid{prods}\;\Conid{A}\;\Conid{B})\;\to\;\Varid{bound}\;\mathbin{\le}\;\Varid{chainLength}\;\Varid{cs}\;\mathbin{+}\;\Varid{n}\;\to\;{}\<[E]%
\\
\>[5]{}\hsindent{2}{}\<[7]%
\>[7]{}\forall\;\Varid{y}\;\to\;\Conid{RecOrder}\;\Varid{prods}\;\Varid{y}\;(\Conid{A}\;\Varid{,}\;\Varid{str})\;\to\;\Conid{Acc}\;(\Conid{RecOrder}\;\Varid{prods})\;\Varid{y}{}\<[E]%
\ColumnHook
\end{hscode}\resethooks

Our next goal is that \ensuremath{\Conid{RecOrder}} is a variant for \ensuremath{\Varid{fromProds}}, as abbreviated by the \ensuremath{\Varid{prodsVariant}} type.
We cannot follow the definitions of \ensuremath{\Varid{fromProds}} as closely as we did for the partial correctness proof;
instead we need a complicated case distinction to keep track of the left-recursive chain we have followed in the proof.
For this reason, we split the \ensuremath{\Varid{parseStep}} apart into two lemmas \ensuremath{\Varid{parseStepAdv}} and \ensuremath{\Varid{parseStepRec}}, both showing that \ensuremath{\Varid{buildParser}} maintains the variant.
We also use a \ensuremath{\Varid{filterStep}} lemma that calls the correct \ensuremath{\Varid{parseStep}} for each production in the nondeterministic choice.
\begin{hscode}\SaveRestoreHook
\column{B}{@{}>{\hspre}l<{\hspost}@{}}%
\column{5}{@{}>{\hspre}l<{\hspost}@{}}%
\column{7}{@{}>{\hspre}l<{\hspost}@{}}%
\column{9}{@{}>{\hspre}l<{\hspost}@{}}%
\column{E}{@{}>{\hspre}l<{\hspost}@{}}%
\>[5]{}\Varid{prodsVariant}\;\mathrel{=}\;\Varid{variant'}\;(\Varid{pts}\;\Varid{prods})\;(\Varid{fromProds}\;\Varid{prods})\;(\Conid{RecOrder}\;\Varid{prods}){}\<[E]%
\\[\blanklineskip]%
\>[5]{}\Varid{parseStepAdv}\;\mathbin{:}\;\forall\;\Conid{A}\;\Varid{xs}\;\Varid{str}\;\Varid{str'}\;\to\;\Varid{length}\;\Varid{str'}\;\mathbin{<}\;\Varid{length}\;\Varid{str}\;\to\;{}\<[E]%
\\
\>[5]{}\hsindent{2}{}\<[7]%
\>[7]{}\Varid{prodsVariant}\;\Conid{A}\;\Varid{str}\;(\Varid{buildParser}\;\!\!\;\Varid{xs})\;\Varid{str'}{}\<[E]%
\\
\>[5]{}\Varid{parseStepRec}\;\mathbin{:}\;\forall\;\Conid{A}\;\Varid{xs}\;\Varid{str}\;\Varid{str'}\;\to\;\Varid{length}\;\Varid{str'}\;\mathbin{\le}\;\Varid{length}\;\Varid{str}\;\to\;{}\<[E]%
\\
\>[5]{}\hsindent{2}{}\<[7]%
\>[7]{}\forall\;\Varid{ys}\;\!\!\;\to\;\Varid{prod}\;\Conid{A}\;(\Varid{map}\;\Conid{Inr}\;\Varid{ys}\;\plus \;\Varid{xs})\;\Varid{sem}\;\in\;\Varid{prods}\;\to\;{}\<[E]%
\\
\>[5]{}\hsindent{2}{}\<[7]%
\>[7]{}\Varid{prodsVariant}\;\Conid{A}\;\Varid{str}\;(\Varid{buildParser}\;\!\!\;\Varid{xs})\;\Varid{str'}{}\<[E]%
\\
\>[5]{}\Varid{filterStep}\;\mathbin{:}\;\forall\;\Varid{prods'}\;\to\;(\!\!\;\Varid{x}\;\in\;\Varid{prods'}\;\to\;\Varid{x}\;\in\;\Varid{prods})\;\to\;{}\<[E]%
\\
\>[5]{}\hsindent{2}{}\<[7]%
\>[7]{}\forall\;\Conid{A}\;\Varid{str}\;\Varid{str'}\;\to\;\Varid{length}\;\Varid{str'}\;\mathbin{\le}\;\Varid{length}\;\Varid{str}\;\to\;{}\<[E]%
\\
\>[5]{}\hsindent{2}{}\<[7]%
\>[7]{}\Varid{prodsVariant}\;\Conid{A}\;\Varid{str}\;{}\<[E]%
\\
\>[7]{}\hsindent{2}{}\<[9]%
\>[9]{}(\Varid{foldr}\;(\Varid{choice}\;\!\!)\;(\Varid{fail}\;\!\!)\;(\Varid{map}\;\Varid{fromProd}\;(\Varid{filterLHS}\;\Conid{A}\;\Varid{prods'})))\;{}\<[E]%
\\
\>[5]{}\hsindent{2}{}\<[7]%
\>[7]{}\Varid{str'}{}\<[E]%
\ColumnHook
\end{hscode}\resethooks
In the \ensuremath{\Varid{parseStepAdv}}, we deal with the situation that the parser has already consumed at least one character since it was called.
This means we can repeatedly use the \ensuremath{\Conid{Left}} constructor of \ensuremath{\Conid{RecOrder}} to show the variant holds.

In the \ensuremath{\Varid{parseStepRec}}, we deal with the situation that the parser has only encountered non-terminals in the current production.
This means that we can use the \ensuremath{\Conid{Right}} constructor of \ensuremath{\Conid{RecOrder}} to show the variant holds until we consume a character,
after which we call \ensuremath{\Varid{parseStepAdv}} to finish the proof.

The lemma \ensuremath{\Varid{filterStep}} shows that the variant holds on all subsets of the production rules,
analogously to the \ensuremath{\Varid{filterStep}} of the partial correctness proof.
It calls \ensuremath{\Varid{parseStepRec}} since the parser only starts consuming characters after it selects a production rule.
\begin{hscode}\SaveRestoreHook
\column{B}{@{}>{\hspre}l<{\hspost}@{}}%
\column{5}{@{}>{\hspre}l<{\hspost}@{}}%
\column{7}{@{}>{\hspre}l<{\hspost}@{}}%
\column{9}{@{}>{\hspre}l<{\hspost}@{}}%
\column{E}{@{}>{\hspre}l<{\hspost}@{}}%
\>[5]{}\Varid{filterStep}\;\Conid{Nil}\;\Conid{A}\;\Varid{str}\;\Varid{str'}\;\Varid{lt}\;\Varid{subset}\;\mathrel{=}\;\Varid{tt}{}\<[E]%
\\
\>[5]{}\Varid{filterStep}\;(\Varid{prod}\;\Varid{lhs}\;\Varid{rhs}\;\Varid{sem}\;::\;\Varid{prods'})\;\Varid{subset}\;\Conid{A}\;\Varid{str}\;\Varid{str'}\;\Varid{lt}\;\Keyword{with}\;\Conid{A}\;\mathbin{\smash{\overset{\raisebox{-0.2em}{\tiny ?}}{=}}}\;\Varid{lhs}{}\<[E]%
\\
\>[5]{}\Varid{...}\;\mid \;\Varid{yes}\;\Varid{refl}\;{}\<[E]%
\\
\>[5]{}\hsindent{2}{}\<[7]%
\>[7]{}\mathrel{=}\;\Varid{variant-fmap}\;(\Varid{pts}\;\Varid{prods})\;(\Varid{fromProds}\;\Varid{prods})\;(\Varid{buildParser}\;\Varid{rhs})\;{}\<[E]%
\\
\>[7]{}\hsindent{2}{}\<[9]%
\>[9]{}(\Varid{parseStepRec}\;\Conid{A}\;\Varid{rhs}\;\Varid{str}\;\Varid{str'}\;\Varid{lt}\;\Conid{Nil}\;(\Varid{subset}\;\in\!\mathit{Head}))\;{}\<[E]%
\\
\>[7]{}\hsindent{2}{}\<[9]%
\>[9]{}\Varid{,}\;\Varid{filterStep}\;\Varid{prods'}\;(\Varid{subset}\;\circ\;\in\!\mathit{Tail})\;\Conid{A}\;\Varid{str}\;\Varid{str'}\;\Varid{lt}{}\<[E]%
\\
\>[5]{}\Varid{...}\;\mid \;\Varid{no}\;\Varid{¬p}\;\mathrel{=}\;\Varid{filterStep}\;\Varid{prods'}\;(\Varid{subset}\;\circ\;\in\!\mathit{Tail})\;\Conid{A}\;\Varid{str}\;\Varid{str'}\;\Varid{lt}{}\<[E]%
\ColumnHook
\end{hscode}\resethooks
As for partial correctness, we obtain the proof of termination by applying \ensuremath{\Varid{filterStep}} to the subset of \ensuremath{\Varid{prods}} consisting of \ensuremath{\Varid{prods}} itself.

\fi

\end{document}